\documentclass[12pt,preprint,deluxe]{aastex}

\usepackage{epsfig}

\begin{document}
\voffset-1cm

\newcommand{\gsim}{\hbox{\rlap{$^>$}$_\sim$}}
\newcommand{\lsim}{\hbox{\rlap{$^<$}$_\sim$}}
%\flushright{CERN-TH/2003-137}
\title{Towards a complete theory of Gamma Ray Bursts}

\author{Arnon Dar\altaffilmark{1} and
A. De R\'ujula\altaffilmark{2}}

\altaffiltext{1}{arnon@physics.technion.ac.il, 
dar@cern.ch.\\
Physics Department and Space Research Institute, Technion, Haifa 32000, 
Israel and\\Theory Division, CERN, CH-1211 Geneva 23, Switzerland}
\altaffiltext{2}{alvaro.derujula@cern.ch.
Theory Division, CERN, CH-1211 Geneva 23, Switzerland.\\
Physics Dept. Boston University, USA{\flushright{CERN-TH/2003-137}}}
%\altaffiltext{0}{\flushright{CERN-TH/2003-137}}

% \maketitle

\begin{abstract} 

Gamma Ray Bursts (GRBs) are notorious for their diversity. Yet, they have
a series of common features. The typical energy of their $\gamma$ rays is
a fraction of an MeV.  The energy distributions are well
described by a ``Band spectrum'', with ``peak energies'' spanning 
a surprisingly narrow range. The time structure of a GRB consists of pulses,
superimposed or not, rising and decreasing fast. The number of photons in
a pulse, the pulses' widths and their total energy vary within broad but
specific ranges. Within a pulse, the energy spectrum softens with increasing
time. The duration of a pulse decreases at higher energies and its peak
intensity shifts to earlier time. Many other correlations between pairs of GRB
observables have been identified. Last (and based on one measured event!) 
the $\gamma$-ray polarization is very large. A satisfactory theory of GRBs
should naturally and very simply explain, among others, all these facts.
We show that the ``cannonball" (CB) model does it.  In the CB model the
process leading to the ejection of highly relativistic jetted CBs in
core-collapse supernova (SN) explosions is akin to the one observed in
quasars and microquasars.  The prompt $\gamma$-ray emission ---the GRB---
is explained extremely well by inverse Compton scattering of light in the
near environment of the SN by the electrons in the CBs' plasma. We have
previously shown that the CB-model's description of GRB afterglows as
synchrotron radiation from ambient electrons ---swept in and accelerated
within the CBs--- is also simple, universal and very successful.  The only
obstacle still separating the CB model from a complete theory of GRBs is
the theoretical understanding of the CBs' ejection mechanism in SN
explosions.

\end{abstract}


\section{Introduction}

Once upon a time there were ---literally--- some five score and seven
theories of gamma-ray bursts (GRBs, see e.g.~Nemiroff 1994). In this {\it
Dark Era}, the observations were scarce, and they were the exclusive realm
of a few satellites (VELA, GRANAT, SMM, GINGA).  During a
good fraction of the last decade of the past century, the {\it BATSE Era},
a single detector ---the Burst And Transient Satellite Experiment aboard
the Compton Gamma Ray Observatory (CGRO) satellite--- dominated the
data-taking effort (see e.g.~Fishman \& Meegan 1995). The BATSE team
determined that the distribution of GRB arrival directions was extremely
isotropic in the sky, a very strong hint that their sources were
``cosmological'' (Meegan et al.~1992). This reduced the number of tenable
GRB theories by a very large factor.

In parallel, or subsequently to the very successful CGRO mission, various
$\gamma$- and X-ray satellites ---BeppoSAX, Rossi, HETE II, Integral and 
the
Inter-Planetary Network of spacecrafts (Wind, PVO, Ulysses, Mars Odyssey 
and
RHESSI)--- were operational and capable of performing faster and more
precise directional localizations of GRBs. Following the consequent
discovery that the sources of GRBs continue to shine after their transient
high-energy pulses (Costa et al.~1997), a quantum leap of information took
place. Indeed, a GRB ``event'' does not end as the $\gamma$-ray flux
becomes undetectably small, for there is an ``afterglow'' (AG): the source
continues to emit light at all smaller observable frequencies, ranging
from X-rays to radio waves, 
and to be observable for months, or even years
(van Paradijs et al.~1997;  Frail et al.~1997). In the {\it Afterglow Era}
(see e.g.~Proc.~{\it GRBs in the Afterglow Era,} 1999, A\&AS, Vol.~138),
the fact that these remaining emissions can be very well localized in the
sky led to the discovery of the GRBs' host galaxies (Sahu et
al.~1997); to the measurement of their redshifts (Metzger et al.~1997)
that verified their cosmological origin; to the identification of their
birthplaces ---mainly star formation regions in normal galaxies (Holland
\& Hjorth 1999)--- and to the first evidence for a possible physical
association between GRBs and supernova explosions:
that of GRB 980425 and SN1998bw  (Galama et al.~1998).

The possibility to observe the relatively intense early optical AGs, even
with rather small telescopes, has had the beneficial effect of enlarging
the ``GRB community'' well beyond its previous bounds. A recent, serendipitous
 and most-welcome newcomer to the GRB observational community
was the satellite RHESSI (Reuven Ramaty High
Energy Solar Spectroscopic Imager satellite).
Looking with this instrument close to the Sun's direction,
Coburn and Boggs (2003) discovered GRB 021206,
and measured a very large linear polarization of
its prompt $\gamma$-rays: $\Pi= (80\pm 20)\,\%$. 
This polarization is much
higher than the few per cent
values observed in the optical AG of GRBs
(GRB 990123: Hjorth et al.~1999;  
GRB 990510: Wijers et al.~1999; Covino et al.~1999;
GRB 990712: Rol et al.~2000;  
GRB 010222: Bjornsson et al.~2002;  
GRB 011211: Covino et al.~2002; 
GRB 020405: Bersier et al.~2003a; Masetti et al.~2003; Covino et al.~2003a;
GRB 020813: Covino et al.~2003b; 
GRB 021004: Rol et al.~2003; Wang et al.~2003; 
GRB 030329: Efimov et al.~2003; Magalhaes et al.~2003; Covino et al.~2003c).

The influence of the discovery of AGs on the theory of GRBs has been
enormous, and not only because we now know that their sources are
cosmological, and must ---somehow--- be related to stars. It is very very
difficult to imagine a GRB theory in which there is no sequel to a GRB. In
the {\it fireball model} 
(the FB model, in its many variants: Paczynski 1986; Goodman 1986; Shemi \&
Piran 1990;  Narayan, Paczynski \& Piran 1992;  Rees \&
M\'esz\'aros 1992, 1994; Katz 1994a,b; M\'esz\'aros \& Rees 1997; 
Waxman 1997a,b;  Dermer \& Mitman 1999;
for reviews see, e.g.~Piran 1999, 2000; M\'esz\'aros 2002; Hurley, Sari \&
Djorgovski 2002; Waxman 2003a), long a leading contender for consideration
as {\it the} theory of GRBs, the existence of AGs declining in intensity
as an inverse power of time\footnote{AGs do not decline as a simple (or
single) power of time, one of the reasons why the fireball models have 
evolved
into ``firecone'' models (Rhoads 1999; Sari, Piran \& Halpern 1999).}
had been anticipated (Katz 1994b; M\'esz\'aros \& Rees 1997;
M\'esz\'aros, Rees 1997 \& Wijers 1998).
In the AG
era, the FB model rose in consideration to become generally accepted as
{\it the standard model} of GRBs. Radically different models, such as our
``cannonball" (CB) model (Dar \& De R\'ujula 2000a,b;  Dado, Dar \& De
R\'ujula 2002a, 2003a), were received with considerable skepticism (De
R\'ujula 2003). Even small deviations from the prevailing credo (Dermer \&
Mitman 1999) met a similar fate (Dermer 2002).

In the FB models, both the prompt $\gamma$-rays and the AG are due to
synchrotron radiation from shock-accelerated electrons moving in a chaotic
magnetic field. Thus, their very different polarizations were not
expected (see, e.g.~Gruzinov 1999; Gruzinov \& Waxman 1999). A large 
polarization 
requires at first sight an ad-hoc magnetic
and/or jet structure (Lyutikov, Pariev \& Blandford 2003; Waxman 2003b;
Eichler \& Levinson 2003; Nakar, Piran \& Waxman 2003) and is ---to say 
the least--- quite a surprise.

The $\gamma$-rays of a GRB may not be produced by synchrotron radiation
and their polarization may not necessarily imply a strong, large-scale,
ordered magnetic field in their source.  In fact, Shaviv and Dar (1995)
suggested that highly relativistic, narrowly collimated jets ejected near
the line of sight in accretion-induced collapse of stars in distant
galaxies may produce cosmological GRBs by inverse Compton scattering (ICS)
of stellar light.  If the Lorentz factor of the jet is $\gamma\sim 10^3$,
ICS of isotropic unpolarized stellar light by the electrons in the jet
boosts the photons to $\gamma$-ray energies, beams them along the
direction of motion of the jet, and results in a large polarization
($\Pi\approx 100\%$ when the jet is viewed close to the most probable
viewing angle, $\theta\sim 1/\gamma$). But the density of the radiation
field, even in the most dense star-burst regions, was found to be
insufficient to explain the $\gamma$-ray fluence of the most powerful
GRBs, such as GRB 990213. We shall see that, in the CB model, this
problem ---the dearth of ``target'' light--- does not arise.

In the CB model, long-duration GRBs are made by core-collapse supernovae
(SNe). As we asserted in Dar \& De R\'ujula (2000a) ``the light from 
the SN shell is Compton up-scattered to MeV energies, but its contribution
to a GRB is sub-dominant''. That assertion is correct: the light
from the SN shell is too underluminous and too radially directed
to generate GRBs of the observed fluence and individual-photon
energy. With our collaborator Shlomo Dado, we have developed a
very complete, simple and ---we contend--- extremely successful analysis of
GRB AGs (Dado, Dar \& De   
R\'ujula 2002a,b,c, 2003a,b,c,d,e,f). This thorough
analysis has taught us that there should be another, 
much more intense and more isotropic, source of scattered light:
the SN's ``glory''. The glory is the ``echo'' (or {\it ambient}) light 
from the SN, permeating the ``wind-fed''
circumburst density profile, previously ionized 
by the early extreme UV flash accompanying a SN explosion, or by the
enhanced UV emission that precedes it. In Sections 2 and 3
we summarize the observations of pre-SN winds, early SN luminosities, 
and the UV flashes of SNe, to obtain the reference values of the
very early quantities  of interest: the wind's density $\rho(r)$,
and density profile (roughly $1/r^2$), and the very early SN luminosity.
These, and other quantities of interest here, are listed in Table 1.

The CBs of the CB model are inspired by the ones observed in quasar and
microquasar emissions. One example of the latter is shown in the upper
panel of Fig.~(\ref{CBGlory}), showing two opposite CBs emitted by 
the microquasar XTE
J1550-564 (Kaaret et al.~2003). The winds and echoes of GRB-generating SNe
are akin to those emitted and illuminated by some very massive stars. The
light echo (or glory) of the stellar outburst of the red supergiant
V3838 Monocerosis in early January 2002 is shown in the lower panel of
Fig.~(\ref{CBGlory}), from Bond et al.~(2003). In a sense all we are doing
in this paper is to superimpose the two halves of Fig.~(\ref{CBGlory}), 
and to work out in detail
what the consequences ---based exclusively on Compton scattering--- are.

The varied time structures of GRB $\gamma$-ray number-counts
generally consist of fast rising and declining
isolated or partially superimposed pulses. 
We show here that ---in a CB model
in which ICS of the wind's {\it ambient light}
is the dominant $\gamma$-ray-generating mechanism---
the following observed properties of long-duration GRBs naturally follow
(for the reader's convenience, we also list here the equations
and figures corresponding to the respective predictions):
\begin{itemize}
\item{}
The large polarization of the $\gamma$ rays (Coburn \& Boggs 2003),
Eq.~(\ref{polSN})  and Fig.~(\ref{f2}).
\item{}
The characteristic energy $E={\cal{O}}(250)$ keV of the $\gamma$ rays
(Preece et al.~2000; Amati et al.~2002), Eq.~(\ref{boosting}).
\item{} 
The narrow distribution of the ``peak'' or ``bend'' energies of the GRB
spectra (e.g.~Preece et al.~2000), Eq.~(\ref{boosting}) and Fig.~(\ref{figEpobs}).
\item{}
The duration of the single pulses of GRBs: a median $\Delta t\sim 1/2$ s 
full width at half-maximum (McBreen et al.~2002), 
Eqs.~(\ref{ttrans}) and (\ref{FWHM}).
\item{}
The characteristic (spherical equivalent) number of photons per pulse, 
$N_\gamma\sim 10^{59}$
on average, which, combined with the characteristic 
$\gamma$ energy, 
yields the average total (spherical equivalent) fluence of a GRB 
pulse: $\sim 10^{53}$ erg,
Eqs.~(\ref{Niso}), (\ref{Eiso}), (\ref{NUMTOT}), (\ref{Eprime}). 
\item{}
The general {\it FRED} pulse-shape: a very ``fast rise'' followed by a fast 
decay $N(t)\propto 1/t^2$,
inaccurately called ``exponential decay", perhaps
because {\it FRPD} is unpronounceable 
(Nemiroff et al.~1993, 1994;         
Link \& Epstein 1996; McBreen et al.~2002), Eq.~(\ref{pheno})
and Fig.~(\ref{fig3}).
\item{}
The $\gamma$-ray energy distribution, $dN/dE\sim E^{-\alpha}$,
with, on average, $\alpha\sim 1$ exponentially evolving into $\alpha\sim 2.1$
 and generally well fitted by the ``Band function''
(Band et al.~1993), our derived version of which is Eq.~(\ref{totdist}).
The theoretical and Band spectra are compared in Fig.~(\ref{figband}).
\item{}
The time--energy correlation of the pulses: the pulse duration 
decreases like $\sim E^{-0.4}$ and peaks earlier the higher the energy 
interval
(e.g.~Fenimore et al.~1995; Norris et al.~1996; Ramirez-Ruiz \& 
Fenimore 2000; Wu 
\& Fenimore 2000); the spectrum gets softer
as time elapses during a pulse (Golenetskii et al.~1983;  
Bhat et al.~1994), Eqs.~(\ref{pevol}) to
%,\ref{dEdt1},\ref{dEdt},
(\ref{fenimore}) and 
Figs.~(\ref{Noft1}) to (\ref{figSpectra2}).
\item{}
Various correlations between pairs of the following observables:
photon fluence, energy fluence, peak
intensity and luminosity, photon energy at peak intensity or luminosity,
and pulse duration (e.g.~Mallozzi et al.~1995; Liang \& 
Kargatis 1996;
Crider et al.~1999; Lloyd, Petrosian \& Mallozzi~2000; Ramirez-Ruiz \& 
Fenimore 2000;
McBreen et al.~2002; Kocevski et al.~2003), Eqs.~(\ref{fenimore}) to
%, \ref{epwidth},\ref{epint}, \ref{epflu}, \ref{epeiso}, \ref{epeiso}, 
(\ref{epilum}) and Figs.~(\ref{figFWRise},\ref{figalphabeta}) and
(\ref{figLogWLogE}) to (\ref{figEpVar}). 
\item{} The completion of the
demonstration that both GRB 980425 and its associated SN1998bw were in no
way exceptional, summarized by Figs.~(\ref{fig425spectra},\ref{fig425}).
\end{itemize}

We have organized this paper in order of the increasing amount
of algebra required to derive the results. But for the last two items,
the ensuing order is that of the above list. By far the largest amount
of algebra, but the one leading to one of our most detail-independent, 
---i.e. ``first-principled''--- results, is the one involved in the theoretical
derivation of the Band spectrum. 

For the sake of hypothetical readers not very familiar with the field, we
also include various comparative discussions of the FB models and the CB
model. Section 19 is a brief review of the FB-models' results on the
$\gamma$-rays of GRBs.  Shocks are a fundamental building-block of the FB
models, while in the CB model they play no role. The substance of the
shells responsible GRBs is, in the FB models, an $e^+e^-$ plasma with a
fine-tuned ``baryon load''. The substance of CBs is ordinary matter. In
Appendix I we review the observational situation regarding these two
issues in the realm of the other relativistic jets observed in nature: the
ejecta of quasars and microquasars.  We devote Appendix II to a short
commentary on GRB AGs ---as described by the FB and CB models. 
But for the last item above, we have nothing new to add in this paper
on the subject of the association of GRBs
and SNe. But the question is important because ---in the CB model--- SNe
are {\it the progenitors} of GRBs, and because ---in the FB models--- the
GRB/SN association is gaining importance, subsequent to the discovery of
the pair GRB 030329/SN2003dh (Stanek et al.~2003; Hjorth et al.~2003). We
review very briefly this subject, within the CB model, in Section 4.4. The
history of the idea and its observational support are summarized in
Appendix III.

\section{The ``wind" environment of SNe}

Two SNe play a particularly important role in this chapter: SN1987A,
famous for its proximity to us and for the neutrinos its core-collapse
emitted, and SN1998bw, famous for
its association with GRB 980425, and also for its relative proximity.

Massive stars lose mass throughout their life in the form of slow and fast
winds, and die in SN explosions.  The ejected stellar material has been
detected as circumstellar nebulae around Wolf--Rayet stars, luminous blue
variables, and blue and red supergiants, such as
V3838 Monocerosis, shown in the lower panel of Fig.~(\ref{CBGlory}).
The ejections feed and compress the
nebulae into dense shells, which are ionized by the UV fluxes emitted by
the stars.  Ionized circumstellar nebulae surrounding young supernova
remnants have also been detected around Cas A, extending to a distance of
$\sim 7$ pc (7' at a distance of 3.4 pc) by Fesen, Becker \& Blair (1987)
and around SN1987A. The nebula of SN1987A, observed as a {\it dust echo},
ends in a patchy shell of radius 4.5 pc (Chevalier \& Emmering 1989).

The mass-loss rate from SN progenitors intensifies in their late
evolutionary stages, which are not fully understood, in
particular shortly before the explosion
(e.g.~Podsiadlowski 1992; Chugai 1997a,b;
Chu, Weis \& Garnett 1999; Chevalier \& Oishi 2003). 
The observations of very narrow
P-Cygni profiles superposed on the broad emission H$\alpha$ and H$\beta$
lines of the SN ejecta in some young SN remnants ---e.g.~in SN1997ab 
(Salamanca et al.~1998), SN1997cy (Turatto et al.~2000), SN1998S (Fassia 
et al.~2001), SN1997eg (Salamanca, Terlevich \& Tenorio-Tagle
2002), SN1994W (Chugai et 
al.~2003) and SN1995G (Chugai \& Danziger 2003)--- indicate very high wind
particle-number densities, 
$n\sim 5\times 10^{7}$ cm$^{-3}$, at the distances of ${\cal{O}}(10^{16})$
cm of interest to the production of GRBs in the CB model. The measured $n$
declines roughly as $1/r^2$ and its corresponding {\it ``surface
density''} is 
\begin{equation}
\rho\, r^2 \sim 10^{16}\; {\rm g\; cm^{-1}}.
\label{surfdens}
\end{equation}
This ``close-by''
result is two orders of magnitude larger than the one observed for
``canonical'' winds of massive stars with a typical mass loss rate
$\dot{M}\sim 10^{-4}\, M_\odot$, and a typical wind velocity $V\sim 100$
km s$^{-1}$, which yield $\rho\, r^2 = \dot{M}/4\, \pi\, V\,\sim 5\times
10^{13}$ g cm$^{-1}$ at distances of ${\cal{O}}(10)$ pc 
(for a recent review, see Chevalier 2003),
the ones relevant to the CB model predictions for GRB AGs (Dado et al.~2003e). 
This alterity of surface densities may be understood if
the star's $\dot M/V$ increases to a much higher value in the final stages
of its pre-SN evolution. This intensified wind may blow continuously
or in a series of ejection episodes, and may also be  non-isotropic,
as the one shown in Fig.~(\ref{CBGlory}) is.
We will refer to the circumstellar matter distribution simply as {\it the
wind}.

\section{The ambient light  around  SNe}

The wind environment of SNe may be ionized prior to the
SN explosion by the light of the progenitor star, which becomes intense
even at Extreme Ultraviolet (EUV) frequencies prior to the explosion
(the recombination time at the densities characteristic of the wind
is very long). Even if that prior ionization did not occur, the wind
is ionized by the EUV flash from the SN explosion: the observations
of SN1987A indicate that SNe  shine briefly
at EUV frequencies a few hours after their core
collapses, when the blast wave ---presumably produced by the ``bounce" 
of the
collapse, and re-energized by neutrino energy deposition--- reaches the
surface of the progenitor star (e.g.~Arnett et al.~1989; Leibundgut 1995)
and/or, presumably, when the jet of CBs is ejected.

The fast, transient and hard initial 
rise in luminosity has been observed only in SN1987A: 
 the International Ultraviolet Explorer satellite,
which began observations of this SN a day after its neutrino burst,
detected a strong UV continuum with a colour temperature exceeding 14000
K, which declined to $\sim$ 5000 K within 20 days. The luminosity of the
EUV flash is expected to exceed $10^{43}$ erg s$^{-1}$ over a good
fraction of an hour (see, e.g.~Arnett et al.~1989 for a review), 
which provides a sufficient number of
photons to fully ionize a 10--20 $M_\odot$ wind environment. 
For the observed ``close by'' wind densities,
the ionized wind is semitransparent at visible and UV frequencies: neither
optically thin nor thick. 

Subsequent to the early ionizing UV radiation,
Compton scattering of the SN light in the ionized wind,
line emission and thermal bremsstrahlung, result in a locally 
quasi-isotropic light environment within the wind.  Since we are
more interested in the light permeating the wind than in the emitted
``echo'' light seen from afar, we shall give a name to the former embedded 
light: {\it the ambient light.}  Since the wind is semitransparent,
the  expectation is that the ambient light should have a
thin-bremsstrahlung spectrum: $dn_\gamma(E)\sim exp(-E/T)/E$. The wind
may be highly structured by non-uniform emission in both time and angle.
But, in a quasi-stationary situation, the photon density of the ambient
light must be:
\begin{equation}
        n_\gamma (r)\approx {L_{SN}\over
                            4\, \pi r^2\, c\, E_i}\, ,
\label{echolight}
\end{equation}
where $E_i\sim 1$ eV is the typical
energy of an ambient-light photon, and $L_{SN}$ is the SN's luminosity
at early times. This estimate, which corresponds to a flux 
$c\,n_\gamma$, is accurate for the net outward
flux, but may be an underestimate of the total flux, though a slight one,
since the wind is semitransparent.

The observed initial bolometric luminosity of SN1987A, uncorrected for
extinction, was $L_{_{SN}}\sim 3\times 10^{41}$ erg s$^{-1}$; it declined
by one magnitude within a day (Arnett et al.~1989).  The explosion energy
of SN1987A, estimated from the observed velocity of the ejected shell, was
smaller than that estimated for SN1998bw by an order of magnitude. This
may imply that in its early phase SN1998bw was also approximately ten
times more luminous. Indeed, this estimate is consistent with the early
time observations of SN1998bw, started on April 26.60 UT, 1998 (Galama 
et
al.~1998), 0.7 days after its associated GRB was detected with Beppo-SAX
(Soffitta et al.~1998; Pian et al.~2000) and by BATSE (Kippen et al.~1998)
on April 25.90915 UT, 1998. The V-band and R-band light curves of SN1998bw
showed a slow initial decline, or a ``plateau'', after which they rose at
a rate of 0.25 mag per day. This plateau may have been the late signature
of the expected sharp initial peak in the light curve as the blast wave
reached the surface of the progenitor star, but lack of early-time data
prevented establishing its existence in the EUV and UBI bands. Assuming a
Galactic foreground extinction, $Av=0.2$, in the direction of SN1987A, we
estimate its (spherical equivalent) luminosity to be 
\begin{equation}
L_{_{SN}}=5\times10^{42}{\rm \;erg \; s^{-1}}. 
\label{SNluminosity}
\end{equation}
We shall use this estimate of the initial
luminosity of SN1998bw, our {\it ``standard candle''} for
GRB-generating SNe\footnote{The contention that SN1998bw is a standard
candle for the SNe, which are the origin of long-duration GRBs 
---not only an {\it ab initio} hypothesis, but also {\it an observed fact} in a 
CB-model analysis of GRB AGs (see Dado et al.~2002a,b,c, 2003e,f 
and Appendix III)--- is becoming increasingly accepted after the
spectroscopic discovery, in the direction of GRB 030329,
of SN2003dh (Garnavich et al. 2003b),  which looks  identical to SN1998bw 
(Stanek et al.~2003; Matheson et al.~2003).}.

\section{The cannonball model}

In the CB model, {\it long-duration} GRBs and their AGs are produced in
{\it ordinary core-collapse} supernovae by jets of CBs, made of {\it
ordinary atomic matter}, and travelling with high Lorentz factors,
$\gamma$. An accretion disk or torus is hypothesized to be produced around
the newly formed compact object, either by stellar material originally
close to the surface of the imploding core and left behind by the
explosion-generating outgoing shock, or by more distant stellar matter
falling back after its passage (De R\'ujula 1987). A CB is emitted, as
observed in microquasars, when part of the accretion disk
falls abruptly onto the compact object  (e.g.~Mirabel \& Rodrigez 1999; 
Rodriguez \& Mirabel 1999 and references therein).

The $\gamma$-rays of a single pulse in a GRB
are produced as a CB coasts through the  ambient light.
An artist's view of the CB model is given
in Fig.~(\ref{figCB}). The electrons enclosed 
in the CB Compton up-scatter
photons to GRB energies. Each pulse of a GRB corresponds to
one CB. The timing sequence of emission of the successive 
individual pulses (or CBs) in a GRB reflects the chaotic accretion process
and its properties are not predictable, but those of the single pulses 
are. 
%and many follow from simple relativistic kinematic relations
%(Dar and De R\'ujula 2000). 

\subsection{Times and energies}

Let primed quantities refer to a CB's rest system and unprimed ones
to the observer's system, a convention to which we shall adhere throughout
this paper.
Let $\theta$ be the angle between the line of sight and the CB's velocity
vector. Relative to their energy 
in the CB's rest system, $E'$, an observer at a
redhift ``distance'' $z$, sees photons red-shifted by a factor $1+z$, and
Lorentz-boosted, or blue-shifted, by a ``Doppler" factor $\delta$:
\begin{equation}
E = {\delta \over 1+z}\,E' ,
\label{energy}
\end{equation}
\begin{equation}
  \delta = {1\over \gamma\, (1-\beta\, \cos\theta)}\approx 
           {2\gamma\over 1+\gamma^2\, \theta^2}\; ,       
\label{Doppler}  
\end{equation}
where the approximation is valid for $\gamma^2 \gg 1$ and $\theta^2 \ll 1$, 
the domain of interest here, 
for which $\beta\equiv 1/\sqrt{1-\gamma^2}\approx 1-1/(2\gamma^2)$.
The observed time intervals, $dt$, are related to those in a CB's rest system,
$dt'$, by:
\begin{equation}
dt= {1+z\over \delta}\, dt' ,
\label{time}
\end{equation}
where, this time, the factor $\delta$ is the literal (relativistic) Doppler 
factor of Doppler's effect. The
distance travelled by a CB in the SN rest frame during  an 
observer time $dt$ is:
\begin{equation}
dx_{_{SN}} = {\gamma\, \delta \over 1+z}\, c\, dt  .
\label{distance}
\end{equation}

\subsection{Angular distributions}

Let $dN_{_{CB}}/d\Omega'$ be the angular distribution of the time-integrated,
total number of photons emitted  by a CB {\it in its rest system}. Let
$d{\cal{E}}'_{_{CB}}/d\Omega'$, likewise, be the distribution of the total 
energy carried by the photons. The relation between the observer's
viewing angle, $\theta$, and the same angle in the CB's rest system, $\theta'$,
both relative to the CB's direction of motion, is:
\begin{equation}
  \cos\theta' = {\cos\theta-\beta \over 1-\beta\, \cos\theta}\, ,
\label{thetaprime}  
\end{equation}
while $\phi'=\phi$ for the azymuthal angles. The photon total
number-fluence and energy-fluence measured by a cosmologically distant observer 
 are, respectively:
\begin{equation}
 f_{_{CB}}\equiv {dN^{obs}_{_{CB}}(\theta,\gamma)\over d\Omega}
 = {(1+z)\,\delta^2
                    \over D_L^2}\; {dN_{_{CB}}\over d\Omega'}\, ,
\label{photonfluence}
\end{equation}
\begin{equation}
F_{_{CB}}\equiv {dF^{obs}_{_{CB}}(\theta,\gamma)\over d\Omega}
 = {(1+z)\,\delta^3
                    \over D_L^2}\; {d{\cal{E}}'_{_{CB}}\over d\Omega'}\, ,
\label{Fluence}
\end{equation}
where $D_L(z)$ is the luminosity distance (7.12 Gpc at $z=1$, for the
current cosmology with $\Omega=1$, $\Omega_\Lambda=0.7$ and $H=65$
km s$^{-1}$ Mpc$^{-1}$). 
% $dcos\theta'/dcos\theta=\delta^2\, .$
For an isotropic emission {\it in the CB rest frame}, the total 
number of photons and the total energy fluence are, respectively:
\begin{equation}
f_{_{CB}} = {(1+z)\,\delta^2
                    \over 4\,\pi D_L^2}\; N^T_{_{CB}}\, ,
\label{Numberapprox}
\end{equation}
\begin{equation}
F_{_{CB}} = {(1+z)\,\delta^3
                    \over 4\pi D_L^2}\; {\cal{E}}'_{_{CB}}\, ,
\label{Fluenceapprox}
\end{equation}
where $N^T_{_{CB}}$ is the total number of photons emitted by the CB,
and ${\cal{E}}'_{_{CB}}$ is their total energy in the CB's rest frame 
(Shaviv \& Dar 1995; Dar 1998).
An observer who assumes ---incorrectly, we contend---
that the pulse is isotropic in the observer's frame
would infer much larger  figures for the
total number of photons and for the total energy  in a GRB pulse:  
\begin{equation}
N^{iso}_{_{CB}}= {4\, \pi\, D_L^2\over 1+z}\, f_{_{CB}}
 = \delta^2\, N^T_{_{CB}}\, ,
\label{Niso}
\end{equation}
\begin{equation}
{\cal{E}}^{iso}_{_{CB}}= {4\, \pi\, D_L^2\over 1+z}\, F_{_{CB}}
 = \delta^3\,{\cal{E}}'_{_{CB}}\, ,
\label{Eiso}
\end{equation}
whereas the proper angular integration of Eqs.~(\ref{Numberapprox})
and (\ref{Fluenceapprox}) yields  $N_{_{CB}}$ for the total
photon number and  ${\cal{E}}_{_{CB}} = \gamma\, {\cal{E}}'_{_{CB}}/(1+z)$ for
their total energy in the observer's frame. 

\subsection{Typical Lorentz factors and viewing angles}

In our first analysis of GRBs in the CB model (Dar \& De R\'ujula 2000a) 
we concluded that the typical Lorentz factors are $\gamma\sim 10^3$
and the typical viewing angles are $\theta\sim 1/\gamma$, so that
$\delta\sim\gamma\sim 10^3$. This was
corroborated by our systematic analysis of GRB AGs
(Dado et al.~2002a, 2003a): we found that the
fit values of $\gamma$ snuggly peak around 10$^3$, while the $\theta$  
distribution peaks around $\theta\sim 1/(2\gamma)$ and decreases fast
thereafter\footnote{The exception is GRB 980425, observed
at an exceptionally high $\theta\sim 8$ mrad, but located at an 
exceptionally low $z=0.0085$.}, in  good agreement with the expectation
for the rate of photons detected above a certain threshold:
\begin{equation} 
{dN_{GRB}\over{d\theta\,dt}}\propto \theta\, \delta^3 \;F(\delta),
\label{Nvstheta}
\end{equation}
where $F(\delta)$ is a complicated, slowly $\delta$-dependent
function that depends on $z$, the geometry of the Universe
and the case-by-case instrumental effects related to
trigger and measuring efficiencies in various energy windows.
That the distribution of $\theta$-values is narrow is the
quintessential selection effect, induced by the very fast
$\theta$-dependence in Eq.~(\ref{Nvstheta}). The narrowness
of the $\gamma$ distribution is physically significant, although
some ``tip-of-the-iceberg'' effect no doubt plays a role.
Three other parameters
of interest here were constrained by our analysis of GRB AGs: 
the typical baryon (or electron) number of a CB, $N_B$;  the total $\gamma$-ray
energy emitted by a CB in its rest system, ${\cal{E}}'_{_{CB}}$;
and the CBs' initial transverse velocity of expansion, $\beta_s\, c_s$,
where $c_s=c/\sqrt{3}$ is the speed of sound in a relativistic 
plasma (Dado et al.~2002a, 2003a). 
These typical values are summarized in Table 1. In Table 2 we list
the values of $\gamma$, $\theta$ and $\delta$ for the GRBs 
of known redshift whose AG we have analysed 
(Dado, Dar \& De R\'ujula 2002a,b,c; 2003a,b,c,d,e,f).

Our confidence in the values of the parameters describing CBs and their
circumstellar density profiles stems from
the excellence of the description of AG light curves and spectra
in the CB model. An example is given in Fig.~(\ref{f1}), the R-band AG
of GRB 021211 (Dado et al.~2003e). 
At a fixed optical frequency the prediction
for the early AG's fluence is simply $F_\nu\propto (n_e)^{3/4}$, with
$n_e$ the circumburst electron density, assumed to be 
a constant plus a wind contribution decreasing as the inverse square
of the distance. This produces the observed early decline
$\propto t^{-3/2}$ and the subsequent flattening at $t\sim 0.025$ d,
at which point the two contributions to the density are equal and the CBs are at
a distance $\bar r\simeq 1.2$ pc away from their parent SN. The fitted value
of the wind's grammage is $\rho\,r^2=(6.8\pm 0.5)\times 10^{13}$
g cm$^{-1}$, in agreement with the typical value  $\rho\,r^2=5 \times 10^{13}$
g cm$^{-1}$ for winds at that
distance.
%\footnote{In the FB models  the scarcity of AGs indicating a windy
%circumburst density distribution is a problem, admitted even by its
%staunchest defenders (Piran 2001; Price et al.~2002). In the CB model all
%the AGs observed early enough to show a windy signature... show it 
%(Dado et al.~2003e).}. 
Thereafter, as the
CBs decelerate significantly, $F_\nu$ steepens. The late bump
is the underlying SN, identical to SN1998bw except for the effects
of redshift ($z=1.006$, in this case). The single expression for the
AG fitting all of this evolution, as well as the wide-band spectrum,
does an excellent job at describing the AGs and spectra of all other GRBs of
known redshift, including GRB 980425 and its associated
SN1998bw, neither of which is ---in the CB model---
exceptional. 
%In the FB models, such claims cannot be made. 
In Appendix II we give a brief comparison of the confrontation 
of the FB models and the CB model with the AG data.

\subsection{The GRB/SN association in the CB model}

In the CB model, by hypothesis, construction and demonstration, 
long-duration GRBs are made in 
core-collapse SN explosions (these SNe comprise all spectroscopic types, 
but Type Ia). What fraction of core-collapse SNe generate GRBs?

From a CB-model analysis of GRBs and their AGs, Dado, Dar \& De R\'ujula
(2002a,b,c; 2003a,b,c,d,e,f) determined
that GRBs more distant than GRB 980425 are observable with
past and current instruments only for $\theta\leq$ 2--3 mrad. 
With two CB jets per GRB,  only a fraction
\begin{equation}
f\sim 2\, {\pi\, \theta^2 \over 4\, \pi}\sim
(2\;\rm{to}\; 4.5)\times 10^{-6}
\label{fraction}
\end{equation}
of SN-generated GRBs are observable. 
The SN rate in the visible Universe, 
$R_{SN}$, is
proportional to the formation rate of massive stars, which is not very well
known as a function of redshift (e.g.~Madau et al.~1998). Using the 
observed
SN rate in the local universe (e.g.~Capellaro 2003),  
estimates cover the range:
\begin{equation}
 R_{SN}\sim \rm{ (1\; to\; 10)\times 10^8\; yr^{-1} }.
  \label{RSN}
 \end{equation}
About
25\%--30\% of all SNe {\it in the local Universe} are of Type Ia, and
the rest are core-collapse SNe: 12\%--15\%
of Type Ib/Ic and 55\%--65\% of Type II (Tammann, Loeffler \& Schroeder
1994, van den Bergh \& McClure 1994). If these rates were representative
of a cosmic average, we would expect a rate of long-duration GRBs:
\begin{equation}
R_{GRB}\sim {\rm(0.7\; to \; 0.75)}\;f\;R_{SN}\sim 
{\rm(140\; to \; 3375)\; yr^{-1}. }
\label{RGRB}
\end{equation}
The value of $R_{GRB}$ inferred from
BATSE observations (Fishman \& Meegan 1995) is $\sim$ 500 to 700 yr$^{-1}$,
compatible with the range in Eq.~(\ref{RGRB}).  

A comparison independent of the unknown star-formation rate at large $z$
yields a similar result: Schmidt (2001) has derived a luminosity function
for GRBs of known redshift from which he estimated a local rate of long
duration GRBs, $(2.5\pm 1.0) \times 10^{-10}\; {\rm Mpc^{-1}\, yr^{-1}}$
for the current cosmology. The local rate of core-collapse SNe is $(7.5\pm
3.8)\times 10^{-5}\; {\rm Mpc^{-1}\, yr^{-1}}$ (Cappelaro 2003). The ratio
of these rates, $(3.3\pm 2.1)\times 10^{-6}$, is consistent with
Eq.~(\ref{fraction}).

In view of the very large uncertainties, the fair
conclusion is that the GRB rate is consistent with being equal, either to
the total rate of core-collapse SNe, or to a fraction of it
that may be as small as $\sim 1/4$. Within errors, it may also be that
these statements apply to only the Type Ib and Ic subclasses.
Spectroscopic information on many more SNe is needed before these
issues can be decisively resolved.

\section{The polarization of a GRB}

%Let primed quantities refer to a CB's rest system and unprimed ones
%to the SN's rest system.
In the CB's system, because of the large value of $\gamma$, the bulk of the photons
of the wind's light (of energy $E_i={\cal{O}}(1)$ eV)
are incident practically along the direction of relative motion,
$\theta_i'={\cal{O}}(1/\gamma)$. Their energy
is $E'_i={\cal{O}}(\gamma\,E_i)\ll m_e\,c^2$,
so that their Compton scattering cross section is in the  low-energy 
``Thomson'' limit. Let $\theta'$ be the angle  at which a photon is Compton
scattered by a CB's electron\footnote{The Thomson cross section
is $\propto 1+\cos^2(\theta')$, so that 
$\theta'\gg\theta'_i\simeq 0$ is an excellent approximation,
but for extremely forward-scattering events that do not result
in observable photons at GRB energies.}, related to the observer's
angle as in Eq.~(\ref{thetaprime}). 
The scattering linearly polarizes the outgoing photons in the direction
perpendicular to the scattering plane by an amount 
(e.g.~Rybicki \& Lightman 1979):
\begin{equation}
    \Pi(\theta')\approx {1-\cos^2\theta'\over 1+\cos^2\theta'}.
\label{polarcb}
\end{equation}
Substitute Eq.~(\ref{thetaprime}) into Eq.~(\ref{polarcb}) to obtain the
value of the (Lorentz-invariant) linear polarization in the observer's 
frame.
In the large-$\gamma$ approximation, the result is:
\begin{equation}
\Pi(\theta,\gamma)\approx {2\;\theta^2\,\gamma^2\over 1+\theta^4\,\gamma^4},
\label{polSN}
\end{equation}
which, for the probable viewing angles,
$\theta\sim 1/\gamma$, is of ${\cal{O}}(100\%)$ (Shaviv \& Dar 1995).
This result is easy to understand: photons viewed at $\theta =1/\gamma$
had, according to Eq.~(\ref{thetaprime}), suffered a scattering of
$\theta'=90^{\rm o}$, which fully polarized them according to 
Eq.~(\ref{polarcb}), as one may also
recall from elementary electron-oscillator considerations.

The polarization $\Pi(\theta,\gamma)$ is a function of only the product 
$\theta\,\gamma$
and has a universal shape, but it is shown in Fig.~(\ref{f2}) as a function
of $\theta$ at $\gamma=10^3$, in order to compare it with the expected
number, roughly $dN_{_{GRB}}\propto \theta\,[\delta(\theta,\gamma)]^3$, 
of GRBs detectable above a given flux threshold, as a function of $\theta$ at the
same typical $\gamma$, see Eq.~(\ref{Nvstheta}). 
Without further ado, the figure shows that it
is very probable, in the CB model, to observe a GRB with a measured 
polarization $\Pi=80\pm 20\%$. To add some ado: the polarization of
Eq.~(\ref{polSN}) exceeds 60\% for $\sqrt{3}>\theta\,\gamma>1/\sqrt{3}$,
which is the case for a third of the CBs listed in Table 2.

\section{The energy of the $\gamma$ rays of a GRB}

\subsection{The typical energy}

Let $E_i\sim 1$ eV be the typical photon energy of the 
wind's ambient light. More precisely, $E_i$ is to be interpreted
as the more sharply defined
energy at which most of the energy of the ambient light
resides (for a thermal or a thin thermal-bremsstrahlung
spectrum that is the energy at which the distribution
$E^2_i\,dn_\gamma/dE_i$ peaks).
Let such a photon move with an angle
$\theta_i$ relative to the direction of motion of the CB. 
Viewed from the CB's system, the energy is 
$E'_i=\gamma\, E_i\,(1+\beta\,\cos\theta_i)$
and the incoming angle is $\theta_i'\sim 0$.
If the photon is deflected to an angle
$\theta'$ in the CB's rest frame by Compton scattering, its energy changes
very little, but in the SN rest frame its energy is boosted by the
CB's motion to 
$E\ = \gamma\, E_i \, (1+\beta\, \cos\theta_i)\, 
                (1+\beta\, \cos\theta')$.  
Rewrite $\theta'$ in terms of
the observer's angle $\theta$ with use of
Eqs.~(\ref{Doppler}) and (\ref{thetaprime}), and take into account the
cosmological redshift to obtain the measured energy of a GRB photon
produced this way: 
\begin{eqnarray} 
E\ &=& {\gamma\,\delta\over 1+z}\,
(1+\beta\, \cos\theta_i)\, E_i\, \nonumber\\ &=& (500\;{\rm keV})\;
\sigma\;(1+\beta\, \cos\theta_i)\,
{E_i\over 1\;\rm eV}\nonumber\\
\sigma&\equiv&{\gamma\,\delta\over 10^6}\,{2\over 1+z}\; , 
\label{boosting} 
\end{eqnarray} 
where we have normalized  to an isotropic
distribution of the ambient photons, $\langle \cos\theta_i\rangle=0$,
and to the typical values of the parameters (the average $\sigma$
of the GRBs listed in Table 2 is 1.08). The
result for $\langle (1+z)\,E\rangle$ is centred at $\sim 1$ MeV.

For typical parameters, the wind is semitransparent at the time a
GRB pulse is emitted: its optical depth is $\sim 1/2$,
as we shall see in detail in Section 7, Eqs.~(\ref{ttrans}) and (\ref{twtrans}).
For a non-transparent
wind $\langle\cos\theta_i\rangle=0$, while for a transparent one, the
ambient light photons would be radially directed away from the parent
SN and $\langle\cos\theta_i\rangle=-1$. For the semitransparent
typical case $\langle 1+\beta\,\cos\theta_i\rangle\sim 1/2$, and the central
numerical prediction of Eq.~(\ref{boosting}) is 250 keV.  This
 prediction precisely coincides with the observed median value of the
peak energy (photon energy at peak $E^2\, dN/dE$) in GRBs, $\langle
E_p\rangle\sim 250$ keV (e.g.~Preece et al.~2000; Quilligan et al.~2002),
or with the peak energy in the rest frame of the GRB's progenitor:
$E'_p=0.5$ MeV, according to Amati et al.~(2002), for GRBs of known 
$z$, for which $\langle z\rangle\sim 1$.

We conclude that GRBs are made of $\gamma$-rays of a few hundred
keV energy because that is the typical energy to which ambient light
is nearly-forward Compton up-scattered by electrons of $\gamma\sim 10^3$.
The first attempt to explain GRB energies in this way was that by
Shaviv \& Dar (1995).

\subsection{The energy distribution}

One of the most puzzling facts about GRBs is the narrowness of the
distribution of their typical energies, shown in 
Fig.~(\ref{figEpobs}) in terms of
the ``break energy'': the fitted energy at which their energy spectrum
$dN/dE$
steepens, which is also, approximately, the energy $E_p$
at which $E^2\,dN/dE$
peaks (Band et al.~1993;
Preece et al.~2000). In the CB model, this distribution is that
of the factor $\sigma\,(1+\beta\,\cos\theta_i)\,\epsilon_i$ in Eq.~(\ref{boosting}),
with $\epsilon_i\equiv E_i/(1\;{\rm eV})$.
We have no a-priori way of knowing what the distribution of values
of $(1+\beta\,\cos\theta_i)\,\epsilon_i$ is, but for the GRBs of known redshift
whose AG we have fitted in the CB model, we know the $\sigma$
distribution.

In Table 2 we give the redshifts and fitted values of $\gamma$,
$\delta$ and $\sigma$ for the quoted GRBs. The corresponding
distribution in Log$_{10}(\sigma)$ is shown in Fig.~(\ref{figEpth}),
where the continuous line is a  (log-normal) fit. Due to the binning
of the modest number of events shown in Fig.~(\ref{figEpth}),
the fit looks rough. But it is not. It was made with the known
binning-independent Kolmogorov-Smirnov (KS) test, and its
KS probability is a very comfortable 92\%.
With so little statistics on GRBs of
known $z$, the fitted width of the Log$_{10}(\sigma)$ distribution
is only determined to $\sim 30$\%.
We have superimposed this KS fit on Fig.~(\ref{figEpobs}),
showing it to be surprisingly compatible with the distribution
of Log$_{10}(E_p)$. This leaves only little room ($< 30$\%) for a further
(unknown) broadening due to the inevitable spread in 
$(1+\beta\,\cos\theta_i)\,\epsilon_i$. In spite of this minor caveat,
we consider the CB-model expectation for the narrow width
of the $E_p$ distribution to be very
satisfactorily consistent with the observations.
For the central value of the theoretical distribution
to agree with the observed one, as in Fig.~(\ref{figEpobs}), it must be that
$\langle(1+\beta\,\cos\theta_i)\,\epsilon_i\rangle\sim 0.5$, as we argued
in the previous subsection.

\section{The width in time of a GRB pulse}

A CB is heated, while crossing the SN's shell and prior wind, by hadronic
collisions between the CB's constituents and those of the circumburst
material. This process stops as the expanding CB becomes transparent
to these collisions. Let the ordinary matter of which the CB and
the material it encounters are made be approximated as hydrogenic, and let the
CB (in its rest frame) be approximated as a sphere of constant
density. At $\gamma\sim 10^3$,
the $pp$ cross section is dominantly inelastic and its value is 
$\sigma_{pp}\approx 40$ mb. The CB's radius of collisional transparency
is $R_{pp}\sim [3 \,\sigma_{pp}\,N_{_{B}}/(4\pi)]^{1/2}$, or 
$\sim 10^{12}$ cm,
for a CB's baryon number $N_{_{B}}= 10^{50}$. Since the CB's 
initial internal radiation
pressure is large, it should expand (in its rest frame) at a radial velocity 
$\beta_s\,c_s$ comparable to the speed of sound,
in a relativistic plasma,  $c_s=c/\sqrt{3}$.

Let $\sigma_T=0.665\times 10^{-24}$ cm$^{-2}$ be the Thomson cross 
section. A CB becomes transparent to its enclosed radiation when it
reaches a radius:
\begin{equation}
R_{tr}\sim\sqrt{3\, \sigma_T\, N_{_{B}}\over 4\, \pi}\approx  
(4\times 10^{12}\,{\rm cm})\; \left({ N_{_{B}}\over 
10^{50}}\right)^{1\over2}\, . 
\label{rtrans}  
\end{equation}
Seen by a cosmological observer, the time elapsed from the CB's
ejection to the point at which $R(t)=R_{tr}$ is:
\begin{equation}
t_{tr}=(0.46\,{\rm{s}})\; {1\over \beta_s}\;{10^3\over\delta}\;
{1+z\over 2}\;\left({N_{_{B}}\over 10^{50}}\right)^{1\over 2}.
\label{ttrans}
\end{equation}
During this time the CB has moved a distance:
\begin{equation}
r_{tr}={\sqrt{3}\,\gamma\,  R_{tr}\over \beta_s}\approx (6.9 
\times 10^{15}
\, {\rm cm})\; {1\over \beta_s}\;{\gamma\over 10^3}\; 
\left({N_{_{B}}\over 10^{50}}\right)^{1\over 2}
\label{disttrans} 
\end{equation}
away from its parent SN, a distance at which it is still embedded in the wind. 

The electrons contained in a CB Compton up-scatter the ambient  photons to
the typical GRB energies of Eq.~(\ref{boosting}). But not all of them escape
unscathed to become observable: they may be reabsorbed by the wind's
material. The probability that a GRB photon produced at a distance $r$ from the SN
evades this fate, in a wind with a density profile $n_e\propto r^{-2}$,
is $A(r)= {\rm exp}[-(r^w_{tr}/r)^2]$ with
$r^w_{tr}=\sigma_T\, \rho\, r^2/m_p$  the distance at which the 
remaining optical
depth of the wind is unity. In the SN rest frame the
``wind transparency time'' is $r^w_{tr}/c$, and it corresponds
to an observer's time:
\begin{equation}
t^w_{tr}=(0.27\,{\rm s})\;{\rho\,r^2\over 10^{16}\,{\rm g \,cm}^{-1}}\;
{1+z\over 2}\; {10^6\over \gamma\,\delta}\; ,
\label{twtrans}
\end{equation}
coincidentally close ---for the typical parameters---
to the CB's transparency time, $t_{tr}$.

Absorption in the wind affects the shape of a GRB pulse by a
multiplicative factor $A(t)={\rm exp}[-(t^w_{tr}/t)^2]$, rising very
fast from zero to unity. After the wind and the CB are both transparent,
the number of photons per unit time in the pulse decreases with time as
$(t_{tr}/t)^2$, simply reflecting the number density of the scattered
ambient photons $n_\gamma\propto 1/r^2\propto 1/t^2$,
as in Eq.~(\ref{echolight}). The net result
---supported by our detailed study of the pulse shape in Section 9--- is
that the full width at half-maximum (FWHM) of a CB pulse is 
$t_{_{\rm FWHM}}\sim t_{tr}$ or, more precisely: 
\begin{equation}
t_{_{\rm FWHM}}\approx 1.8\; t_{tr}, 
\label{FWHM} 
\end{equation} 
with $t_{tr}$ as in Eq.~(\ref{ttrans}). For the standard CB parameters 
Eq.~(\ref{FWHM}) yields $t_{_{\rm FWHM}}\approx 0.83$ s,
in excellent agreement with the (median) observed result,
$t_{_{\rm FWHM}}\approx 0.90$ s for photon energies above 20 keV. 
In the 110--325 keV energy band Lee, Bloom, \&  Petrosian (2000)
report $t_{_{\rm FWHM}}\approx 0.68$ s
and McBreen et al.~(2002) obtain $t_{_{\rm FWHM}}\approx 0.58$ s
(why pulses are narrower at higher energies is explained in Section 12).

\section{The number of photons and the energy of a pulse}

The number of ambient-light photons scattered 
by a CB into a particular observer's
angle can be computed explicitly, but it is a function of the CB's geometry
and the density distribution of electrons within the CB.  Yet, to obtain a
good estimate is fairly simple. The total number of ambient  photons scattered 
by the CB up to its transparency time is:
\begin{equation}
N^{tk}\approx \pi \,  \int^{R_{tr}}_0\, R^2 \,n_\gamma\  dr
   \approx{L_{SN}\,\beta_s\, R_{tr}\over 4\,\sqrt{3}\, c\,
\gamma\,E_i}\; ,
\label{Nuntrans}
\end{equation}
where we have used $R=R_{tr}\,(t/t_{tr})=R_{tr}\,(r/r_{tr})$ and Eqs.~(\ref{echolight},\ref{rtrans},\ref{disttrans}). 
The fraction of these photons that is observable depends on the relative values
of the wind and CB absorption coefficients, which are comparable, according to 
Eqs.~(\ref{ttrans}) and (\ref{twtrans}). It also depends on geometry: consider 
the likely case $\theta\sim 1/\gamma$ (scattering at right angles in the CB's 
rest system, $\theta'=\pi/2$) in two very extreme geometries. Within a 
longitudinally thin, slab-shaped CB, such as the one shown
in Fig.~(\ref{figslab}), photons scattered
onto that direction would all reinteract, and be
degraded to lower energies by adiabatic cooling. A thin rocket-like CB,
contrarywise,
would let most of the photons scattered at right angles escape unscathed.

The total number of photons scattered after the CB 
becomes optically thin to Compton scattering, similarly calculated, is:
\begin{equation}
N^{tn}_T\approx \sigma_T\, N_{_{B}} \int_{R_{tr}}^\infty \, n_\gamma\  dr
   \approx{L_{SN}\,\beta_s\,  R_{tr}\over 3\, \sqrt{3}\, c\,
\gamma\,E_i}\; .
\label{Ntrans}
\end{equation} 
Around $t=t_{tr}$,  photons have a 
roughly 50--50 chance of not being scattered a second
time within the optically thinning CB. The total number of scattered photons is 
somewhere between  $N^{tn}/2$ and $N^{tn}/2$ plus a small fraction of 
$N^{tk}=3\, N^{tn}/4$, so that
$N^{tot}\simeq N^{tn}/2$.
For a given angle of observation, the result is modulated
by the relatively weak angular-dependent factor $G(\theta)$ 
in the Thomson scattering cross section:
\begin{equation}
G(\theta')=1+\cos^2\theta'
\label{G}
\end{equation}
\begin{equation}
{d\sigma\over d\Omega'}= 
{3\,\sigma_{_{T}}\over 16\,\pi}\, G(\theta') \approx
{3\,\sigma_{_{T}}\over 16\,\pi}\;
{2\, (1+\theta^4\,\gamma^4)\over (1+\theta^2\,\gamma^2)^2}\, ,
\label{compton}
\end{equation} 
where in Eq.~(\ref{compton}) we used Eq.~(\ref{thetaprime}) to express $G$ 
in the observer's frame. 
The angular distribution of the number of scattered ambient photons and of
their total energy in the CB's rest system are:
\begin{equation}
{dN_{_{CB}}\over d\Omega'}\approx {3\,G\over 16\,\pi}  \, N^{tot}\approx 
       {G\, L_{_{SN}}\,\beta_s\, R_{tr}\over 32\,\sqrt{3}\, \pi\, c \, 
\gamma \, E_i}\, ,
\label{dNprime}
\end{equation}
\begin{equation}
{d{\cal{E}}'_{_{CB}}\over d\Omega'}
\approx \gamma\,E_i\,{dN_{_{CB}}\over d\Omega'}
             \approx{G\, L_{_{SN}}\,\beta_s\, R_{tr}\over 32\,\sqrt{3}\, 
\pi\, c}\, .
\label{dEprime}
\end{equation}
The total number of scattered photons is:
\begin{eqnarray}
N_{CB}&=&\int d\Omega'\, {dN_{_{CB}}\over d\Omega'}\nonumber\\
&\approx& (5 \times 10^{52})\,
\left({L_{_{SN}}\over 5\times 10^{42}\,{\rm erg}}\right)\,
\left({1\, {\rm eV}\over E_i}\right)\,
 \left({ N_{_{B}}\over 10^{50}}\right)^{1\over2}\,
 \left(10^3\over \gamma\right)\,\beta_s\, ,
 \label{NUMTOT}
 \end{eqnarray}
and the equivalent spherical photon emission that follows from 
Eq.~(\ref{Niso}) is $\sim 10^6$ times larger. 
The energy emitted by a CB in its rest frame 
in the form of scattered ambient  photons,  is:
\begin{eqnarray}
{\cal{E}}'_{CB}&=&\int d\Omega'\, {d{\cal{E}}'_{_{CB}}\over d\Omega'}\nonumber\\
&\approx&(0.8 \times 10^{44}\,{\rm erg})\;
\left({L_{_{SN}}\over 5\times 10^{42}\,{\rm erg}}\right)\,
 \left({ N_{_{B}}\over 10^{50}}\right)^{1\over2}\; \beta_s,
\label{Eprime}
\end{eqnarray}
and the equivalent spherical energy that follows from 
Eq.~(\ref{Eiso}) 
is $\sim 10^9$ times larger, i.e. ${\cal{E}}^{iso}\sim 0.8\times 10^{53}$ 
erg.
The average ${\cal{E}}^{iso}$ 
of the GRBs with known redshift that were measured by BeppoSAX
is $\sim 6\times 10^{53}$ erg  (Amati et al.~2002). There are 6 pulses 
on the average in a single GRB (Quilligan et al.~2002), yielding 
$\sim 10^{53}$ erg mean pulse-energy, in agreement with the expectation.

We have fitted to the CB model the AGs of all GRBs of known redshift (Dado
et al. 2002a,b,c; 2003a,b,c,d,e,f; Dado et al.~in preparation) and
extracted the parameters $\theta$ and (the initial) $\gamma$ for all of
them, in the approximation that the AG is dominated by one CB, or a
collection of similar ones\footnote{Two notable exceptions are GRB 021004
and 030329 whose $\gamma$-ray {\it and} AG light curves are clearly
dominated by two CBs (Dado et al.~2003c; 2003e).}. From these analyses we
extracted, using Eqs.~(\ref{Doppler}) and (\ref{Fluence}), the value of
${\cal{E}}'_{_{CB}}$ for each GRB, in an approximation in which the weak
angular dependence of the Thomson cross section was neglected (an
excellent approximation around the most probable observation angles
$\theta\sim 1/\gamma$, for which $G\approx 1$). The remarkable
result\footnote{In the CB model GRBs are much better standard candles than
in the FB models (Frail et al.~2001; Berger et al.~2003; Bloom et al.
2003).} was that the values of ${\cal{E}}'_{_{CB}}$ span
a very narrow rage: from 0.6 to 2.1
times 10$^{44}$ erg, see Fig.~(41) of Dado et al.~(2002a). Also to be
remarked is how close these results are to the simple prediction of
Eq.~(\ref{Eprime}).

It is at first sight surprising that the parameters in Eq.~(\ref{Eprime})
conspire to give a rather narrow range of ${\cal{E}}'_{_{CB}}$ values. Yet, it is not
unreasonable that $\beta_s$ be narrowly distributed around 1, 
the expectation for an expanding relativistic plasma. 
The baryon number dependence is only
via a square root. But why should $L_{_{SN}}$ be in a narrow range? Besides
SN1998bw, there is only one other SN for which the early
$L_{_{SN}}$ was measured:
SN1987A. Both SNe have the same ratio of $L_{_{SN}}$ to peak optical
luminosity. It is quite conceivable that core-collapse SNe be much less
varied than the observations seem to indicate: much of the variability
could be due to the different observer's angles relative to the jet
axis of bipolar CB emission, so that these SNe would be {\it standard 
torches}, rather than standard, spherically-emitting candles. 
%It is also conceivable that the UV flash be
%associated with the emergence of the jet from the SN's surface.

\section{The shape of a GRB pulse} 

\subsection{Smooth pulse shapes}

Let us first discuss the shape of a pulse for which the quantities
defined in Eqs.~(\ref{ttrans}) and (\ref{twtrans}) satisfy
$t_{tr}\gg t^w_{tr}$, so that $\gamma$-ray absorption in the
wind plays no significant role. We shall also simplify the discussion
to a manageable level
by working out the result for photons scattered only
once within the CB, more collisions ``adiabatically"
degrade the photon energy because of the CB's expansion.

The shape of a GRB pulse depends, albeit quite moderately, on the CB's
geometry. To illustrate this fact, we have worked out the result for
various geometries and observation angles. The simplest case
to present is an unrealistically extreme but ``pedagogical'' geometry: 
a CB consisting of a 
slab that is much larger in the direction transverse to its
motion than in the direction parallel to it, as in Fig.~(\ref{figslab}).
Consider photons suffering backward scattering: exiting the
CB in the direction opposite to the incoming one.
Photons reaching an observer at a fixed time may
originate from different depths into the slab
at which they interacted: they have different times of entry into the CB.
To lighten the notation let $t'$ be temporarily umprimed
and let time be measured in units of $t_{tr}$, so that
$t=1$ is transparency time. Let also the unit of distance
be such that $c=1$. 
Let $n_{_{CB}}(t)\propto 1/t^3$ be the time-varying
CB's density, approximated as uniform within the CB.
As in Fig.~(\ref{figslab}) the photons interact at the various depths $(t-x)$,
and they must escape absorption during their trip
to that point and back.
The pulse's photon number per unit time is then:
\begin{eqnarray}
{dN\over dt}&\approx&K
 \int_0^{Up(t)}
\!n_{_{CB}}(t-x)\,dx\;
e^{-\sigma_T\;\int^t_{t-2\,x}\, n_{_{CB}}(t')\,dt'}\nonumber\\
&\propto&\int_0^{Up(t)}{dx\over(t-x)^3}\;
exp\left[{1\over 2\,t^2}-{1\over 2(t-2\,x)^2}\right]\; ,
\label{slab}
\end{eqnarray}
where $Up(t)={\rm min}[t/2,\,R(t)]$, 
 $K\equiv \pi \,R^2 \,n_\gamma\, d\sigma_T(\pi)/d\cos\theta$, and we have 
used the fact that $R^2\propto t^2$
and $n_\gamma\propto t^{-2}$. We have also simplified
the result by use of
Eqs.~(\ref{echolight}), (\ref{rtrans}) and (\ref{disttrans}). 
The case of a spherical CB seen at any given angle, shown in 
Fig.~(\ref{figsphere}),
is a trivial but very tedious 3-D generalization of Eq.~(\ref{slab}).

The pulse shapes for various geometries and scattering angles in
the CB system are shown in Fig.~(\ref{fig3}), in which time is measured
in units of the observer's transparency time: $t_{tr}$ in Eq.~(\ref{ttrans}).
The curves labelled 1 and 2 are for the slab geometry of Eq.~(\ref{slab}),
with $\beta_s=1/5$ and 1 for the expansion velocities,
respectively. The result for a spherical CB, also at a backward scattering angle, and 
with $R(t)$ fixed ($\beta_s=0$), is labelled 6. The result labelled
5 is for the same geometry and $\beta_s=1/5$. The other two
pulse shapes are for a sphere seen at right angles in the CB rest system
(an observer's $\theta=1/\gamma$), 4 is for $\beta_s=1/5$ and 3
for $\beta_s=1$. 
The difference between the shape of the slab light curves and the
others is only apparent, an increase of $t_{tr}$ by a factor of 2 in
the slab pulse shapes makes them resemble the others. Except for the fixed-radius
case, the large-$t$ pulse shapes and their normalization are the ones implied by 
Eq.~(\ref{Ntrans}), to wit:
\begin{equation}
{dN\over dt}\to c\,\sigma_T\,G\,N_{_{B}}\,n_\gamma\propto 1/t^2\;,
\label{latepulse}
\end{equation}
with $G$ as defined in Eq.~({\ref{compton}).
A  decline $\sim 1/t^2$ is the mean observed late-time dependence 
of GRB pulses (Giblin et al.~2002).

To a rather good approximation, all the expanding-CB pulses in 
Fig.~(\ref{fig3}) have shapes that resemble that of the very simple function:
\begin{equation}
{dN\over dt}=exp\left(-\,{1\over t^2}\right)\,
\left[1-exp\left(-\,{1\over t^2}\right)\right],
\label{naivepulse}
\end{equation}
with time measured in units of $t_{tr}$.

Several complications may affect this result. First, pulses are
observed in certain energy intervals, and their shapes depend
on these, as we discuss in detail in Section 12. Second, absorption
in the wind, if the condition $t_{tr}\gg t^w_{tr}$ is reversed, modifies
Eq.~(\ref{naivepulse}) into 
$exp(-(t^w_{tr}/t)^2)\,[1-exp\{-(t_{tr}/t)^2)\}]$. 
A further complication is that the ``effective number'' of Compton
up-scattered photons is modulated, as is their final energy in
Eq.~(\ref{boosting}), by a factor $1+\beta\,\cos\theta_i$, which
decreases as $1/r^2$ at distances large enough for the wind
to be quite transparent. A pulse's late rate of decline may therefore
evolve in some cases
from $1/t^2$ to $1/t^4$. We embody all of these complications
in the following approximate form of a pulse:
\begin{equation}
{dN\over dt}=exp\left[-\left({t_1\over t}\right)^m\right]\,
\left\{1-exp\left[-\left({t_2\over t}\right)^n\right]\right\}\, .
\label{pheno}
\end{equation}

Even the most naive pulse shape with $m=n=2$, $t_1=t_2$, shown in
Fig.~(\ref{fig3pulses}), does a very good job
at describing individual ``FRED'' shapes as well as results averaged over
all observed GRB shapes. For example, for this pulse shape, the
ratio of the rise-time from  half-maximum to maximum to the total 
width at half-maximum is, $\simeq 0.27$,
while the observed result, reproduced in 
Fig.~(\ref{figFWRise}) is $\sim 0.3 $ (Kocevski et al. 2003).
Also shown in Fig.~(\ref{fig3pulses}) are a very fast-rise case 
with $t_{1}=0.03\,t_2$, $m=n=2$, and a fast-decline case with 
$t_{1}=t_2$, $m=2$ and $n=4$.

\subsection{The fast variability of GRB light curves}

It is often claimed that GRBs have ``variability'' at the few millisecond level
(e.g.~Walker, Shaefer \& Fenimore 2000 and references therein). 
Yet, the Fourier 
transform $P(\omega)$ of
GRB light curves has very little power at $\omega=100$ Hz and above, compared with
the power at
$\omega\sim 1$ Hz ---the frequency at which $P(\omega)$ ``bends''---
reflecting the fact that the structure of most GRBs is
dominated by rather smooth pulses of $\sim 1$ s width, and not by
much narrower features. The fraction of GRBs with significant
variability much faster than 1 Hz must be small, as reflected by the Fourier
properties of GRB ensembles (Shaviv \& Dar 1995; Beloborodov et al.~2000). 
In order to extricate a putative short-time variability from the data, methods
more sophisticated than mere Fourier transforms appear to be necessary.

Two common dictums in high energy physics are: ``Do not trust results
coming from the edge of your distributions" and ``If you need statistics
---rather than your naked eye---
to prove your discovery, you have not made one". These wisecracks
make us a tad uneasy about the subject of this subsection.

In a paper where they studied 20 GRBs, Walker et al.~(2000)
found that two {\bf short-duration} GRBs had spikes of a few ms duration.
In a wavelet analysis of the ensemble, which included 5 
GRBs with short ($<2$ s) duration, they concluded that the majority
had ``spikes or flickers with rise times shorter than 4 ms in the first 
$\sim 1$ s of their light curves".  

At first sight, the CB model cannot accommodate such results, since
the shape of the pulses in Fig.~(\ref{fig3}) is dominated by retardation
effects: it takes a time of ${\cal{O}}(t)$ for a photon to enter the CB,
scatter within it, and exit in a given direction to be observed
at time $t$. This means
that short pulse substructure is, as allegedly observed, 
more likely at early times. But, how can
it be produced? A non-uniform density distribution within a CB is
not the answer, for the volume integration at fixed $t$ would erase
its details. But suppose that the density distribution, quite reasonably,
is significantly peaked towards the centre of a (roughly spherical)
CB. At the transparency time corresponding to the average density,
then, the core of the CB may be quite non-transparent, while the
rest is transparent. Consider a number-density inhomogeneity 
{\it in the ambient  light}, such as would be produced by an over-density
inhomogeneity in the wind, as the ones pictured in Fig.~(\ref{CBGlory}). 
Seen by the CB in its rest system, the inhomogeneity is
longitudinally foreshortened by a factor $\gamma$. Its photons 
are likely to scatter the central overdensity. This would produce
a ``spike" whose duration is of the order of the light-crossing
time of the core overdensity, which may be much smaller than
that of the CB as a whole. It is even likely that the ambient-light temperature
within the overdensity be enhanced, which would make the
spikes' spectrum harder, as claimed by Walker et al.~(2000).

In the FB models the fast variability plays an important role, the width 
of the colliding $e^+e^-$ shells is adjusted to produce it (e.g.~Waxman 2003a).

\section{Shockless acceleration}

In Dado et al.~2002a, we argued that the electrons entering a
CB from the external medium would undergo acceleration
by their successive deflections in its enclosed chaotic and 
turbulently moving
magnetic field. Subsequently, there have been illuminating numerical
studies of this process (Frederiksen et al.~2003).
These authors have investigated what happens when a collisionless plasma of
ions and electrons impinges at large $\gamma$ on a similar plasma at rest. 
The trajectory of each particle is governed by the Lorentz
force that the ensemble of all other particles exert on it,
whose $\vec E$ and $\vec B$ fields are determined by Maxwell's equations.
An infinitesimal seed magnetic field suffices to separate the trajectories
of the different-charge particles, creating an instability
leading to moving electric fields
(that is, extra magnetic fields).  The process induces a turbulent
flow and a turbulent magnetic field that is carried in with the incoming
particles at $v\sim c$. The particles are accelerated extremely fast
by their interactions with these fields to a spectrum hardening
with energy to a  power law, 
\begin{equation}
{dn\over dE}\propto E^{-p},\;\;\;\;\;p\sim 2.2.
\label{acceleration}
\end{equation}
In this acceleration and magnetic-field-generating process
there are no shocks: no surfaces discontinuously separating
two domains and, consequently, no ``shock acceleration'' by
successive crossings of the shock, the mechanism allegedly
responsible for particle acceleration in the FB models,
as discussed in Appendix I. The 
moving ``front'' of the incoming particles within the bath
of the target ones does not leave on its wake neither a kinetic
nor a true thermal distribution. In front of the front nothing
happens.

These numerical studies are approximations, in that the statistics
are limited, the ratio $m_e/m_p$ is not as small as the observed
one, the transverse boundary conditions are periodic, as opposed
to self-regulatory, and radiative processes are ignored. Yet, none of their
input ingredients ---special relativity, Maxwell's equations and the
Lorentz force--- is potentially dubious. That is why we have refrained
from calling these studies {\it simulations}\footnote{We do not know
whether or not the person who first introduced this term to physics was
aware of the fact that the definition 1.a.~of 
{\it Simulation} in the Oxford English Dictionnary (http://dictionary.oed.com/)
is: {\it The action or practice of simulating, with intent to deceive; false 
pretence, deceitful profession.}}.

In the bulk of a non-transparent CB, the shockless acceleration 
process would be quenched by Coulomb interactions with the 
enclosed photon bath. However, in the outer transparent part
of the CB (that at $t\sim t_{tr}$ becomes the whole
object), the acceleration process does take place. 
It is from this region that ambient  photons entering the CB are scattered
out; they are thus subject to Coulomb collisions
with accelerated electrons, not only unaccelerated ones.

While a CB is not yet fully transparent to the strong interactions
of the wind's hadrons that penetrate it, there is
another mechanism endowing it with a high-energy electron constituency.
The entering hadrons lose energy mainly by
Coulomb collisions with the CB's electrons, which lead to
electromagnetic showers initiated by the {\it knocked-on} electrons.
%and hadronic collisions with the CB's nuclei, which also lead to
%high-energy electrons and positrons via pion production, followed by the
%decay chains $\pi^{+}\to \mu^{+}\,\nu$, $\pi^{-}\to \mu^{-}\,\bar\nu$;
%$\mu^{\pm}\to e^{\pm}\,\nu\,\bar\nu$.
In the CB's rest system, a nucleus of charge $z$
 and Lorentz factor $\gamma$ gives rise to knocked-on
electrons (or ``$\delta$-rays'')
with Lorentz factors up to $\gamma'_{e,max}\approx
2\, \gamma^2$.  In a CB of density $n_{_{CB}}$,
the number of electrons  scattered to a
Lorentz factor $\gamma'_e$ is:
\begin{eqnarray}
{d^2 N_e\over d\gamma'_e\, dt}&\approx&
2\, \pi \,r_e^2\, n_{_{CB}}\,c\, z^2 \;{F(\gamma'_e)\over (\gamma'_e)^2}\, ,
\nonumber\\
 F(\gamma'_e)&\approx& 1- {\gamma'_e\over \gamma'_{e,max}}\, ,
\label{dndgamma}
\end{eqnarray}
where $r_e$ is the classical electron radius.
For electrons with Lorentz factors considerably
smaller than $\gamma'_{e,max}$, the shape of the knocked-on
electron distribution of Eq.~(\ref{dndgamma})
is $\approx dn/dE\propto E^{-2}$, very close to that
of the accelerated electrons in Eq.~(\ref{acceleration}), for which $p\sim 2.2$.
For photons scattered by knocked-on electrons closer to the cutoff 
$\gamma'_{e,max}$, the effective value of $p$ may be greater than 2.2.

\section{The GRB spectrum}

The ``final"  energy distribution 
$dN/dE$ of the $\gamma$ rays in a GRB pulse is
that of the ``initial'' ambient-light photons, $dN/dE_i$,
 uplifted by ICS  with the electrons
of a moving CB. Let the initial ambient-light number density be approximated as a
thin thermal bremsstrahlung spectrum of temperature $T_i$:
\begin{equation}
{dN\over dE_i}\propto {1\over E_i}\; e^{-E_i/T_i}\, .
\label{thin}
\end{equation}
The energy (or Lorentz factor) distribution of the electrons within the 
CB has two components. One corresponds to the bulk of the CB's
electrons, comoving with the CB and having non-relativistic motions
in its rest frame. The other corresponds to the electrons that have
been accelerated to an approximate power-law distribution. The
total electron-number distribution as a function of their Lorentz
factor $\gamma_e$ is, in the CB's rest frame:
\begin{equation}
{dN\over d\gamma_e'}\propto \delta(\gamma_e' - 1)
+{b'\over (\gamma_e')^{{\tilde{p}}}}\; ,
\label{electrons}
\end{equation}
where $b'$ is a constant that we do not attempt to determine a-priori.
The ``cooling time'' of these accelerated electrons to Compton scattering 
off the photons enclosed in a semi-transparent CB is of the same
order as the Coulomb transparency time of the CB. Therefore, we
expect $\tilde{p}$ to evolve in such a time from $\tilde{p}=p\simeq 2.2$,
the index expected in the absence of cooling, to 
$\tilde{p}=p+1\simeq 3.2$, the index expected for a completely ``cooled''
spectrum (once again, the contribution of knocked-on electrons may 
result in an evolution to $\tilde{p}>3.2$).
In this section we discuss the GRB spectrum at fixed
$\tilde{p}$, the effects of its evolution are discussed in the next section.

\subsection{ICS convolutions, an approximate treatment}

The exact convolution of $dN/dE_i$ and $dN/d\gamma_e$
via Compton scattering involves the angular distribution of the
latter, as well as the angular distributions of the target photons and
the accelerated electrons within the CB. Even in the approximation
in which these distributions are isotropic (in their different respective 
frames), this convolution is fairly complex. We discuss it in
some of its gory
detail in the next subsection. Here we just outline the derivation of the
final result, the various steps being quite intuitive. The gist of the 
simplification is that all of the distributions being convoluted are very
broad, and it is consequently an extremely good approximation to
substitute a (well chosen) subset of these distributions (essentially
the ICS one) by their averages. This fact is familiar in the study of
ICS (Rybicki \& Lightman 1979) and considerably simplifies the 
discussion.

Let us first study ICS by the electrons at rest in the CB. The discussion
is simplest in the SN rest frame, in which the electrons are comoving
with the CB at a common Lorentz factor $\gamma_e=\gamma$, so that
their distribution is $dN/d\gamma_e=\delta(\gamma_e-\gamma)$.
In this frame, the average energy of a Compton up-scattered electron
---viewed in the final state at an angle $\theta$, corresponding 
to a Doppler boost $\delta$--- is
$\bar E(E_i)=(4/3)\,\gamma\,\delta\,E_i\,\langle 1+\cos\theta_i\rangle$. 
Substituting this average
for the corresponding distribution, we obtain: 
\begin{eqnarray}
{dN\over dE}
&\propto&
\int d\gamma_e\,dE_i\;
\delta(\gamma_e - \gamma)\;
 {e^{-E_i/T}\over E_i}\;
\delta\left[E-\bar E(E_i)\right]
\nonumber\\
&\propto&
{1\over E}\; e^{-E/T} 
\;,
\label{deltadist}
\end{eqnarray}
where, including the effect of
cosmological redshift,
\begin{equation}
T\equiv {4\over 3} \;T_i\;{\gamma\;\delta\over 1+z}\;\,\langle 1+\cos\theta_i\rangle.
\label{Teff}
\end{equation}
The interpretation of Eqs.~(\ref{deltadist}) and (\ref{Teff})
is obvious: the target ambient-photon distribution is simply boosted
by ICS on the comoving electrons to a similar distribution at
a much higher energy scale.

Inverse Compton scattering by the power-law-distributed electrons in 
Eq.~(\ref{electrons}) is simplest to discuss in the CB's rest frame.
In it, the initial ambient  photons are beamed towards the CB in a 
narrow cone of opening $\sim 1/\gamma$. They have an energy
distribution akin to that of Eq.~(\ref{thin}), with $T_i\to\gamma\,T_i$,
that is,
$dN/dE_i'\propto exp[-E_i'/(\gamma\,T_i)]/E_i'$.
They collide with electrons of various $\gamma_e$, moving 
isotropically in this frame, so that the collisions are at various
angles and the relative velocity of the ``beams'' is also varying.
We prove in the next subsection that  ---once again because of the 
smoothing effect of convoluting
broad distributions--- the brutal ``approximation" of
considering only head-on collisions is actually a very good one,
it simply changes a little the energy scale of the $E'$
distribution.
In this approximation, the average energy of a scattered photon is 
$\bar E'(E'_i)=(4/3)\,\gamma_e^2\,E_i'\,\langle 1+\cos\theta_i\rangle$, and: 
\begin{eqnarray}
{dN\over dE'}
&\propto&
\int d\gamma_e\,dE'_i\;
{1\over (\gamma_e)^{{\tilde{p}}}}\;
 {e^{-E_i'/(\gamma\,T_i)}\over E_i'}\;
\delta\left[E'-\bar E'(E_i')\right]
\nonumber\\
&\propto&
{1\over E'}\; 
\int_1^\infty 
{d\gamma_e\over (\gamma_e)^{{\tilde{p}}}}\;
exp\left[-{3\,E'\over 4\,\gamma\,T_i\,\langle 1+\cos\theta_i\rangle}
\;{1\over\gamma_e^2}\right].
\label{powerdist}
\end{eqnarray}
To express this result in terms of the observer's GRB energies, we must
replace in the above expression $T_i\to \delta\,T_i/(1+z)$.
The $\gamma_e$ integral in Eq.~(\ref{powerdist}), for any ${\tilde{p}}>1$,  is an 
incomplete $\Gamma$ function
which, for ${\tilde{p}}\sim 3$ and to an excellent approximation, is:
\begin{equation}
\int_1^\infty 
{d\gamma_e\over (\gamma_e)^{{\tilde{p}}}}\;
exp\left[-{a\over\gamma_e^2}\right]
\approx (1-e^{-a}) \; a^{-{{\tilde{p}}-1\over 2}}\, ,
\label{int}
\end{equation} 
an exact result for ${\tilde{p}}=3$. 

We may now replace Eq.~(\ref{int}) into Eq.~(\ref{powerdist}) and add the
result to that of Eq.~(\ref{deltadist}) to obtain the complete spectrum of
the observed $\gamma$ rays in a GRB:
\begin{eqnarray}
{dN\over dE}
&\propto&
\left({T\over E}\right)^\alpha\; e^{-E/T}+b\;
(1-e^{-E/T})\;
{\left(T\over E\right)}^\beta\nonumber\\
\alpha&=&1\; ; \;\;\;\;\;\;\;\;
\beta={{\tilde{p}}+1\over 2}\approx 2.1\, ,
\label{totdist}
\end{eqnarray} 
with $T$ as in Eq.~(\ref{Teff}), and $b$ a constant that we have made
dimensionless by rescaling the two contributions to $dN/dE$
by appropriate powers of $T$. The quoted $\alpha$ and $\beta$ 
are ``preferred'' values, because the power index of the accelerated
plus knocked-on
electrons may not be exactly ${\tilde{p}}=3.2$ (which affects $\beta$); the
ambient-light distribution may not be exactly
``thin-thermal'' and its effective temperature $T_i$
 may vary along the CB's trajectory
(which affects both $\alpha$ and $\beta$). Notice that the shape of
the spectrum in Eq.~(\ref{totdist}) is independent of the CB's expansion
rate, its baryon number, its geometry and its density profile. Moreover,
its derivation rests only on observations of the properties of the
surroundings of exploding stars, Coulomb scattering, and an
input electron-distribution extracted from numerical studies also
based only on ``first principles''.
 
The predicted $dN/dE$ of Eq.~(\ref{totdist}) bears a striking
resemblance to the Band distribution traditionally used to describe
GRB energy spectra (e.g.~Band et al.~1993: Preece et al.~2000
for an analysis of BATSE data, Amati et al.~2002 for BeppoSAX data,
and Barraud  et al.~2003 for HETE II data):
\begin{eqnarray}
{dN\over dE}&=&
\left({E_{ref}\over E}\right)^\alpha\;
e^{-E/E_0}\;\;\;\;
{\rm for}\;\; 
E\le(\beta-\alpha)\,E_0\,;
\nonumber\\
&=&\left[{(\beta-\alpha)\,E_0\over
E_{ref}}\right]^{(\beta-\alpha)}\;
e^{(\beta-\alpha)}\;
\left({E_{ref}\over E}\right)^\beta\;\;\;\;
{\rm otherwise.}
\label{band}
\end{eqnarray}
In this Band spectrum, $E_0$ plays the role of $T$ in Eq.~(\ref{totdist}).
The energy at which $E^2\,dN/dE$ is maximal is often called 
the {\it peak energy}, $E_p$.
Its value is $E_p\simeq E_0$, an exact result for $\alpha=1$, $\beta=2$.

The similarity of the prediction of Eq.~(\ref{totdist}) and the
consuetudinary  spectrum of Eq.~(\ref{band})
is demonstrated in Fig.(\ref{figband}), where we have plotted
the two distributions for $\alpha=1$, $\beta=2.1$, and $T$, 
$b$, $E_0$ and $E_{ref}$ all set to unity. One cannot tell which curve is which!
%In this figure, the parameters
%in the prediction were set to their central {\it predicted} values, while the
%parameters in the Band phenomenological
%spectrum were set to their average {\it observed} values.
Considering that the
prediction is based  on first principles, the agreement is rather
satisfying. 

The distributions of
values of $\alpha$ and $\beta$ extracted from fits to the GRB data 
(e.g.~Preece et al.~2000; Amati et al.~2002; Quilligan et al.~2002) peak 
close to the values expected in the CB model 
$\alpha\approx 1$, $\beta\approx 2.1$, as we show in Fig.~(\ref{figalphabeta}).
The result for $\alpha$, which does dot depend on an adopted power
for the spectrum of accelerated electrons, is more satisfactory than the
result for $\beta$, which does. In particular, the events for which 
$\beta > 2.1$ may reflect, as we have discussed, the cutoff energy
of the knocked-on electrons in Eq.~(\ref{dndgamma}).

In the CB model, spectra with the shape given by Eq.~(\ref{totdist})
are not the exclusivity of GRBs. All plasmas subject to an intense
flux of cosmic rays (such as a CB in its rest system) have analogous
spectra.  Two examples are given in Fig.~(\ref{figspectrum}),
borrowed from Colafrancesco, Dar \& De R\'ujula (2003). 
One of them is that of the SN remnant SNR MSH 15-52, the other
that of the galaxy cluster A2256.

\subsection{ICS convolutions, a sketch of the exact treatment}

Consider an electron belonging to the ``thermal'' (i.e.~unaccelerated)
constituency of a CB, travelling, in the SN rest frame, with a Lorentz
factor $\gamma_e=\gamma$, and Compton scattering ambient-light
photons with the energy distribution of Eq.~(\ref{thin}) and an
approximately isotropic initial directional distribution
(the inclusion of a non-trivial angular dependence with
$\langle \cos\theta_i\rangle\neq 0$ can be made along
identical lines). The exact calculation
of the energy distribution of the photons scattered by the electrons 
---which is also the exact calculation of the relativistic Sunayev--Zeldovich
effect--- can be paraphrased from Cohen, De R\'ujula and Glashow (1998).
Let $d^2N(E,E_i)/dE_i\,dE$ denote the
number of photons transferred by one electron  from the energy interval
$dE_i$ to the interval $dE$. Define:
\begin{equation}
{d^2 N(E,E_i)\over dE_i\,dE}=
{d^2N(E,E_i)\over dE\,dn}\;
{dn(E_i)\over dE_i}\;.
\label{transfer}
\end{equation}
The function $d^2N(E,E_i)/dE\,dn$ may be regarded as the
spectral distribution of struk photons of energy $E$  produced
as the energetic electron Compton-scatters the light of an isotropic,
monochromatic photon gas of unit density and energy $E_i$. 

Let $d\Omega_i(\theta_i,\phi_i)$ be the differential solid
angle about the initial photon direction, and $v_i$ be the relative
speed of the colliding particles. We choose to measure angles
relative to the total momentum direction of the colliding particles.
The function $d^2N/dE\,dn$ is
obtained by 
averaging the differential transition rate over target photon directions:
\begin{equation}
{d^2N\over dE\,dn}=
\int\frac{d\Omega_i}{4\pi} \, v_i \,
\frac{d\sigma(E_i,E)}{dE\,d\Omega_i} \ ,
\label{oneel}
\end{equation}
where we have neglected the tiny effect of stimulated emission.

Since
$\gamma \,T_\gamma \ll m_e$, the Thomson limit applies:
the exact expression for ${d\sigma/dE\,d\Omega_i}$
is relatively simple.
After a little algebra, the integrand in Eq.~(\ref{oneel})
can be rewritten as:
\begin{eqnarray*}
v_i\,\frac{d\sigma}{dE} =
\frac{3\,\sigma_T}{16\, \mu^4\, \beta^5 \,\gamma^{10}\, E_i}
\biggl\{ \mu^2\gamma^2\,
(1+2\gamma^2)(1-2\gamma^2 \mu) + (3-4\gamma^2)\mu^4\gamma^4 +4\mu^6\gamma^6\\
  + r\,(r - 2\,\mu\,\gamma^2)\, 
\left[3 - 6\,\mu\,\gamma^2 + \mu^2\gamma^2\,\left(1 + 2\,\gamma^2 \right)  
   \right]  \biggr\} \, \Theta \left(\frac{\mu}{1+\beta} < r <
        \frac{\mu}{1-\beta}\right) \ ,
\end{eqnarray*}
where $r \equiv E/E_i$ and $\mu\equiv 1-\beta 
\cos\theta_i$.

Carrying out the integrations in Eq.~(\ref{oneel}) gives a
relatively simple result for $d^2 N/ dE\,dn$, a function of only
two variables: $E_i$ and $E$. This can then be introduced
into Eq.~(\ref{transfer}), and integrated in initial photon energies $E_i$.
The overall result of this exercise is extremely well approximated by
the simple and intuitive expression in Eq.~(\ref{deltadist}).

The detailed discussion of ICS by the accelerated electrons is
entirely analogous to the above, though somewhat lengthier,
not because of the need to integrate over their energy distribution, but mainly
because of the extra angular sum over electron directions, akin to that in 
Eq.~(\ref{oneel}). Rather than giving the complete discussion, we outline
the reason why the angular sum ``does not matter'', in the same sense
in which the detailed
angular sum in Eq.~(\ref{oneel}) insignificantly affected the result of 
Eq.~(\ref{deltadist}), in which this average was skirted in an apparently
cavalier fashion.

Let an accelerated electron, in the CB rest system, be moving
with a Lorentz factor $\gamma_e$ and velocity $\beta_e$,
at an angle $\alpha$ relative
to the direction of the ambient photons, travelling in this system
practically along $\alpha=\pi$ (we are using units
in which $c=1$). The relative $e\,\gamma$ velocity
is $v_i=(1+\beta_e^2+2\,\beta_e\,\cos\alpha)^{1/2}$.
The average energy of the photons struk by the electrons, 
upscattered
from an energy $E_i'$ , is
$E'\sim (4/3)\gamma_e^2\,E_i'\,(1+\beta_e\,\cos\alpha)$.
The net result of taking the distribution in $\alpha$ into account
is to modify Eq.~(\ref{int}) to
\begin{equation}
\int_1^\infty 
{d\gamma_e\over (\gamma_e)^{{\tilde{p}}}}
\int^1_{-1} d\cos\alpha\; v_i\;
exp\left[-{a\over\gamma_e^2\,(1+\beta_e\,\cos\alpha)}\right].
\label{alpha}
\end{equation}
This function, to an excellent approximation, has the same shape
as that of the r.h.s.~of Eq.(\ref{int}), simply rescaled by $a\to 0.8a$,
tantamount to a 20\% modification of $T$ in Eq.~(\ref{Teff}).
In an entirely analogous fashion one can demonstrate that,
to an excellent approximation, a deviation from an assumed
isotropic ambient-light bath ($\langle\cos\theta_i\rangle=0$) simply
results in a modification, $T\to T\,\langle 1+\cos\theta_i\rangle$,
of the final ``temperature''.
 
 \section{The time--energy correlation}

Such as we have treated it so far, the distribution $dN/dt\,dE$
of the $\gamma$ rays in a GRB pulse ---as a function of both time and
energy--- is a product of a function of only time, Eq.~(\ref{pheno}), and a 
function of only energy, Eq.~(\ref{totdist}). One reason for this
is that we have
not yet taken into account the fact that the cooling time $t_c$
of the accelerated electrons in a CB ---by Compton scattering--- is of the 
same order of magnitude as the (Compton-scattering) transparency time 
of the CB. Consequently, the index of the power-law
electron energy distribution, $\tilde{p}$ in Eq.~(\ref{electrons}), ought to evolve
in a time $t_c\sim t_{tr}$ from $\tilde{p}\sim 2.2$ to
$\tilde{p}\sim 3.2$, or a bit larger. Equivalently, the index $\beta$ in 
Eq.~(\ref{totdist}) is expected to vary from $\beta=\beta_1\sim 1.6$ to 
$\beta=\beta_2\sim 2.1$, or ``half a bit" larger.
Since the $(e\,\gamma)$ interaction probability
within a CB
varies exponentially with time, we characterize this evolution as follows:
\begin{equation}
\beta(t)=\beta_1\,exp(-t/t_c)+\beta_2\,[1-exp(-t/t_c)],
\label{pevol}
\end{equation}
with $t_c={\cal{O}}(t_{tr})$, given in Eq.~(\ref{ttrans}).
The energy and time distribution within a pulse is then:
\begin{equation}
{dN\over dt\,dE}\propto 
{dN\over dt}\times {dN[\beta(t)]\over dE}\, ,
\label{dEdt1}
\end{equation}
with $dN/dt$ as in Eq.~(\ref{pheno}) and $dN/dE$ as in 
Eq.~(\ref{totdist}). 

To illustrate the time--energy correlation embedded in Eq.~(\ref{dEdt1})
we present results for the times in units of the observer's $t_{tr}$, 
with all parameters fixed at 
their ``central'' values: $t_1=t_2$, $m=n=2$ in Eq.~(\ref{pheno}),
$b=1$ in Eq.~(\ref{totdist}), and $t_c=t_{tr}$, $\beta_1=1.6$, $\beta_2=2.1$
in Eq.~(\ref{pevol}). The parameter $T$ of Eqs.~(\ref{pheno}) 
and (\ref{dEdt1})
is chosen such that the characteristic GRB energies are in the typical
observed range, as in the prediction of Eq.~(\ref{Teff}).
In Fig.~(\ref{Noft1}) we show the pulse shapes (arbitrarily normalized
for presentation) of a single GRB pulse in three energy intervals. 
The pulses are seen to rise faster and be narrower the higher the
energy interval, as in GRB observations (e.g.~Norris et al.~1996 ). In 
Fig.~(\ref{figSpectra1bis})
we show the energy distributions  $E^2\,dN/dE$
(arbitrarily normalized for presentation)
at three time intervals within a pulse. The spectrum is seen to become
softer as time evolves, as in GRB observations (e.g.~Norris et
al.~1996, Frontera et al.~2000).

There is another fact contributing to a non-trivial
correlation between energy and time within a GRB pulse.
The relation between the energy of a struck ambient
photon, $E_i$ ---or the temperature $T_i$ characterizing
its initial distribution--- and those of the resulting GRB photons,
$E$ anf $T$, is that of Eqs.~(\ref{boosting}) and (\ref{Teff}). 
As the CB reaches the more transparent
outskirts of the wind, its ambient light distribution is bound to become
increasingly radial, meaning that the average $1+\cos\theta_i$ in 
Eqs.~(\ref{boosting}) and (\ref{Teff}) will depart from $\sim 1$ and tend 
to $0$ as $1/r^2$: the point-source long-distance limit. Since this transition
depends on the integrated absorption by a wind with 
$\rho\propto 1/r^2\propto 1/t^2$,
we characterize it by a simple time-dependence
of the effective temperature in Eq.~(\ref{totdist}):  
\begin{equation}
T\to T(t)\sim T\,\{1-exp[-(t_T/t)^2]\},
\label{Tevol}
\end{equation}
with $t_T={\cal{O}}(t^w_{tr})={\cal{O}}(t_{tr})$, see 
Eqs.~(\ref{ttrans}) and (\ref{twtrans}).
To investigate the incidence of a varying $T$ by itself (i.e.~separately
 from that of an evolving $\tilde{p}$) we study the distribution:
\begin{equation}
{dN\over dt\,dE}\propto 
{dN\over dt}\times {dN[T(t)]\over dE}\; ,
\label{dEdt}
\end{equation}
with $dN/dt$ as in Eq.~(\ref{pheno}) and $dN/dE$ as in 
Eq.~(\ref{totdist}). 

To illustrate the time-energy correlation embedded in 
Eq.~(\ref{Tevol}) and (\ref{dEdt}) we set all parameters to their
reference values: $t_T=t_{tr}$ in Eq.~(\ref{Tevol}),
 $t_1=t_2=t_{tr}$ in Eq.~(\ref{pheno}),
$\alpha=1$, $\beta=2.1$, $b=1$ in Eq.~(\ref{totdist}).
In Fig.~(\ref{Noft2}) we show the pulse shapes (arbitrarily normalized
for presentation) of a single GRB pulse in two energy intervals. 
Once again, the narrower and 
faster-rising curve is that corresponding to the higher energy interval, 
so that a varying temperature enhances the effect of a varying 
$\beta$, as in Eq.~(\ref{pevol}) and Fig.~(\ref{Noft1}).
In Fig.~(\ref{figSpectra2}) we plot $E^2\,dN/dE$
for a single pulse in three time intervals.
Naturally, the spectral shape is invariant, but
the spectrum gets softer as time elapses within a pulse, adding
to the similar effect induced by a time-dependent $\beta$.
This effect, equivalent to a reduction with time within a pulse
of its fitted peak energy in a Band fit, is also observed
(e.g.~Norris et al.~1996, Frontera et al.~2000).

We shall not embark here in a thorough analysis of GRB 
$(E,t)$ correlations,
embodied in the combination of 
Eqs.~(\ref{pevol}), (\ref{dEdt1}), (\ref{Tevol}) and (\ref{dEdt}), but we
offer two simplified examples of their predictions.

The effect described by Eqs.~(\ref{pevol}) and (\ref{dEdt1}) changes
the shape of GRB spectra with time, but does not affect their
energy scale, as the effect described by Eqs.~(\ref{Tevol}) 
and (\ref{dEdt})
does. The dominant $(E,t)$ correlation is thus the one of the
latter effect, which implies that $E$ and $t$, in $dN/dE$
appear in the combination $E/T(t)$. Since the (exponential) rise of a 
typical pulse, Eq.~({\ref{naivepulse}), is much faster than its
(power) decay, the width of a peak is dominated by its late
behaviour at $t>t_{tr}$. At such times, $T\propto 1/t^2$ in Eq.~(\ref{Tevol}),
so that $dN/dE$ is, approximately, a function of the combination $E\,t^2$.
Consequently the width of a GRB
pulse in different energy bands is:
\begin{equation}
 \Delta t\propto E^{-\kappa},\;\;\;\;\;\;\; \kappa\; \lsim \;0.5,
\label{fenimore}
 \end{equation}
where $\kappa=0.5$ is the limiting value for an exact dependence 
on $E\,t^2$. This result is 
in agreement with the observationally inferred  relation
$t_{_{\rm FWHM}}\propto E^{-0.43\pm 0.10 }$ 
for the average FWHM of peaks as a function of the energies
of the four BATSE channels (Fenimore et al.~1995, Norris et al.~1996),
as shown in Fig.~(\ref{figLogWLogE}).

The width of successive pulses within a given multipulse GRB has also
been studied by e.g.~Ramirez-Ruiz \& Fenimore~(1999,2000), 
Ramirez-Ruiz, Fenimore \& Wu~(1999), Quilligan et al.~(2002) and
McBreen et al.~(2002). 
The result is 
remarkably
simple: the width of pulses of similar $E_p$ is independent of
the time within the GRB's duration at which they are located.
In the CB model there is no reason for an ``ageing'' of the pulses:
the ambient light that successive CBs scatter is time-independent,
since the CBs do not ``make a hole'' in it.

The correlations between time and energy in a GRB pulse 
discussed in this section are dependent on the details
of the CB model. Other correlations, which we proceed to review,
are not ``detail-dependent'', and have  been discussed
before (Dar \& De R\'ujula~2000b; Plaga 2001).

\section{More on correlations}

\subsection{``Relativistic'' correlations between pulse properties} 

The CB model predicts very
simple approximate correlations between pulse properties that 
depend only on special relativity (the various relations between
times and energies reviewed in section 4.1 and the 
relativistic light-beaming effects discussed in section 4.2)
and on general relativity (in the sense of involving the ubiquitous
factor $1+z$ of an expanding Universe). Naturally, the correlations
should be better satisfied if one can correct for the latter effect,
as one can for the GRBs with known redshift, 32 at the time of writing.

If core collapse SNe and their environments were all identical, and
if their ejected CBs were also universal in number, mass,
Lorentz factor and velocity of expansion, all differences
between GRBs would depend only on the observer's
position, determined by $z$ and the angle
of observation, $\theta$. For a distribution of Lorentz factors that,
as observed, is very narrowly peaked around $\gamma\simeq 10^3$,
the $\theta$-dependence is in practice the dependence on $\delta$,
the Doppler factor. This dependence is strong in various observables,
e.g.~cubic in the fluences of Eqs.~(\ref{Fluence}), (\ref{Fluenceapprox}).
Therefore, the correlations between these observables and others
that are only linear or antilinear in $\delta$ ---such as the energies
in Eq.~(\ref{energy}) and the times in Eq.~(\ref{time})--- are 
``strong correlations'' and they might
overwhelm much of the case-by-case variability induced by the
distributions of the other parameters.

Let $\Delta t\propto 1/\delta$ be any measure of time, such as the
width of a pulse, its rise time, or the ``lag time'' (the difference 
between
the peak times of a given pulse in two different energy intervals).
A measure of energy, such as the peak-energy $E_p$ in Band's
spectrum, is $\propto \delta$. The photon-number fluence is 
$f\propto \delta^2$, as in Eq.~(\ref{photonfluence}). 
The peak photon intensity $f_p$ (number of photons
per unit time), the energy fluence, $F$ in 
Eq.~(\ref{Fluence}), and the ``isotropic'' energy of 
a pulse, ${\cal{E}}^{iso}$ in Eq.~(\ref{Eiso}),  
are all proportional to $\delta^3$. Finally, the
peak luminosity $L_p$ (energy fluence per unit time)  
is proportional to $\delta^4$. 
All this implies, among others, the following correlations
(Dar \& De R\'ujula 2000b):
\begin{equation}
E_p\propto [\Delta t]^{-1}\, ,
\label{epwidth}
\end{equation}
\begin{equation}
E_p\propto [f_p]^{1/3} \, ;~~~~ \Delta t\propto [f_p]^{-1/3}\, ,
\label{epint} 
\end{equation}
\begin{equation}
E_p\propto F^{1/3}\, ; ~~~~~ \Delta t\propto [F]^{-1/3}\, ,
\label{epflu}
\end{equation}
\begin{equation}
E_p\propto [{\cal{E}}^{iso}]^{1/3}\, ; ~~~~
\Delta t\propto[{\cal{E}}^{iso}]^{-1/3}\, ,
\label{epeiso}
\end{equation}
\begin{equation}
E_p\propto [L_p]^{1/4}\, ;~~~~ \Delta t\propto [L_p]^{-1/4}\, .
\label{epilum}
\end{equation}
The correlation of Eq.~(\ref{epwidth}) is independent of redshift, all
others should be better satisfied for pulses of GRBs with known
redshift, after correction for the $z$-dependence. 

In Figs.~(\ref{figEpFp}) and (\ref{figEpFot}) 
we show two examples of the above correlations:
the average peak energy, $E_p$, versus 
peak photon intensity, $f_p$, and versus total fluence, $F_{tot}$, in bins
containing  20 GRBs of similar peak intensity (Lloyd et al.~2000).
The respective lines are the prediction of Eqs.~(\ref{epint}) and
(\ref{epflu}).
Another version of the $(E_p,f_p)$ correlation is that in
Fig.~(\ref{figEpInt}), based on a data analysis by
Mallozzi et al.~(1995), where
the line is again the prediction of Eq.~(\ref{epint}).
In Fig.~(\ref{figLogALogW}) we show the prediction of Eq.~(\ref{epint}) for
the normalized ratio $ f_p/{\langle f_p\rangle}$ of peak
pulse fluences versus the pulse's FWHM. The data analysis is
from Ramirez-Ruiz \& Fenimore (2000), who state:
{\it If we were
to use the average point of all the normalized amplitudes in each selected
range, the result is a power law: $f_p / {\langle f_p \rangle} \sim
[{\rm FWHM}]^{-3.0}$}. This is precisely the prediction.

A correlation that ---to our knowledge--- has not been investigated, is
the following. The time delays between the pulses of a GRB are
simply stretched by a factor $1+z$  relative to the emission times
of the corresponding CBs at the location of the parent SN.
The same is the case for the total duration of a GRB. On the other hand,
the GRB energies and the time intervals within pulses, relative
to their values in a CB's rest system, are related as in 
Eqs.~(\ref{energy}) and (\ref{time}), which invlove the combination
$\delta/(1+z)$.
The consequent relations in Eqs.~(\ref{fenimore}) and
(\ref{epwidth}) are $z$-independent, while all of the
other relations in Eqs.~(\ref{epint}) to (\ref{epilum}) have the explicit
$\delta$ and $z$ dependences that can be read from 
Eqs.~(\ref{photonfluence}) to (\ref{Nvstheta}). All this opens a host
of obvious combinatorial possibilities from which one could
extract, in a statistical sense, GRB distributions in $z$ and $\delta$,
which are explicitly testable for GRBs of known redshift.
To give an example, the time intervals, $t_{int}$, between pulses increase
with $z$ as $1+z$, while the widths of the pulses, $t_{FWHM}$, increase as
$(1+z)/\delta$, with their ``power'' ($E_p\,\Delta t$) remaining constant.
The ratio $t_{int}/t_{FWHM}\propto\delta$
is independent of $z$. Since at higher $z$ GRBs with higher $\delta$
(narrower pulses)
are favoured by selection effects, this may in part explain the ``Cepheid-like''
correlation between variability and redshift advocated by 
Fenimore and Ramirez-Ruiz (2000) and by  Reichart et al.~(2001).

\subsection{Correlations between global GRB properties}

The correlations in Eqs.~(\ref{epwidth}) to (\ref{epilum}) apply
to individual pulses and to pulse averages over a GRB. When applied
to global GRB properties, some of these correlations are expected
to have larger scatter than they have for individual or averaged pulses.
An example is the correlation of any quantity with
the total duration $W$ of a GRB, often defined as $t_{50}$
or $t_{90}$ for the per cent of total energy measured
in a given time interval $t_i$. These correlations mix the pulse durations
with the durations of the inter-pulse intervals which have, as discussed 
in the previous subsection, a different $\delta$ dependence. 
In Fig.~(\ref{figT90}) we show the trend of the
observed GRB durations ($t_{90}$) versus the total energy fluence
in the 7-400 keV band of 35 GRBs that were measured by HETE II
(Barraud et al. 2003). The theoretical continuous line is the average trend 
expected in CB model for the width of the individual
pulses. The theoretical dashed line is the expectation for the time
intervals between pulses.

For GRBs
with several pulses ($N_p\gg 1$), $W$ is approximately proportional to $N_p$,
since the mean time interval between pulses is independent on the GRB's
duration (e.g.~McBreen et al.~2002) and $N_p-1\approx N_p$. One may thus
use this proportionality to obtain approximate correlations between global
GRB properties (indicated by a GRB subindex), such as 
${\cal{E}}^{iso}_{_{GRB}}\propto W\, {\langle
E_p\rangle}^3 $ and $F_{_{GRB}}\propto W\, {\langle E_p\rangle}^3$,
with $\langle E_p\rangle$ the average of the peak energies of the
pulses in a given GRB. Such relations are also well satisfied
by the observations.

The  time variability of a GRB is a ``global'' measure of an inverse
time,  $V\propto L_p^{-1/4}$, according to
Eq.~(\ref{epilum}). This variability--luminosity relation
is shown in Fig.~(\ref{figLV}), for GRBs of known redshift 
(Reichart et al.~2001). An example of the variability--peak energy
correlation in Eq.~(\ref{epilum}) is given in Fig.~(\ref{figEpVar}),
in which the data analysis is from 
Ramirez-Ruiz \& Lloyd-Ronning (2002).

\section{A detailed example: GRB 980425}

This GRB and its associated supernova, SN1998bw ---both
traditionally considered entirely exceptional in the FB models--- 
are the battle-horses of the CB model. Neither of them is
---in the CB model--- a special class onto itself: 
\begin{itemize}
\item{}
In Dar \& De R\'ujula (2000a) we argued that the only 
peculiarity of GRB 980425 was its nearness ($z=0.0085$), 
which allowed for  its detection at an angle, $\theta\sim 8$ mrad,  
unusually large with respect to that of all other GRBs of known redshift,
for which $\theta\sim 1$ mrad. 
These facts conspired to produce a ``normal'' GRB fluence
---given the strong $\delta$-dependence of Eq.~(\ref{Fluence})--- and 
resulted in an optical AG dominated by the SN. 
\item{}
In Dado et al.~(2002a)  we  demonstrated 
that the X-ray AG of this GRB was also ``normal'': it has precisely the  
light curve (in shape and normalization) expected in the CB model
if the X rays are produced by the CBs and {\it not}, 
as the observers assume (Pian et al.~2000), by the associated SN.  
Moreover, subsequent X-ray data from XXM Newton  and 
Chandra (Pian et al.~2003) agree exactly with the prediction
in Dado et al.~2002a; 2003a). 
\item{} In Dado et al.~(2003a) we demonstrated that
the radio AG of this GRB was also ``normal'': it has precisely the
normalization, spectrum and fixed-frequency light curves  
expected in the CB model
if the radio emission is produced by the CBs and {\it not}, 
as the observers assume (e.g.~Kulkarni et al.~1998), by SN1998bw.
\item{}
Deprived of the X-ray and radio emissions that it did not emit,
SN1998bw loses almost all of its alleged exception.
Its only peculiarity was that it was viewed very near its 
axis,
in comparison with ordinary SNe. This is no doubt the reason why
exceptionally high velocities ($v\sim 23,000\pm 3,000\, {\rm km\, 
s^{-1}}$) 
for its expanding shell were
deduced from observations of its line emissions (Patat et al.~2001). 
Indeed,
the exiting jets of CBs are surely accompanied by a fast outward
motion of the SN shell in its ``polar caps''. Viewed almost on-axis,
such motions should result in highly Doppler-boosted line emissions. 
\item{}
Since GRB 980425 was an ordinary GRB\footnote{In the FB models,
all long-duration GRBs ---but GRB 980425--- are dubbed ``classical''
(e.g.~the discoverers of SN2003dh: Stanek et al. 2003) to distinguish 
them from that
``exceptional''  one.} it makes sense ---{\it in the CB model}--- to use
SN1998bw as a potential standard candle, or {\it standard-torch,} 
associated with other
GRBs (Dar 1999a,b). This naive hypothesis has met an incredible
success. In all GRB AGs for which such a SN contribution could in
practice be discerned, the contribution was discernible, with various
degrees of significance (Dado et al.~2002a). In four cases 
Dado et al.~2002a,b; 2003e,f)  we
{\it predicted} the presence of a SN1998bw-like contribution from AG
data taken {\it before} the SN was observable. The last and
quite impressive case\footnote{For the sake of fairness we
must report that, according to Anonymous (2003): ``Astronomers
point out that the CB theory did not predict the event from first 
principles''.} is that of GRB 030329 and SN2003dh (Dado et al~2003f;
Stanek et al.~2003). 
\end{itemize}

In this section we complete our argument regarding the ``normality''
of GRB 980425 by studying in more detail the $\gamma$-rays of its burst.

The total observed fluence of this GRB was (Kippen et al.~1998;
Frontera et al.~2000):
\begin{equation}
{dF\over d\Omega}={1\over 2\pi}\,{dF\over d\cos\theta}
\simeq0.44\times 10^{-5}\;\rm erg\; cm^{-2},
\label{F425}
\end{equation}
comparable to that of other GRBs of known redshift, but corresponding
to a spherical-equivalent energy $8.1\times 10^{47}$ erg, some five
orders of magnitude smaller than average.
The energy-integrated spectrum was analysed in the Band
model by Yamazaki, Yonetoku \& Nakamura (2003), with the result
that $\alpha=1\pm 0.3$, $\beta=2.1\pm 0.1$, both perfectly compatible
with the central expectations of Eq.~(\ref{totdist}). The peak energy
was found to be: 
\begin{equation}
E_p=54.6\pm 20.9\; \rm keV,
\label{peak}
\end{equation}
anomalously small by average standards. The energy spectrum in
various energy bands is shown in Fig.~(\ref{fig425spectra}), from
Frontera et al.~2000. The shape of the single pulse (or single dominant CB)
of this GRB, in the 50--300 keV energy band (Kippen 1998), is shown in 
Fig.~(\ref{fig425}). 

The relative smallness of the fluence of this GRB can be trivially understood
(Dar \& De R\'ujula 2000a). For a rough estimate, 
take the single CB's rest-system energy output 
to be that of Eq.~(\ref{Eprime}), with all parameters fixed at their
reference values. We may then use $dF/d\Omega$ as
in Eq.~(\ref{Fluence}) and the known distance to the progenitor (39 Mpc
for $H_0=65$ km s$^{-1}$ Mpc$^{-1}$) to obtain the value of $\delta$
that would reproduce the observation in Eq.~(\ref{F425}): $\delta\sim 20$.

With the above value of $\delta$ and all other parameters fixed at their
reference values, we may proceed to investigate whether or not 
the $\gamma$-ray pulse of GRB 980425 was in any sense exceptional.
From Eq.~(\ref{boosting}) we obtain a peak energy of 20 keV, fairly close
to the 1$\sigma$ lower limit of Eq.~(\ref{peak}). For the full width at
half-maximum of the pulse's peak we obtain 11.6 s, by use of 
Eq.~(\ref{ttrans}),
in agreement with the observation reported in Fig.~(\ref{fig425}).

In Dado et al.~(2003a) we obtained the parameters
$\theta$ and $\gamma$ for this GRB from the observations of its
radio AG: $\theta=7.83$ mrad, $\gamma=495$, corresponding to
$\delta=62$. These parameters are not as well determined as
for other GRBs, since we have no information on the optical
AG, which, given the large value of $\theta$,
 was overwhelmed by the light of SN1998bw. Moreover, the
 X-ray data for this GRB are quite poor, and the
 radio AG, which is well measured is (at the
large $\theta$-value of this GRB) most sensitive to details such
 as the CB's geometry (the radio emission originates
 in the CB's surface, as opposed to the higher-frequency AG
 emissions, to which the CB is transparent). We therefore
 do not consider the ``large'' discrepancy between $\delta=62$ and
 $\delta=20$ to be a problem, particularly since the $\gamma$-ray
 properties of this GRB could be easily described with either
 $\delta$, should we be willing to moderately depart from
 the use of the reference values for the rest of the parameters
 in Eqs.~(\ref{boosting}), (\ref{ttrans}). It is clear that ---in the CB
 model--- GRB 980425 was not exceptional, and we may proceed
 to study its spectrum and light-curve in more detail.
 
 To describe the spectrum of GRB 980425 in various time
 brackets, we use Eq.~(\ref{totdist}),
 with $\alpha=b=1$, their reference values and $T=E_p$, given
 in Eq.~(\ref{peak}). The quantity $\beta$
 in Eqs.~(\ref{totdist}) and (\ref{pevol}) is allowed to evolve from a
 value $\beta_1=1.6$ to $\beta_2=2.3$ ---the observed $\beta$ 
 peak value, as in Fig.~(\ref{figalphabeta})--- according to Eq.~(\ref{pevol}),
 with $t_c=t_{tr}^w/2$ (this is the only parameter we choose not
 to have its precise reference value).
 We also allow for the effect of a varying $T(t)$, as in Eq.~(\ref{Tevol}),
 with $t_T=t_{tr}$, its reference value. For each time bracket 
 we present results for the corresponding average time, since the brackets are
 narrow and a time-integration insignificantly affects the results.
 The result of this exercise is
 shown in Fig.~(\ref{fig425spectra}) and is quite satisfactory, considering
 that we are ``describing'', rather than fitting, the spectral evolution.
% It is clear by inspection
% of the figure that a slight increase in $\beta_2$ suffices to obtain 
% an even better ``description''. 
 
In Fig.~(\ref{fig425}) we repeat the above exercise, with the same
parameter choices, to describe the very good data on the 
pulse-shape of GRB 980425 in the 50--300 keV bracket. This time the
bracket is wide, and we have integrated $dN/dt\,dE$ in the
quoted energy interval. The result, naturally, is sensitive to the
width of the pulse, which we have taken to be the observed
$\sim 25$ s full width at half-maximum of the pulse integrated
over {\it all} observed energies (Frontera et al.~2000).
 
Our results prove that GRB 980425 was a perfectly normal GRB
that just happened to occur nearby, and to be observed at a
comparatively slant angle. This makes the use of its associated SN 
---as a putative standard candle, or standard torch,
for other SNe associated with GRBs--- a reasonable
undertaking. The next best-measured GRB-associated SN
is the next closest one: SN2003dh, associated with GRB 030329
(Stanek et al.~2003). SN2003dh and SN1998bw look like
identical twins, as shown in the comparison of their spectra
in Fig.~(\ref{figSN2003dh}), from Matheson et al.~(2003). 

\section{The parameters of the cannonball model}

More often than not, the description of astrophysical phenomena involves
parameters that are chosen ``just so'' that the desired result is obtained.
As we discuss in Section 19 and Appendices II and III, the
FB models  of GRBs are no exception. The CB model is.

CBs decelerate in the interstellar medium by 
``collisionless'' interactions with the ISM's heavy constituents:
protons and nuclei, which bounce off their enclosed magnetic field. 
The particles recoiling from a single such collision
have energies extending up 
to $E_{max}=2\,\gamma^2\,M\,c^2$ (or even larger in the case of multiple 
collisions).
The particles thus accelerated are {\it the}
cosmic rays of the GRBs' host galaxy.
These $M$-dependent values of $E_{max}$ coincide with the ``knees'' in 
the cosmic-ray spectra  
for $\gamma\simeq10^3$, and qualitatively explain their features 
(Dar \& Plaga 1999). Thus the {\it ab initio} choice of the typical 
$\gamma$ (Dar \& De R\'ujula 2000a).

The initial expansion velocity of a CB is naturally chosen to be of the order of the
speed of sound in a relativistic plasma, i.e. $\beta_s\sim 1$. The photon-number 
density of the ambient light
and the winds' density we use here originate in observations.
The observer's angle $\theta$ is, like the redshift $z$, not a property of
the CBs themselves.

The large peculiar velocities of neutron stars, 
$ \langle v \rangle\!\sim\! 450$ km s$^{-1}$,
 are attributed, in the CB model, to an
imbalance in the momentum of the two opposite jets of CBs that accompany their
birth (Dar \& Plaga 1999). Neither that imbalance (perhaps 50\%) nor the 
average number of
significant CBs per GRB (perhaps 10) can be easily ascertained. The choices in
parenthesis and an initial $\gamma=10^3$, for stars of mass 1.4 $M_\odot$, yield
an {\it a priori} rough estimate, $N_{_{B}}=5\times 10^{50}$, 
for the CB baryon number\footnote{Our
original estimate $N_{_{B}}=6\times 10^{50}$ (Dado et al.~2002a)
was based on a 10\% asymmetry and fewer cannonballs, but it took into account
that CBs lose a fraction of their original Lorentz factor as they cross the SN shell.
Not knowing precisely the jet asymmetry, nor the shell's properties in the
polar directions, we have to live with some a-priori uncertainty, which 
would look
very small if we wrote it as: $N_{_{B}}=10^{51\pm1}$. }. 
In the study of AGs, a deceleration parameter $x_\infty$ is
introduced. It governs the slow-down of CBs by collisions with the ISM.
Its value is fixed by $N_{_{B}}$, the ISM density, and the calculable
fixed asymptotic value of the CB's radius, reached within a few minutes of
observer's time.  The values of $x_\infty$ that the AG fits return are in
the anticipated range (Dado et al.~2002a). In Dado et al.~(2003a)
we found that our ab initio parameter choices resulted in a one
order of magnitude overestimate of the AGs' absolute luminosity. Whence 
our
slight reduction of the originally adopted $N_{_{B}}$, by a factor of
$\sim 6$.

In the discussion of the wide-band spectrum of a GRB AG, only {\it one}
extra independent parameter must be introduced, a free-free absorption
frequency $\nu_a$, describing the attenuation of radio waves at a CB's
surface. This is the only parameter for which we do not have a solid
a-priori estimate.

In this paper we have introduced a series of parameters pertinent to the
description of a GRB itself: $t_1$ and $t_2$ in Eq.~(\ref{pheno}), 
$\alpha$, $\beta$ and $b$ in Eq.~(\ref{totdist}), $\beta_1$, $\beta_2$
and $t_c$ in Eq.~(\ref{pevol}). These are quite a few parameters,
but their central predicted values coincide
with those deduced from observations. Here the only exception
of a parameter for which we have not attempted to derive an
a-priori value is $b$ in Eq.~(\ref{totdist}), which sets the relative
contributions of accelerated and unaccelerated CB electrons to
the GRB energy spectrum.

The conclusion of this section is that it is fair to say that the 
overwhelming majority of the parameters 
of the CB model either have theoretically predicted central values
in agreement with observations or ---since CBs play a role in areas of 
astrophysics other than GRBs--- are not 
specifically fine-tuned to the GRB observations.

\section{Limitations of our current analysis}

\subsection{The timing of the CB emission}

Two time scales characterize the infall of material onto the newly formed
compact object resulting from the collapse of the core of a star.
One is the infall time of matter in the immediate vicinity of the pre-collapsed
core, of the order of tens of seconds. The second, of the order of hours,
is the fall-time of material from the outer layers of the pre-collapsed
star. There is no a-priori way of knowing whether the accreting disk 
or torus, whose further successive
episodes of violent accretion generate the CBs, is formed on the first
or the second time scale. Current models of SN explosions do not explode
SNe\footnote{Neither do they produce highly relativistic
CBs or shells of $e^+e^-$ pairs. In the
jetted emission of CBs and the more isotropic ejection of a SN shell,
we do not know who is the chicken and who is the egg.},
unless some ad-hoc physics is added to them: the position where
their ``shocks stall" may not be a serious indication of the location
of matter that may ``fall-back". The only indication of a ``double bang"
in a SN (De R\'ujula 1987) was provided by SN1987A, which generated 
two neutrino bursts, one irrefutable in its statistical significance
(Bionta et al.~1987; Hirata et al.~1987) the other much
less so (Aglietta et al.~1987). These bursts, both lasting a few seconds, 
were separated by some 5 h.

In our original model, we assumed that the CBs were emitted a few hours
after core-collapse (Dar \& De R\'ujula 2000a). 
We have implicitly made here the same assumption,
for it also takes hours for the information that the core has collapsed to
reach the surface of the parent star, the moment at which the star's
luminosity rises sharply to ionize and illuminate its prior wind. But we
do not know whether or not processes occurring prior to the collapse, such
as the violent wind-illuminating episode shown in Fig.~(\ref{CBGlory}) may
be the ones responsible for generating the ambient echo light. In this
latter case there would be only one time-scale associated with GRBs: the
shorter one. This, if the second neutrino bang of SN1987A is simply
ignored, is an attractive possibility.

The only methods to tell apart the one-time and two-time scenarios
with the help of GRB observations are to measure the timing
of a GRB relative to the prompt flash of its associated SN
(difficult, because the signal is superimposed on the
AG ``background''), or relative to the thermal neutrino flash (a dream).
The launching of CBs from the SN's central region ought to
be accompanied by the emission of gravitational waves,
which are much more isotropic than the GRB  (Segalis and Ori 2001).
An observation of these waves in coincidence with a GRB (most unlikely)
or with a SN explosion (not out of the question) would also
help. Finally, it is conceivable that a convincing double bang
be observed in the form of two separate thermal neutrino bursts
from a SN explosion in the Galaxy or the Magellanic Clouds.

\subsection{The GRB energy emitted by a single CB}

In Section 8, we compared the prediction for the total energy, 
${\cal{E}}'_{CB}$,
emitted by a single CB in its rest system ---in the form of photons that
an observer sees as $\gamma$-rays--- with the range of values
(0.6 to 2.1 times $10^{44}$ erg) that we had previously extracted
from analyses of GRB AGs (e.g.~Dado et al.~2002a).
The question is: Does this ``observed'' range correspond to the emission 
by a 
{\it single} CB?

Two cases in which the answer is affirmative are: GRB 030329
(Dado et al.~2003f), 
which we discuss in further detail in Appendix I,
and GRB 021004 (Dado et al. 2003c).
These cases have various common properties: the GRB's $\gamma$-ray
light curve has a very clear two-peak (two-CB) structure,
as can be seen in Fig.~(\ref{fig329NC}); the AG fluence
evolution has two humps which correspond, in the CB model, to 
the
individual contributions of the two CBs, as in Fig.~(\ref{fig329red}); 
and the AG of both  GRBs
was ``caught'' very early: good measurements of it were taken before
$t\sim 1$ day. This allowed us to perform analyses in which the extracted
${\cal{E}}'_{CB}$ indeed corresponds to each single CB.

Several other cases we have studied (such as GRB 980425, discussed
in detail here and in Dado et al.~2003a) the GRB light
curve has a single or a single-dominant CB. For these cases
the answer is also affirmative.
But there are cases, such as that of the fluence record-breaking
GRB 980123, in which the GRB light curve has several clear pulses.
In most of these cases, however, the AG was caught at a relatively
late time, so that it is quite possible that it was dominated (as the
relatively late time AG of GRB 021004 or GRB 030329 is) by a single CB.
In that case, the value we extracted for ${\cal{E}}'_{CB}$ would also
correspond to just one CB.

Finally, there may be a small fraction of cases in which the AG,
analysed by us as if due to a single CB, received comparable
contributions from various CBs, with similar Lorentz factors
and emission angles. For those cases, our extracted ${\cal{E}}'_{CB}$
corresponds to a sum of CBs, and the ``real'' single-CB number 
would be smaller by a factor of a few. This does not 
detract much from the striking coincidence between the range of ``observed''
values of ${\cal{E}}'_{CB}$ and the prediction of Eq.~(\ref{Eprime}).  

\subsection{The coalescence of CBs}

Two CBs emitted roughly in the same direction, at times differing by
$\Delta t$ in the SN rest frame, may coalesce on their way to the observer,
if the Lorentz factor, $\gamma_2$, of the second one is larger than
that of the first, $\gamma_1$. Is this a complication that needs to be
addressed in detail? The short answer is: No. Thus, we shall discuss
the issue with all parameters set at their reference values, but for 
%the expansion velocities, which play an important role and (necessarily)
one of the Lorentz factors.

The typical separation of pulses in a long-duration GRB is of
${\cal{O}}(1\, \rm s)$ which, at a typical $z=1$, corresponds to
$\Delta t=2$ s in the local SN rest frame. At the CB-transparency
time at which the bulk of its GRB pulse is emitted,
a CB has moved to a distance $r_{tr}\sim 7\times 10^{15}$ cm
from the SN, see Eq.~(\ref{disttrans}). Consider
first two CBs emitted {\it with the same $\gamma$ and in the same
direction $\theta$}
at times 0 and $\Delta t$. Their centres would travel at a constant 
distance $d=c \, \Delta t=6 \times 10^{10}$ cm, but could their
expansion make them touch? The CBs' longitudinal expansion rate 
in the SN frame is a factor $\gamma$ smaller than in the CBs'
rest frame, so that their longitudinal dimension at transparency time is 
$l=R_{tr}/\gamma\sim 4\times 10^9$ cm, where we have used 
Eq.~(\ref{rtrans}). Since $l\ll d$, the answer is negative,
unless $\Delta t$ happens to be much smaller, but in that case
the pulses of the two CBs would be superimposed anyway.
This takes us to the next case in point: $\gamma_2 > \gamma_1$.

The time it takes the centres of two CBs emitted in the same
direction, at an interval $\Delta t$, to have their centres touch is:
\begin{equation}
t_{merge}={\Delta t\over \beta(\gamma_1)-\beta(\gamma_2)}
\simeq2\,\Delta t \;{\gamma_1^2\;\gamma_2^2 \over \gamma_2^2-\gamma_1^2}.
\label {merge}
\end{equation}
For $\gamma_2=2\, \gamma_1=10^3$, the corresponding
merger distance is $c\,t_{merge}=4\,\times 10^{16}$ cm, about
one order of magnitude larger than $r_{tr}$ in Eq.~(\ref{disttrans}), the distance
at which the GRB pulse is emitted, so that, typically, a CB--CB
coalescence would not be observed in the GRB phase, but
in the very early AG phase. 

The ratio $c\,t_{merge}/r_{tr}$ is sufficiently close to unity that 
we may ask: What happens in the 
unlikely case that two CBs do merge? The electron
constituency of the merged object will be boosted to a somewhat
higher energy, by the shockless acceleration process we have
advocated (at the relatively small relative Lorentz factor). The merger 
takes a time $\sim 2\,R_{tr}/c$ in the merged CB's rest frame, and 
a cosmological observer sees it shortened by a factor
$(1+z)/\delta$ to $\sim 1/2$ s, for typical parameters.
Since at the time of merging the two CBs are already transparent,
or nearly so, their total ICS signal, proportional to 
the sum of their baryon numbers, as in Eq.~(\ref{latepulse}),
is not significantly affected,
but the sudden modest hardening of the electron spectrum would give
rise to a hardening of the spectrum during a pulse.

Finally, we may ask how close the angles, $\theta_1$ and $\theta_2$,
of the emission of two CBs must be, in order for a merger 
at the GRB-emission (or CB-transparency) time to be at all possible.
At the merger time, the
 CBs are at a distance $r_{tr}=\sqrt{3}\, \gamma\, R_{tr}/\beta_s$ away
from the SN, and at a transverse distance 
$d_t=|\theta_2-\theta_1|\,r_{tr}$ from each other. For that distance
to be smaller than $R_{tr}$ (so that they touch), it is required that
$|\theta_2-\theta_1|<\beta_s/\sqrt{3}\, \gamma$. For $\beta_s$
close to its upper limit 1 that is fairly probable,
so that our previous conclusions stay put.

If two or more CBs merge during the late GRB phase, the number of
CBs that give rise to the AG may be reduced, relative to the
number of significant pulses in the GRB. This would go in the
direction symplifying the analysis of AGs and reducing the spread in the values of
${\cal{E}}'_{CB}$ extracted from fits to the AGs, which is observed
to be very small.

\subsection{Other $\gamma$-ray emitting mechanisms}

The shell of ejecta of a SN should not be spherically symmetric, for various
reasons. The angular momentum of the parent star implies that its
collapse should be faster close to the poles than to the equator. The exiting
jet of CBs no doubt affects the shell around its polar caps. Assuming these
effects not to be very important ---that is, in  the approximation of a
spherically symmetric shell--- we estimated in Dar \& De R\'ujula (2000b)
how the collisions of the CBs with the shell material would
heat the CBs. We conjectured that a GRB pulse would be the radiation
emitted by the heated CBs, visible after the CBs enter the transparent
outskirts of the SN shell, and would diminish in temperature (and 
$\gamma$-ray
energy) as the shell's material thins down. We found that two models
along this line (a ``surface-'' and a ``volume-" heating model) were 
also very successful in predicting all the properties of a GRB pulse for CB 
parameters in the range we have rediscussed here. Why then do these 
quasi-thermal (QT)
variants of the CB model not imply that there is
an additive contribution to the ICS process discussed in this paper?

The simplest answer to this question is that such a contribution may
indeed be there. The second simplest
answer is that the assumption of spherical symmetry may be very wrong:
the polar regions of the SN shell may be depleted in density, in which case
the quasi-thermal GRB-generating process would be subdominant. We
subscribe to the second possibility for the following reasons:
\begin{itemize}
\item{}
The QT models predict a vanishing GRB polarization.
\item{}
In each QT model we had to make one rough approximation.
In the simplest ``surface model'', for instance, we assumed that
all the heat hadronically deposited in a GRB within a photon's
interaction length from its surface was instantaneously
re-emitted. The beauty of the ICS process that we have discussed
here is that no such approximations have 
to be made: all the physics is simple, transparent and unquestionable.
\end{itemize}

\section{Are short GRBs generated by Type Ia SNe?}

We contend that, in the CB model, it is very natural to answer the
question affirmatively (Dar \& De R\'ujula 2003):
the observed properties of short GRBs are
understood in much the same way as those of the long-duration
ones, which we discussed so far. The information we have about Type 
Ia SNe, however, is very meager relative to that on core-collapse SNe,
which encompass all other spectroscopic types. Thus, our 
considerations stand on less firm grounds. Yet, as we shall see,
the model explains
why the AG of short-duration GRBs should be hard to detect.
This means that observers should look in the GRB's direction 
not only as fast as possible after the GRB time (in case an AG
is discernible) but also about one month later,
when it should be possible to discover the
allegedly associated Type-Ia SN around its peak luminosity.  
Such a SN, if it is there, may not look identical to the 
standard-candle Type Ia SNe so useful in cosmology, for
it would be viewed nearly on-axis.

Our suggested association of short-duration GRBs with Type Ia SNe
(Dar \& De R\'ujula 2003) appeared to be in conflict
with the possible association (Germany et al.~2000)
of the short-duration ($\sim 0.2$ s)
GRB 970514 with SN1997cy at $z=0.07$, which is 
Type IIn, and not Type Ia. 
But this situation was recently reversed.
Hamuy et al.~(2003) observed that SN2002ic at 
almost the same redshift, $z=0.0666$, 
---whose Type is Ia--- had a spectrum almost identical
to that of SN1997cy. Moreover, unexpectedly, its spectrum
also featured a narrow H$\alpha$ line of
FWHM $<300$ km s$^{-1}$, on top of a broader line
of FWHM $\sim 1800$ km s$^{-1}$, see Fig.~(\ref{figSN2002ic}).
Hamuy et al.~(2003) suggested that the identification of
SN1997cy as Type IIn could have been wrong, and that most probably it was of 
Type Ia, like SN2002ic. They also concluded from the light-curve 
and the lines of SN2002ic that the explosion's energy, like that of
SN1997cy, was unusually high, 
$ \sim 3\times 10^{52}$ erg. Since the thermonuclear explosion of 
a carbon--oxygen Chandrasekhar white dwarf is very unlikely to produce 
more 
than $2\times 10^{51}$ erg (Arnett 1996), we must conclude that Type Ia
SNe are more axially than spherically symmetric and that
SN1997cy and SN2002ci were observed close to their jet axis, like SN1998bw.
There, the expansion velocity is naturally much higher than 
near the equator, leading to an overestimate of the explosion's energy.

\subsection{Type Ia SNe}

Little is known for sure about the progenitors or the production
mechanisms of Type Ia SNe. The prevailing theory is that accretion onto a
$C/O$ White Dwarf (WD) from a companion star in a close binary
system causes their collapse ---accompanied by a thermonuclear
explosion--- when the accreting WD's mass exceeds the Chandrasekhar limit
(Whelan \& Iben 1973).  In the case of a WD--WD binary, the trigger may
also be a merger, the end-result of a shrinking of the orbit due to
gravitational-wave emission (Iben \& Tutukov~1984; Webbink~1984). 

In every one of the quoted scenarios, the specific angular momentum of the
collapsing system is likely to be large. It is natural to expect that the
collapsing object may have an axial symmetry leading to the bipolar
ejection of jets of CBs, as in quasars, microquasars and the core-collapse
SNe responsible ---in the CB model--- for long-duration GRBs. About 
$70\%\pm10\%$ of all SN explosions in the local Universe are of the
core-collapse types, the rest being Type Ia SNe (Tamman,  Loeffler \&
Schroder 1994; van den Bergh \& McClure 1994). Intriguingly, $\sim 75\%$
of all GRBs are long and the rest are short. The coincidence may not be
accidental.

\subsection{The environment of Type Ia SNe}

The narrow lines and light curves of SN1997cy and SN2002ic have been
interpreted as evidence for a circumstellar wind ejected from the WD's
companion, or from prior successive thermonuclear explosions on the WD's
surface (Chevalier 2003).  In all other Type Ia SNe there is no evidence
for circumstellar wind but the mass-loss rate is believed to be one to two
orders of magnitude smaller than that from the progenitors of
core-collapse SNe. The theoretical models of Type Ia SNe suggest that
their wind intensifies as the collapse approaches, generating a surface
density $\rho(r)\, r^2\sim 10^{15}$ g cm$^{-1}$ in their near environment
(Chevalier 2003), an order of magnitude smaller than that for
core-collapse SNe, Eq.~(\ref{surfdens}).  Such a wind becomes optically
thin to Compton scattering at $r_{tn}^w\simeq \sigma_{_{T}}\, \rho(r)\,
r^2/m_p\sim 4\times 10^{14}$ cm, a distance much smaller than the
corresponding one for core-collapse SNe, Eq.~(\ref{disttrans}).

The wind of Type Ia SNe may be ionized prior to the SN explosion by the 
EUV flux from the SN progenitor. If not, it is ionized by the EUV flash
from the SN explosion: the estimated  initial UV flash from the thermonuclear
explosion of a Chandrasekhar-mass WD contains more than $10^{57}$ EUV
photons, sufficiently many to fully ionize a wind of a few 
$M_\odot$ mass.
This ionized wind scatters the light from the SN, whose initial bolometric
luminosity may reach $10^{43}$ erg s$^{-1}$ (e.g.~Blinnikov 
\& Sorokina 2002; Sorokina \& Blinnikov 2003).

\subsection{Short GRBs}

The observed distribution of the total duration of GRBs has a trough
at about 2 s, separating ``long-duration'' from
``short-duration'' ones (e.g.~Dezalay et al.~1992; 
Kouveliotou et al.~1993; Belli 1995; Mukherjee et al.~1998; Horvath
1998). The light-curves of short GRBs look like a time-contracted
version of those of the long ones: they are made of shorter pulses 
with shorter time intervals between them (e.g.~MacBreen et al.~2001, 2002). 
The energy spectrum of short GRBs may be 
slightly
harder than that of the long ones (e.g.~Paciesas et al.~2001).  For short
GRBs, neither an optical AG nor a host galaxy have been observed to date.
Thus their redshifts and origin remain unknown, but their isotropic
distribution in the sky suggests that, like long GRBs, they stem
from cosmological distances.

As we discussed in the two previous subsections, the properties of the 
environment of Type-Ia SNe (wind's surface density and
wind's extension) appear to be scaled-down versions of the corresponding
properties of core-collapse SNe. The stage is pleasantly set to
conclude that, consequently, short GRBs should be
a time-contracted ---but otherwise very similar--- version of the
long GRBs. But there is an obstacle along this path: we do not
have, for short GRBs, any convincing argument to estimate, a-priori,
what the baryon number and Lorentz factor of the CBs emitted
by Type Ia SNe ought to be.

The progenitor stars of core-collapse SNe, particularly those
supposed to result in black holes of mass larger than that
of a Chandraseckar WD or a neutron star (both $\sim 1.4\,M_\odot$)
are much more massive than WDs, and have much larger radii.
It is quite plausible that the CBs emitted by the implosion of a
WD  be less massive and less spaced in time than the
ones of core-collapse SNe. If that were the case, we could
repeat word-by-word ---and with similar success---
our analysis of the properties of long
GRBs, the only difference being that we would have to choose
by hand a value of $N_{_{B}}$ about two orders of magnitude
smaller than the long-GRB reference value in Table 1, whose
order of magnitude was chosen a-priori, as discussed in
Section 15.  

\subsection{The AG of short-duration GRBs}  

The AG of a short GRB should, in the CB model, have the same
origin as that of a long one: synchrotron radiation from the motion
of the CB's accelerated electrons in its chaotic magnetic field. 
The decline of
the AG's fluence is due to the deceleration of the CB in the ISM.
In an approximately hydrogenic ISM of constant number density 
$n$, the function $\gamma(t)$ is determined by energy-momentum
conservation to be the real root of the cubic:
\begin{equation}
{1\over\gamma^3}-\rm{1\over\gamma_0^3}
+3\,\theta^2\,\left[{1\over\gamma}-{1\over\gamma_0}\right]=
{2\,c\, t\over 3\, (1+z)\, x_\infty}\; ,
\label{cubic}
\end{equation}
where $t$ is the observer's time and
\begin{eqnarray}
&& x_\infty\equiv{N_{_{B}}\over\pi\, R_\infty^2\, n}
\simeq (165\,{\rm kpc})\times\nonumber\\
&&\left[{N_{_{B}}\over 10^{50}}\right]^{1\over 3}
\left[{10^{-3}\,{\rm cm}^{-3}\over n}\right]
\left[{n_{_{SN}}\over 1\;\rm{cm}^{-3}}\right]^{2\over 3}
\left[{\gamma_0\over 10^3}{1\over\beta_{s}}\right]^{4\over 3},
\label{xinf}
\end{eqnarray}
with $R_\infty$ the calculable asymptotic radius of a CB, reached
within minutes of observer's time. We have distinguished between
the ISM density very close to the SN, $n_{_{SN}}$,
and the ISM density further away, $n$. The AGs decline fast with $\gamma$
(Dado et al.~2003e):
\begin{equation}
F_\nu\propto n^{(1+\hat\alpha)/2}\, R_\infty^2\, \gamma^{3\hat\alpha-1}\, 
\delta^{3+\hat\alpha}\sim 
n^{(1+\hat\alpha)/2}\, R_\infty^2 \;\gamma^{4\hat\alpha+2}\, , 
\label{afterglow}
\end{equation}
with $\hat\alpha$ changing from $\sim 0.5$ to $\sim 1.1$ 
as the emitted frequency crosses the {\it ``injection bend''}
(Dado et al.~2003a). 
The time-scale governing the decline of $\gamma(t)$
 is $x_\infty/(\gamma_0\,c)$ (Dado et al.~2002a).
Given the possible variations in the actual values of 
$x_\infty$, the AG of some long-duration GRBs may be hard to detect:
this is the CB-model's explanation of these ``dark" GRBs.

The progenitors of core-collapse SNe are short-lived massive stars.  
Consequently, most of their explosions take place in star-formation regions, in
supperbubbles  produced by the winds of
massive stars and the ejecta from previous SNe. The ISM density in these
bubbles is $n\sim 10^{-2}$--$10^{-3}$ cm$^{-3}$. 
The progenitors of Type Ia SNe are long-lived and are not confined to 
star-formation regions. Their explosions take place in a normal ISM 
of typical density  $n\sim 0.1$--$1.0$ cm$^{-3}$.  As one can read from 
Eq.~(\ref{xinf}), for CBs with a baryon number 100 times smaller in short
GRBs than in long ones, and even for a density around a short GRB
as low as $n=10^{-2}$ cm$^{-3}$, the characteristic time of decline of
the AGs of short GRBs is $\sim 50$ times shorter than for long ones.
Moreover, the smaller CB's radius ($R_\infty^2 \sim [N_{CB}/n]^{2/3}$; 
Eq.~(16) of 
Dado et al.~2002a) also reduces
the intensity of the AGs  considerably.  
When the CBs enter the ISM
(within a few minutes of observer time), the combination
of these effects makes the AGs of short GRBs  
much harder to detect than those of long ones.
 The only chance to detect the AG of short GRBs
is at the very early time when the CBs plough through the 
short-range circumstellar wind. 
Indeed, very early X-ray AGs of short GRBs, declining rapidly with time,
have actually been detected tens to hundreds of seconds after burst
(e.g.~Frederiks et al.~2003).  

In long GRBs the AG is a ``background'' that makes it difficult
for the GRB community to consider the possibility that they are all
associated with SNe, as they are in the CB model (in which this background
is very well understood). One redeeming value of the fact that
the AGs of short GRBs decline so fast is that there will be no background
to the detection of a potentially associated Type Ia SN. Moreover,
the peak bolometric luminosity of such a SN is much larger than
that of a core-collapse SN, to wit $ L_{Ia}\approx
10^{43.35}$ erg s$^{-1}$, reached around $t\sim(1+z)\times 20$ days after
burst (e.g.~Leibundgut \& Suntzeff 2003). If these SNe were to be found 
in the directional error boxes of short
GRBs, they could be used to localize them, to identify their host
galaxies and their location within them,  and to measure their redshifts. 
This may significantly increase the detection rate of Type Ia
SNe at cosmological distances.

\section{X-ray flashes}

X-ray flashes (XRFs) are bursts whose peak flux, $E_p$, is well below 40
keV, i.e.~they are relatively poor in $\gamma$-rays but rich in X-rays.  
They were discovered with the Beppo-SAX Wide-Field Camera, and they were
not seen above 40 keV with the Beppo-SAX GRB Monitor (Heise et al.~2001).
They were detected by the same satellite at a rate of 4 per year,
indicating a population not very much smaller than that of GRBs.
Re-examining the BATSE data, Kippen et al.~(2002) have found some 10 XRFs.  
A few more have been detected by HETE II. To date, about 30 XRFs have been
reported. These bursts are distinguished from Galactic transient sources
by their isotropic spatial distribution. They are softer and weaker than
GRBs but have a Band spectrum, and their durations are similar to those of
GRBs. Their afterglows were first discovered by Harrison et al.~(2001) in
the X-ray band, by Taylor et al.~(2001) in the radio band, and by
Soderberg et al.~(2002) in the optical band. A host galaxy of an XRF has
been found by Fruchter et al.~(2002). The few available redshift and
photometric informations on their hosts indicate that XRFs are
cosmological in origin, but not all of them have redshifts high enough to
explain their relatively low peak flux and low peak energy.

The simplest CB-model interpretation of XRFs (and of X-ray rich GRBs) is
that they are ordinary GRBs with either a high redshift, a large viewing
angle, or both. Thus, all the CB model results are also applicable to
XRFs. In particular, their durations should be similar to those of
``classical'' GRBs, but the duration of their pulses should be longer,
resulting, on the average, in much smoother light-curves for multi-pulse
XRFs. Their AGs should be dimmer than those of long-duration GRBs,
but they should also be described by
Eq.~(\ref{afterglow}) and steepen with time to its
asymptotic behaviour, $F_\nu\sim \nu^{-1.1\pm 0.1}\, t^{-2.13\pm 0.1}$
(Dado et al.~2002a). The AGs of the relatively nearby
XRFs should include a visible SN contribution akin to SN1998bw displaced
to the XRFs position. Such a ``smoking gun'' signature ---as well as,
perhaps, a detectable superluminal motion of their CBs--- can provide 
the best proofs that XRFs are nothing but GRBs viewed at relatively large 
angles, like GRB 980425 (Dar \& De R\'ujula 2000a), which would be
an intermediate case between GRBs and XRFs.

Fynbo et al.~(2003) have just reported the detection of a
rebrightening in the optical AG of XRF 030723 (Fox et al.~2003b; Dullighan
et al.~2003a,b; Smith et al.~2003, Bond et al.~2003) 14 days after the XRF
(Prigozhin et al.~2003) that may be due to the contribution of a SN. This
is demonstrated in Fig.~(\ref{figXRF}) which presents a preview of
a CB model fit to the AG of XRF 030723 (Dado et al.~in preparation).
The normalization of the 1998bw-like SN contribution has been
adjusted: without extinction corrections in the host galaxy or ours,
it corresponds to a redshift of $z\sim 0.75$. The fitted value of the viewing
angle is $\theta\sim 2.8$ mrad, larger than that of any other GRB
listed in Table 2, but 
the very close-by GRB980425.

\section{GRBs in the FB models}

We discuss here the properties of GRBs in the ``standard'' FB scenarios, 
in the same order as we have discussed them in the preceding
text on the CB model.

\subsection{Polarization}

In the most popular variant of the FB models,
a GRB pulse is due to synchrotron radiation from  the
collision of two shells, that generates a magnetic field 
(e.g. Medvedev \& Loeb 1999)
and accelerated-electron distribution. 
Electrons with a power-law distribution $dN_e/dE\sim E^{-p}$, 
immersed in a unidirectional magnetic field,
radiate light with a linear polarization, $\Pi =(p+1)/(p+7/3)$ 
(Rybicki \& Lightman 1979). Thus, the polarization of
photons from Fermi- or shock-accelerated electron distributions 
($ 2\leq p \leq 3.2 $)  in a highly ordered magnetic field is typically 
between 70 and 75\%, although even for optimal geometries 
the polarization cannot exceed $\sim 50\%$ (e.g.~Lyutikov et al.~2003).  

A disordered magnetic field generated in collisions of shells of 
a conical jet with a narrow opening angle $\theta\sim \gamma^{-1}< 10$ mrad
---which is much smaller than those inferred in previous standard FB 
models (e.g.~Waxman 2003a, Frail et al.~2001; Berger et al.~2003; 
Bloom et al. 2003)---  
can generate a high polarization if viewed from an angle 
$\theta\sim 1/\gamma$ (Waxman 2003b), and not from the traditional 
FB models' on-axis viewing angle (e.g.~Rhoads 1997,1999; Sari, Piran \&  
Halpern 1999; Frail et al.~2001;
Berger, Kulkarni and Frail 2003; Bloom et al. 2003). But the light curve 
of GRB 020106
has many pulses, requiring in a FB model a variety of values of $\gamma$ 
for the colliding and merged shells. With these varying
$\gamma$ values and a fixed observer's angle, it is quite difficult to imagine 
how the polarization, integrated over pulses, can be as high as the 
$80\pm 20 \% $ value observed by Coburn and Boggs (2003)
in this GRB.

To date, the most detailed FB analysis of the polarization of GRB 021206
is that of Nakar, Piran \& Waxman (2003). They find that, by adequately
maximizing all effects, and for a single value of the Lorentz factor of
a merged shell, they can raise the polarization to 45--50\%, significantly
below the observation. This 
requires a magnetic field structure which is both random (to accelerate the
synchrotron-emitting electrons) and uniform in space and time
(to give rise to the polarization). Alternatively, a model is considered
in which the magnetic field is uniform over a small region observed
at the edge of the jet. This model requires a large amount of fine-tuning,
between the observational angle, the jet opening angle, the inverse of the 
Lorentz factors of the various merged shells, the Lorentz factors 
themselves (which must be
very similar for the many different merged shells giving rise to the
GRB) and the polarization direction of the emission from each shell. 
But fine-tunings and exceptional cases are the rule in the FB 
models, not the exception. 

The conclusion {\it we} would extract from the above is that the FB
models where GRBs are produced by synchrotron radiation
cannot accommodate a significant GRB polarization: they are ruled 
out by the large polarization measured by Coburn and Boggs (2003) 
in GRB 021206.

In one of its multitude of variants, a ``fireball'' model was 
studied in which a relativistic wind blowing in a funnel within a star
would generate a GRB by ICS off the photons emanating from
the funnel's walls (Ghisellini et al.~2000). The authors noticed that, for
a large $\gamma$, the characteristic GRB energies could be explained
(Shaviv \& Dar 1995); but ---perhaps because at the time fireball
ejecta were still considered to be broad and to point at the observer---
they overlooked the crucial prediction of Shaviv and Dar: a large polarization 
at the most probable viewing angles $\theta\sim 1/\gamma$.

\subsection{Typical energies} 

In the FB models, the typical energy of GRB photons is given by a
combination of parameters that have no reason
to conspire to yield the very narrow observed range (e.g.
Ghisellini 2001). But it has been argued (Waxman 2003a)
that this parameter combination can be rewritten as:
\begin{equation}
E=(1\,{\rm MeV})\;\xi_B^{1/2}\,\xi_e^{3/2}\;
\left(L_{_{GRB}}\over 10^{52}\,{\rm erg\,s^{-1}}\right)^{1/6}\;
\left({10^{-2}\,{\rm s}\over \Delta t}\right)^{2/3},
\label{w1}
\end{equation}
where $\xi_B$ and $\xi_e$ are efficiencies for transforming collisional
energy to magnetic and accelerated-electron energies, $L_{_{GRB}}$
is the GRB's luminosity and $\Delta t$ is the scale of GRB variability.
The problem with this argument, it appears to us, is in the use of
a {\it fixed, universal} time scale $\Delta t=10^{-2}$ s. 
Attributing a fixed variability scale to {\it all} GRBs ---particularly
the dominant fraction without significant narrow spikes--- yields the desired
result, but seems to be unwarranted.
If the more typical variability scale $\Delta t=1$ s were to be used,
Eq.~(\ref{w1}) would fail by over an order of magnitude. Moreover,
for individual GRBs, the correlation between energy and variability 
implied by (our reading of) Eq.~(\ref{w1}) is not observed.
Finally, it is not clear how Eq.~(\ref{w1}) explains a
narrow spread of observed $E$ values.
The CB-model's prediction for $E$ and its distribution,
discussed in Section 6,
seem to us to be more satisfactory.

\subsection{The width of a pulse}

In the FB models, there is no particular reason for the width 
and the total energy of a typical GRB pulse to be what they
are, these results must be imposed by fiat:  adequate
choices of the relevant parameter combinations. Reviews of
the subject, such as those of Piran~(2000), M\'{e}sz\'{a}ros~(2002),  
Hurley et al.~(2002)  and Waxman~(2003b), do not discuss these points,
for which the CB-model predictions are given by Eqs.~(\ref{ttrans})
and (\ref{FWHM}).

\subsection{The shape of a pulse}

An example of the shape of a pulse in an FB model is given in
Fig.~(\ref{FMshape}), from Kobayashi, Piran and Sari (1997). The
combination of parameters determining the rise-time must be tuned to give
the observed range of results. The result in Fig.~(\ref{FMshape}) is for
the pulse's luminosity evolution, $E\,dN/dt$,
not its $\gamma$-ray number evolution.
The result for the latter initially rises as $dN/dt\propto t$ and declines
after the discontinuity as $dN/dt\propto 1/t$, in disagreement with
observations.

\subsection{Spectral shapes}
 
The distribution of values of $\alpha$ and $\beta$ extracted from fits to
the GRB data peak close to the values predicted in the CB model
$\alpha\approx 1$, $\beta\approx2.1$. In the FB models, wherein
synchrotron radiation is responsible for the GRB emission, the observed
$\beta$ can be accommodated, but not the results for $\alpha$: the
``synchrotron limit'' spectrum (Katz 1994a) corresponds to $\alpha=2/3$;
some 10--15\% of bursts are harder than that (Preece et al.~1998). Even
worse, the electron cooling times are very short, and the resulting
synchrotron spectrum should have $\alpha=3/2$, the overwhelming majority
of GRB spectra being harder than that. This is occasionally admitted to be
an unsurmountable problem, as all known alternatives ---synchrotron
self-absorption, a smooth cutoff of the low-energy electron distribution, an
anisotropic electron distribution with a small mean pitch angle,
unsaturated thermal comptonization, Compton drag, synchrotron
self-Compton,  and the incidence of
instrumental effects, are all equally problematic
(e.g.~Liang 1997; Ghisellini \& Celotti 1998; Ghisellini, Celotti \&
Lazzati 1999; Ghisellini 2001;  Preece et al.~2002; Ghirlanda, Celotti \&
Ghisellini 2002).

\subsection{Time--energy correlations}

The correlations between time and energy in a GRB pulse ---expected on
very simple grounds in the CB model--- coincide with the ones observed. We
are not aware of a similar conclusion within the FB models, as reviews of
the subject, such as those of Piran (2000), M\'{e}sz\'{a}ros (2002),
Hurley et al.~(2002)  and Waxman 2003b do not discuss this point. The only
exception is the correlation $\Delta t\propto E^{-1/2}$ of
Eq.~(\ref{fenimore}), obtained in an FB model by Kazanas, Titarchuk and
Hua (1998). But this correlation requires a very slow cooling process,
implying magnetic fields orders of magnitude smaller than the ones
required to accommodate the large observed GRB fluxes. Since the field
cannot be simultaneously large and small, Wu \& Fenimore (2000) conclude
that the agreement with observation (Fenimore et al.~1995, Norris et
al.~1996) is a mere coincidence.

 \subsection{GRB progenitors}

The nature of the progenitors of GRBs is,
in the FB models, undecided. In Hurley et al.~(2002),
as in innumerable other instances, the bursts possibly associated 
with SNe may be of a special class, exactly the opposite of what 
{\it all} GRBs are in the CB model.
 
From our outsiders' perspective, as we just argued, the FB models have 
very serious difficulties accommodating the observations of GRBs.

\section{Conclusions and discussion}

We have shown how simply and successfully the CB model explains the bulk
of the properties of the $\gamma$-rays of  GRBs. The key is simple:
ICS of the circumburst ``ambient'' light by the electrons in a CB. We have 
commented on how unpredictive the FB models are regarding GRBs, their 
only robust predictions being that of the power behaviour of the 
$\gamma$-ray spectrum, which fails  
---unless magnetic fields and the ensuing
electron cooling times can be both long and short--- and the polarization
of the GRBs, which should be tiny (e.g. Medvedev \& Loeb)
---unless magnetic fields can be both
ordered and disordered, or the shells' Lorentz factors, magnetic-field
orientations, opening and viewing
angles are all eagerly fine-tuned ``just so'' (Ghisellini 2003, Nakar et al.~2003).
For the GRB AGs, briefly discussed
in Appendix II, the situation is similar: the CB model does an excellent 
job, while the FB models, which are more predictive for AGs than they are for the 
GRBs themselves, are also more often in disagreement with the observations.

From an insider's perspective, an asset of the study that we have presented
here is the unassailability of its inputs. The properties of the CBs originating
GRBs ---such as their very substance--- are borrowed from those of other
CBs observed in nature. The properties of the ambient light that the CBs
scatter to GRB energies are also based on observations. Although
these inputs cannot be said to emanate from first principles, they are solid.
Given these phenomenological inputs, the rest of the ingredients are not
in doubt: the fundamental electrodynamics on which the expected spectrum
of accelerated electrons within a CB is based, and Compton scattering,
one of the most relevant processes in cosmology and astrophysics,
and the ``Eve'' of all particle-physics processes.

Seen from our outsiders' perspective, 
the FB models look like a baroque and continuously buttressed 
edifice. This is not to say that these models have not recently progressed
in the direction that we consider correct: that of espousing features of the 
CB model. {\it What used to be vices have become fashions}
(Seneca, ${\cal{O}}(30)$ C.E.):
\begin{itemize}
\item{}
From the 8$^{\rm th}$ of April,  
2003, the discovery (Garnavich et al.~2003b, Stanek et 
al.~2003) of 
SN2003dh ---associated with GRB 030329--- has transformed the GRB
association with fairly unconventional SNe from a minor and doubtful
issue (e.g.~Hurley et al.~2002; Waxman 2003a) 
into something crucial that {\it everybody always knew}. This association
is the basis of the CB model; this SN, its properties and the date of
its discovery (that is, the time when the predicted AG would no longer
overwhelm the SN signal) were foretold in the CB model (Dado et 
al.~2003f). The history of the GRB/SN association is reviewed
in Appendix III.
\item{}
Following the CB model (Dar \& De R\'ujula 2000a,b;
Dado et al.~2002a) the observer's angle, 
once upon a time set to zero by fiat in the FM model 
(e.g.~Rhoads 1997, 1999; Sari et al.~1999; 
Kumar \& Panaitescu 2000; Frail et al.~2001; 
Moderski, Sikora \& Bulik 2000;
Panaitescu \& Kumar 2001; 
Kumar \& Panaitescu 2001: Berger et al.~2003; Bloom et al.~2003) is
gaining a non-negligible role (e.g.~Rossi, Lazzati \& Rees 2002;
Granot et al.~2002; Zhang \& M\'esz\'aros 2002;
Panaitescu \& Kumar 2003;  Salmonson 2003).
In the CB model only this
angle plays a role, the jet opening angle, 
$\theta_v=\beta_s/(\sqrt {3}\, \gamma)$,
is smaller than the boosted-light opening angle $1/\gamma$, and its
effects can be neglected. In the AG phase, as the CBs cease to expand,
 $\theta_v\to 0$. 
\item{} 
The ejecta's angular spread has progressively diminished from an original 
$4\,\pi$ solid angle, to a jet opening angle of tens of degrees 
(Frail et al.~2001), to
$\theta_v=10$ mrad (Waxman 2003a). This last value, for a typical FB 
model's 
$\gamma=100$, is getting dangerously close to being negligible relative 
to $1/\gamma$.
\item{}
The correlations discussed in Dar \& De R\'ujula (2000b) and in  Section 13 
(but not those of Fig.~(\ref{figFWRise}) and Section 12)
should be approximately valid for any jets seen off-axis, as it is beginning
to be realized (e.g.~Rossi, Lazzati \& Rees 2002;
Granot et al.~2002; Zhang \& M\'esz\'aros 2002;
Panaitescu \& Kumar 2003;  Salmonson 2003; Waxman 2003b; Perna, Sari \& 
Frail 2003; Yamazaki, Yonetoku \& Nakamura 2003).  
No doubt that these correlations will soon
be fully exploited as a success of forthcoming off-axis FB models. 
\item{}
With their current, very small $\theta_v$ values, GRBs, once 
systematically 
publicized as {\it ``the 
biggest explosions after the Big Bang''}, have become a small fraction of 
the energy budget of a conventional
SN, that is, what they always were in the CB model.
\end{itemize}

The FB-models' ejected shells have lost the elegance of a sphere to become 
rather contrived, very thin and laterally-small waffles, racing along 
a radial track, and emitting GRB photons as they bang
against one another ---like clanging chads. The mystery-clad 
``generic fireballs'',
with their spherical innuendos and their unconventional
composition ($e^+e^-$ pairs) fit ---rather uncomfortably---
into the centre of a fairly conventional exploding star.

The CB model may of course be wrong, but it is successful. It is not at all 
inconceivable that the FB models may
continue to incorporate and ``standardize'' other aspects of the CB 
model.
Three large stumbling blocks lie along this path. One is the substance of
which the FB-models' ejecta are made: a delicately baryon-loaded (that 
is, highly
fined-tuned) plasma of $e^+e^-$ pairs. Such a fancy substance may not be
so difficult to forsake, in comparison with good old ordinary matter. 
Another hindrance is synchrotron emission, the traditional GRB-generating
mechanism in most FB  models. But then, synchrotron and Compton scattering
are the same process: off virtual or real photons.  The
main obstacle may be the magic wand of FB models: shocks. If and when 
these obstacles are overcome, the fireballs may turn out to 
have always been cannonballs for, after all, in the CB model,... SNe
fire balls.

The phenomenological simplicity of the CB model may make 
long and short GRBs, as well as XRFs,
useful in the study of SNe and of the cosmos at large $z$.
Perhaps Type Ia SNe, as we
have briefly discussed, are the originators of short-duration GRBs,
in which case the spotting of these SNe may be greatly facilitated.
The association of long-duration GRBs with core-collapse SNe
will, at the very least, help in ascertaining the death rate of massive stars
at large $z$, to which their birth rate is indisputably proportional.
 
One serious drawback of the CB model is that it makes GRBs become
very uninteresting, in comparison with what they used to be:
one of the biggest mysteries of astrophysics and the biggest of explosions
of all times (the Big Bang, in our current understanding of it, 
was not an explosion in any sensible sense). Fortunately, and
 independently of the ``peripheral'' GRB- and AG-generating
 physics, the biggest conundrum remains:
 How does a SN manage to sprout mighty jets? In the CB model the guidance
 along this path is better than simulations: the CBs responsible for GRBs 
 are akin to the increasingly well--studied ejecta of quasars
 and microquasars. The CB model tends bridges to other observational fields
 as well: cosmic rays (Dar \& Plaga 1999), the gamma background
 radiation (Dar \& De R\'ujula 2001), cooling flows (Colafrancesco, 
 Dar \& De R\'ujula 2003) and, perhaps, astrobiology (Dar \& De R\'ujula 2002). 

The CB model is very predictive, so that its limitations may be readily
found. One example might be that of the superluminal motion of CBs in the
sky. This motion may be observable, indirectly via scintillations 
(Dado et al.~2003a),
directly in the case of GRBs at sufficiently small $z$
(Dar \& De R\'ujula 2000a), or even independently of a GRB signal
 in the case of truly close-by SNe, such
as SN1987A, whose two opposite CB jets are shown in
Fig.~(\ref{figCostas}). 
So far the only putative problem the CB model has encountered has to
do with the superluminal motion of the CBs of GRB 030329. 
Indeed, we learned by reading the e-version of NYT 030529
(the New York Times of that date, in GRB's parlance)
 that, according to Dale Frail {\it ``[Our observations] are sufficient to 
rule out predictions of the cannonball model"}. The {\it predictions}
of a model cannot be ruled out, but a (good) model can... if its predictions are
decisively proved wrong. We are only now beginning to have sufficient
data on the radio AG of this GRB to make these predictions sufficiently
specific. When that is done, and when we see the data,
we shall be able to ascertain whether or not, 
concerning the death of the CB model,
we are ---or we are not--- led to reminisce a similar
announcement: that of the death of Mark Twain.  

%http://www.nytimes.com/2003/05/29/science/29ASTR.html?ex=1055224261&ei=1&en=ea6537b42bf73592

\noindent {\bf Acknowledgements:} This research was supported in part by
the Helen Asher Space Research Fund for research at the Technion.  
Comments and suggestions by Shlomo Dado and Rainer Plaga are gratefully
acknowledged.  We are thankful to Felix Mirabel and Mario Hamuy  for sending us
Figs.~(\ref{figFelix}) and (\ref{figSN2002ic}), respectively.
One of us, Arnon Dar, is also grateful for the hospitality 
extended to him at the CERN Theory Division.

\section{Appendix I: Collisionless shocks or shockless collisions?}

Shocks are a fundamental building-block of the FB models,
while in the CB model they play no role whatsoever in the
generation of GRBs or their AGs. The substance of the shells
responsible for GRBs is, in the FB models, an $e^+e^-$ plasma
with a tuned ``baryon load''. The substance of CBs is ordinary
matter. We comment here on the
observational situation regarding these two issues in the realm of the
other relativistic jets observed in nature: the
ejecta of quasars and microquasars. 

According to the standard lore, the impact of a moving shell on
the ISM produces a forward shock in the ISM
and a reverse shock in the shell.  For 
non-relativistic shocks, such as those generated by the 
expansion of SN shells into the ISM, the temperature of the particles
behind the shocks is $T_i \sim (3/16)\, m_i\, v_s^2$, where
$v_s$ is the velocity of the forward or reverse shock in the ISM's or shell's
rest frame, respectively (e.g.~McKee \&
Hollenback, 1980).  This relation is not well satisfied in young
SN remnants. A typical shock velocity of 4000 km s$^{-1}$
should give rise to a plasma temperature of 19 keV
(e.g.~Hughes et al.~2000). The observed temperatures are
in all cases smaller than 5 keV.
Moreover, high resolution spectral measurements show
that the X-ray line widths of metals in the reverse-shock region do
not have the mass dependence expected for thermal widths (Vink et 
al.~2003).

The above unsuccessful
fluid-dynamic picture has been adopted in the relativistic FB
models of GRBs (Rees \& M\'{e}sz\'{a}ros 1992, 1994; Paczynski \& 
Rhoads 1993;
Katz 1994a,b; M\'{e}sz\'{a}ros \& Rees 1997; Waxman 1997a,b; Sari, Piran 
\& Narayan 1998), using the Blandford--McKee (1976) formalism for 
relativistic
shocks. In this picture, the ``kinetic" temperatures\footnote{It is difficult 
to imagine how collisionless processes can give rise to thermal-like
distributions, for which the occupation numbers are dictated
by equilibrium considerations, and the $e$ and $p$ temperatures
would coincide. In any case, the ``temperatures'' have to be ``kinetic", for
otherwise, at $T\sim 1$ GeV, the stuff of matter is a quark--gluon plasma
with an energy density orders of magnitude higher than the average
density of a neutron star. Since the shocked domain has a volume
many orders of magnitude larger than that of a neutron star,
the rest-mass of the shock would be astronomical by astronomic
standards.} of the random motion of
the protons and of the electrons behind relativistic shocks rise to
$T_p\sim \gamma\, m_p\, c^2$ and $T_e \sim \gamma\, m_e\, c^2$ (e.g.
Waxman 1997a,b). 
The forward shocks become promptly relativistic, and after $\sim\! 1$ s,
when the shell has swept up sufficient ISM, so do the reverse ones 
(e.g.~Wang, Loeb, 
\& Waxman 2002) and so does the ``temperature'' 
of the particles in the shell's rest frame. 
 
Applied to the relativistic jets launched by quasars and microquasars, 
this formalism implies that very high temperatures are produced in the jets 
by reverse shocks. This conflicts with the observations of atomic-line
emission from the jets. The velocity of the 
CBs repeatedly ejected from the microquasar SS433, for instance, is 
$\sim$ 0.26c (Margon 1984).  The
spatially resolved optical (e.g.~Eikenberry et al.~2001), UV (e.g.~Gies et
al.~2002), and X-ray spectra (Marshall, Canizares \&  Schultz 2002;
Migliari, Fender \& M\'endez 2002; Namiki et al.~2003) of the approaching
and receding CBs show blue- and red-shifted
emission lines of H, He, metals and heavy elements. In particular, the 
Balmer lines 
and other forbidden lines indicate that the ambient plasma temperature 
is less than 1 eV, i.e.~a few orders of magnitude  smaller
than expected from a reverse shock (e.g.~McKee \& Hollenback 1980).

One of the arguments often brandished against the CB model 
is that CBs travelling in the ISM should produce bow shocks impeding
the penetration of further ISM particles into them (e.g.~A. Loeb,  
private communication). A bow shock is created
by a bullet travelling in air because neither air nor the bullet are collisionless,
nor relativistic (a beam of high-energy particles does penetrate lead targets).
A collisionless bow shock is made by the {\it low-energy} protons emanating
from the Sun as they encounter the Earth's extensive (i.e.~bipolar) magnetic
field. But high-energy protons do reach the atmosphere, and so do most
protons in the polar regions where the field is weak, witness the aurora.
The ISM protons and electrons impinging on a CB do it with a very large
relative Lorentz factor, and the magnetic field of a CB is chaotic: multipolar
and of very short range outside the CB. Lastly, bow shocks are not 
observed
in the CBs emitted by quasars (e.g.~Pictor A in Fig.~(\ref{Pictor});
M87: Harris et al.~2003; 3C273: 
Sambruna et al.~2001), microquasars (e.g.~GRS 1915+105: Dhawan, 
Mirabel \& Rodriguez 2000, XTE J1550-564: Corbel et al.~2002),  
and pulsars (e.g.~Vela: Pavlov et al.~2003).
In Fig.~(\ref{figFelix}) we reproduce the May 1998
VLBA 2-cm radio images of the microquasar GRS 1915+105
and the superluminal CB it ejected $67 \pm 7$ h before,
at 75 AU resolution. 
%Contours are at -2, 2.3, 4, 6, 
%8, 16, 32, 64, and 96 \% of the peak intensity. The blue and red contours show 
%time-resolved images 4.5 hr apart. 
No bow shocks are observed, neither in
this image nor in higher resolution (2.5--7.5 AU) images, 
which are the most spatially resolved images   
of a relativistic jet ever obtained (Dhawan et al.~2000).

Strong X-ray emission lines with large equivalent widths and Doppler
factors $\delta\sim\! 2$ to 3, interpreted as Fe ${\rm  K\alpha}$ lines, 
were
found with the ASCA (e.g.~Yaqoob et al.~1999) and Chandra 
(Wang et al.~2003) satellites in radio-loud quasars.  Such emission
cannot be produced if the Fe kinetic temperature in the quasar's
jets is well above a few tens of keV.  Yet, if reverse shocks are
generated in these jets by their interaction with the ISM of the host
galaxy, the observed $\delta$ values imply relativistic temperatures. 

The observed atomic lines from the relativistic jets of quasars and 
microquasars imply that the jets are made
of ordinary atomic matter and not of $e^+e^-$ pairs. The lines can be
emitted from relatively cold ordinary matter, by collisional excitation 
of the jet atoms by the incoming ISM particles (as in R\"ontgen's 
experiment). The 
observed non-thermal line widths may be due to velocity dispersion of
the motion of matter in the CBs of the jets. 

In the CB model the CBs responsible for GRBs are assumed to be made of
ordinary matter, like the CBs emitted by quasars and microquasars. Even
after a CB is no longer significantly heated by hadronic collisions with
the SN shell and the wind, it is kept partially ionized by synchrotron
self-absorption (Dado et al.~2003a,b). Its atoms should emit light at
characteristic line energies $E_l$, seen by an observer at
$E_l\,\delta(t)/(1+z)$, with $\delta(t)$ the Doppler factor at the
observation time $t$, a decreasing function of time, given the fact that
$\gamma(t)$ diminishes, in the way described by Eq.~(\ref{cubic}),
 as the CB ploughs  through the ISM (a significant
effect at the time of X-ray AG observations, insignificant in the GRB
phase). For GRBs with measured AGs, the CB model fit to the AG determines
$\delta(t)$, so that the line positions can be predicted. There are
marginal but intriguing observations of X-ray lines in GRB AGs (Piro et
al.~1998, 2000; Antonelli et al.~2000; Yoshida et al.~2001; Reeves et 
al.~2002; Watson et al.~2002a,b; Butler et al.~2003; Watson et al.~2003). 
All of these putative observed lines are, auspiciously, at boosted
line energies expected in the CB model. An example of the comparison of
the predicted boosted Ly$\alpha$ line and the observations is given in
Fig.~(\ref{figXray}).

\section{Appendix II: AGs in the FB models and in the CB model}

Unlike the FB models, the
CB model offers a very simple, successful and complete description of the
AGs of {\it all} GRBs of known redshift, including the all-important GRB 980425,
associated with SN1998bw (Dar \& De R\'ujula 2000a, Dado et al.~2003a).

One key ingredient of the CB model in the AG phase is that, we contend,
CBs stop expanding after minutes of (observer's) time, to reach a stable
situation in which (in the CB's rest system)
their internal magnetic pressure is balanced by the
inward pressure of the ISM electrons and nuclei that they gather, isotropize 
and re-emit into the ISM. The re-emitted particles, accelerated in the CB
by a single ``collisionless elastic scattering'' to energies up 
to $2\,\gamma^2\,M\,c^2$ (and by multiple scatterings to higher energies)
subsequently constitute the cosmic rays of the GRBs' host galaxy.

To quote one of our referees: ``The majority of the [GRB] community 
believes this [CB] model is wrong, as the jet {\bf should} spread'';
the emphasis is ours. In Fig.~(\ref{Pictor}) we show, in the upper panel,
a Chandra X-ray image of the radio Galaxy 
Pictor A, showing a non-expanding jet that emanates from the centre 
of the galaxy and extends
across some 360 thousand light years towards a brilliant hot spot at least 800
thousand light years away from where the jet originates (Wilson, 
Young \& Shopbell 2001), and, in the
lower panel, an XMM/p-n image of Pictor A in the 0.2--12 keV energy 
interval,
superimposed on the radio
contours from a 1.4 GHz radio VLA map (Grandi et al. 2003).
What we intend to illustrate with this figure is that uncanningly
narrow jets are observed in nature, when looked at the X-ray frequencies
at which we see the synchrotron radiation or the Compton up-scattered
photons of the cosmic microwave background,
 emanating from within the CBs.
The extensive and somewhat biconical radio emission is, in our
interpretation, due to synchrotron radiation by the mentioned cosmic
rays in the ambient magnetic field. The hot spot is the place where
the jetted CBs finally have lost all of their kinetic energy to cosmic-ray
acceleration and, no longer suffering an inwards pressure, they
simply expand. 

Rather than deciding {\it by fiat}
that the CB model is wrong, we would conclude that it also offers
an explanation for the extremely narrow jets observed 
at X-ray frequencies, not only in Pictor A, but in more than twenty other quasars, blazars and radio galaxies that were observed with Chandra (Gelbord et 
al.~2003, Marshall et al. 2003). In several cases the narrow jets were
resolved into plasmoids (CBs), such as in
M87, Cygnus A and 3C279, or in microquasars 
such as GRS 1915+105 (Dhawan
Mirabel \& Rodriguez 2000), XTE J1550-564 (Corbel et al.~2002), 
4U 1755-33 (Angelini \& White, 2003), 
and even in pulsars (e.g.~Vela: Pavlov et al.~2003).

We cannot compare one to one the analyses of GRB AGs in the FB and
CB models, and we shall limit ourselves to a few illustrative examples.
In Section 4.3 we mentioned the CB-model interpretation of AGs, with the
example of the R-band AG of GRB 021211, shown in Fig.~(\ref{f1}). This AG
has also been analysed in an FB model (Fox et al.~2003a and references
therein).  The early AG is attributed to a ``reverse shock'', whose fast
time decline and normalization are fit to the data.  The latter behaviour
is, as in other GRBs, not measured at an early time, due to a ``forward
shock'', whose normalization and time decline are independently fit to the
data, with two successive power laws, when a steepening with time (a
``break'' at a fit break time) is observed. The SN signature at $t>10$
days is not discussed.  Moreover the FB model, unlike the CB model, fails
to describe the evolution of the radio flux. Fox et al.~(2003a) conclude
that {\it ``the burst may have suffered substantial radiative
corrections''}. They offer no explicit support (i.e.~a comparison of the
data and the predictions of such a modified FB model) for their
conclusion.
 
The FB-models'  analysis of the AG of GRB 991208 is more elaborate. 
Analysing 
its
early broad--band data, Galama et al.~(2003) are forced to conclude that 
{\it ``the jet
model can account for the observed evolution... provided the jet transition has
not been fully completed in the first two weeks after the event''.}  The predictions
of this paper disagree with subsequent radio data, gathered and 
re-analysed
in Galama et al.~(2003). This time the conclusions, also unsupported, 
change to {\it 
``the relativistic blast-wave entered a non-relativistic expansion
phase several months after the burst''}. In sharp contrast, the parallel successive
analyses of this AG in the CB model  ---in the same simple terms that
describe all other measured AGs--- are predictive and successful (Dado et 
al.~2002a; 2003a,d).

The defence of the FB models in the analysis of AGs is sometimes less 
vague. An example
is GRB 030329, whose light curves for the GRB photon number-count and R-band 
fluence (along with its CB-model fit) are shown in Figs.~(\ref{fig329NC})
and (\ref{fig329red}). In their FB 
model analysis,
Granot, Nakar \& Piran (2003) conclude that the fluctuations in the R-band fluence
relative to a smooth curve are best explained by {\it refreshed shocks.} The logic
of this conclusion is unassailable: fluctuations in the energy output are due to
fluctuations in the energy input. 

In the CB model, GRBs are less interesting, in that not many of them are
---as they are in the FB models---
indicative of new effects (ordered magnetic fields, radiative corrections,
incomplete jet transitions, non-relativistic expansions, refreshed shocks, 
to quote only the ones we have mentioned in connection with AGs).
The CB model fit to the AG of GRB 030329, shown in Fig.~(\ref{fig329red}),
relies on the observation that the GRB light curve is dominated by two clear pulses, 
or CBs, as shown in Fig.~(\ref{fig329NC}), where we have described
the two pulses with the naive shape of Eq.~(\ref{naivepulse}). The AG is thus fitted 
(admittedly with similarly unassailable logic) with two distinct CB contributions.
Notice that the data determining the properties of one or the other CB are different:
before $t\sim 1$ day the AG is overwhelmingly dominated by the first
CB, while the later data is overwhelmingly dominated by the second\footnote{The 
``residua'' of the two-CB fit have ups and downs of $\sim 1/2$ magnitude. 
These are to be expected, since the fluence depends on 
the instantaneous density of the ISM and not (as in the FB models) on its time 
integral.}. Naturally, the fit returns very different values for the parameters
of the two CBs, with one very significant exception: the observation angles are
$\theta_1=2.00$ mrad and $\theta_2=1.95$ mrad.  Values of $\theta$ for other
data sets (other GRBs) expand over a much larger range. The chance 
probability of 
obtaining $\theta_1=\theta_2$ to such precision is negligible unless, that is,
the CB model has some truth in it (the same ``coincidence''
takes place for the other GRB with two clear CB contributions, GRB 021004 
(Dado et al.~2003c). Incidentally, the prediction of a SN contribution,
shown in Fig.~(\ref{fig329red}), was subsequently found to be correct 
(Stanek et al. 2003).

The FB model has been used, thrice to our knowledge, to {\it predict}
the late-time evolution of AGs on the basis of the early data: those
pertaining to the GRBs 021004, 030226 and 030329.
The method was to use the ``empirically demonstrated'' relation between
the total isotropic energy of the
GRB and the ``break'' time, $t_b$, at which the AG light curves
steepen (Frail et al.~2000; Berger et al.~2003; Bloom et al.~2003)
to predict the latter in terms of the former, and to extrapolate
the early AG data to later times with the use of a typical
steepening at the break. In the case of GRB 021004,
Malesani et al.~(2003) predicted a break at
$t_b\sim13.7$ days. A first break took place at $t_b< 0.1$ days after
burst (Weidinger et al.~2002) and a second one at $t_b\sim 5$ days
(Holland et al.~2003).
For GRB 030226, Rhoads et al.~(2003)
predicted $t_b\sim 10$ days. The observation was $t_b\sim 0.8$
days (e.g.~Greiner et al.~2003). Finally, for GRB 030329, the expectation
(Uemura et al.~2003) was $t_b>60$ days, and the observation
$t_b\sim 0.57$ days (Burenin et al.~2003).
So, in each and everyone of these cases
the prediction failed by a considerable margin.

Even when applied
to the same GRB's AG, the extraction of $t_b$ is problematic.
In the case of GRB 020813, for instance, Covino et al.~(2003b) extract
$t_b=0.59\pm 0.03$ days from data in the interval between
3 hours and 4 days, while Li et al.~(2003)
obtain $t_b=0.13\pm 0.03$ days, 
for data in the 1.7 hours to 1.2 days period. The precision with which
the fit values of $t_b$ are quoted is impressive, when compared with
the difference between the central values.

Frail et al.~(2000), Berger et al.~(2003) and Bloom et al.~(2003) have deduced
values of the FB-model's jet opening angle from the 
$t_b$-values of  a  sample of GRBs with good 
AG follow-up and known $z$, and inferred that
the ``true'' $\gamma$-ray energy release in these GRBs is narrowly distributed 
around $1.3\times 10^{51}$ erg. In view of the fate of other uses of
the break-time concept, we do not consider the many-orders-of-magnitude
disagreement between this result and that of Eq.~(\ref{Eprime}) to be
significant. The
CB model relation for the equivalent isotropic energy, Eq.~(\ref{Eiso}), 
with a very narrow range of ${\cal{E}}'_{CB}$ values ---and the value
of $\delta$
determined in each case by fits to the AG data--- is much better
satisfied  than the FB-model-inspired relation,
%of Frail et al.~2000;  Berger et al.~2003 and Bloom et al.~2003, 
and this with no exceptions
(e.g., Dado et al.~2002a; 2003c,e,f).

The absence of AGs with evidence for a wind-fed circumburst medium is a
problem recognized by even the staunchest defenders of the FB models
(e.g.~Price et al.~2002, Nakar et al.~2003). In the CB model all AGs
caught early enough to trace the wind-density profile ---GRB 990123, GRB
021004 and GRB 021211, illustrated in Fig.~(\ref{f1})--- show an impressive
agreement with the very explicit prediction of the theory for the early
optical AGs at fixed frequency:  $F_\nu\propto (n_e)^{2/3}$ 
(Dado et al.~2003e).

The wide-band spectrum of AGs in FB models (e.g.~Sari, Piran \&
Narayan 1998) ---which paraphrases
the theoretical spectrum used to describe synchrotron
emission from quasar jets and lobes (e.g.~Meisenheimer et al.~1989;  
Longair 1994)--- evolved through a series of papers to its current version
(Granot \& Sari 2002). This spectrum is
extremely complicated: it has a handful of frequency breaks with different time
behaviours, which may be ordered in a variety of ways, and must all be fit
to the observations. In the CB model, the spectral shape involves no
independent parameters other than a single number: an absorption frequency
$\nu_a$, relevant only in the radio domain (Dado et al.~2003a). This stark
contrast does not detract the CB model from working well where the FB
models fail (see e.g.~Galama et al.~2000b,2003; Dado et al.~2003d).

Two highly relativistic FB-model shells of equal mass $m$ and Lorentz factors
$\gamma$ and $\gamma/2$ merge into an object of invariant mass
$2\,m\,[1+1/16+{\cal{O}}(1/\gamma^2)]$: the efficiency for creating
new forms of energy is 1/16. The merged object, if colliding with
something much more massive and at rest (the ISM), converts {\it all} of its
kinetic energy into new internal energy. Thus, in the favoured 
internal--external-shock FB models ---and as occasionally recognized
(e.g.~Ghisellini 2001)--- the total energy in a GRB ought to
be at least one order of magnitude smaller than that in the AG.
The observations are the other way around\footnote{It is sometimes stated
(Piran, private communication) that the ``calorimetry'' of GRB AGs,
as performed in the case of GRB 970508 (Frail, Waxman \& Kulkarni 2000)
solves this problem. But it aggravates
it: the calorimetric energy is smaller than the usual estimate. Moreover,
this GRB has the record smallest equivalent spherical energy after GRB 980425,
not a very fair choice for a comparison of its fluence to that of
its AG.}.

We mentioned in the previous appendix the extremely simple and predictive
CB-model interpretation of the X-ray lines allegedly
seen in some GRB AGs (Dado et al.~2003b). In the realm of FB models, these 
lines are
attributed to transitions in ionized Fe and other metals. Accommodating
them calls for heavy metal envelopes with a large variety of ad-hoc
properties and ``designer'' shapes (for reviews see, for instance, 
Boettcher 2002;
Lazzati 2002). In the CB model the lines do not require any extra
assumption, and their positions are predictable.

From our perspective, as we have argued and illustrated with a few
examples, the FB models have very serious difficulties in predicting or even
accommodating a posteriori the observations of GRB AGs.

\section{Appendix III: The GRB/SN association}

In Dado et al.~2003f we foretold that on 8 April, 2003 a SN
akin to SN1998bw would be bright enough to be discovered spectroscopically
as a contribution to the AG of GRB 030329. The spectroscopic discovery of
SN2003dh in the AG of GRB 030329 on the expected date, with a 
luminosity and spectrum remarkably similar to those of SN1998bw 
(Garnavich et al.~2003b; Stanek et al.~2003), has provided convincing evidence that,
undoubtedly, most long-duration GRBs are produced in SN explosions akin to
SN 1998bw, as advocated in the CB model (Dar \& De R\'ujula 2000a; Dado et
al.~2002a), see Figs.~(\ref{figSN2003dh},\ref{figSN2002ic}).

The possible association of GRBs with SN explosions was suggested long
before the first observational evidence was serendipitously found. Colgate
(1968, 1975) conjectured that the breakout of the shock wave from the
stellar surface in a core-collapse SN produces a GRB, but this implied a
GRB rate orders of magnitude larger than observed and a GRB from SN1987A
which was not detected (Chupp et al.~1987). Dar \& Dado
(1987) considered the possibility that radiative decay of neutrinos from
core-collapse SN explosions would produce GRBs, they used the 
disparity
between the cosmic SN and GRB rates to derive bounds on $\nu$ radiative
decay. Goodman, Dar \& Nussinov (1987) suggested that $e^+e^-$ pairs
produced by $\nu\,\bar\nu$ annihilation in accretion-induced collapses of
white dwarfs and/or neutron stars in binary systems, or from neutron-star
mergers, may produce GRBs\footnote{This idea was later adopted by Eichler,
Livio, Piran \& Schramm (1989).}. Yet, the authors found that baryon
contamination of the fireball (now known as ``baryon-load'') poses a
severe problem for this mechanism. Shaviv \& Dar (1995)  proposed that
GRBs may be produced by ICS of light by collimated
relativistic jets ejected in the birth of neutron stars and black holes in
SN explosions in distant galaxies.

The first evidence for a possible GRB-SN association came from the
discovery by Galama et al.~(1998)  of the very bright SN1998bw, at 
redshift $z=0.0085$, within the Beppo-SAX error circle around GRB 980425
(Soffitta et al.~1998), whose light curve indicated that the time of
explosion was within $-2$ to 0.7 days of the GRB (Iwamoto et al.~1998).  
This evidence did not fit at all into the framework of the 
FB model of GRBs.  The total equivalent isotropic $\gamma$-ray energy
release, $\sim 8\times 10^{47}$ erg, was some 5 orders of magnitude
smaller than that expected from a ``classical" GRB at $z=0.0085$. 
The FB community concluded that either SN1998bw and GRB 980425
were not physically connected or that, if they were, they represented a 
new
subclass of rare events (e.g.~Bloom et al.~1998; Norris, Bonnel \&
Watanabe 1998;  Hurley et al.~2002). These would be
associated with what Iwamoto et al.~(1998) and Paczynski (1999)
called  ``hypernovae":
super-energetic explosions with kinetic energy exceeding $10^{52}$ erg, 
as was inferred for SN1998bw from its high
expansion velocity and luminosity (Patat et al.~2001), and from the
very strong radio emission from its direction (Kulkarni et al.~1998).
Hofflich, Wheeler \& Wang (1999) retorted that core-collapse SNe 
may not be spherical-symmetric and the inferred kinetic energy of 
SN1998bw could have been overestimated.

SN1998bw was initially classified as Type Ib (Sadler et al.~1998) and
later as a peculiar Type Ic (Filippenko 1998; Patat and Piemonte 1998, 
Patat et al.~2001). Its discovery initiated intensive searches of 
positional and approximate
temporal coincidence between GRBs and SN explosions (e.g.~Kippen 
et al.~1998), in particular Type Ib and Ic SNe (Woosley, Eastman \&
Schmidt 1999). The search yielded two inconclusive associations, one
of them, discussed in Section 17,
between the peculiar Type II SN1997cy, at $z=0.063$, and the short-duration
($\sim 0.2$ s) GRB 970514 (Germany et al.~2000) and
another one between the Type Ic SN1999E at $z=0.0261$ and the long-duration 
GRB 980910 (Thorsett \& Hogg 1999; Rigon et al.~2003).

SNe of Types II/Ib/Ic are far from being standard candles. But if they are 
axially as opposed to
spherically symmetric ---as they would be if a fair fraction of them
emitted bipolar jets--- much of their diversity could be due to the angle
from which we see them.  Exploiting this possibility to its extreme, i.e.~using
SN1998bw as an ansatz standard candle, Dar (1999a) suggested that
the AGs of
all GRBs may contain a contribution from a SN akin to SN1998bw,
placed at the GRB's position.  Dar \& Plaga (1999) and Dar \&
De R\'ujula (2000)  advocated the view that most core-collapse SN
explosions may result in GRBs.

Possible evidence for a SN1998bw-like contribution to a GRB AG (Dar 1999a;
Castro-Tirado \& Gorosabel 1999) was first found by Bloom et al.~(1999)
for GRB 980326, but its unknown redshift prevented a definite conclusion.
The AG of GRB 970228 (located at redshift $\rm z=0.695$) appears to be
overtaken by a light curve akin to that of SN1998bw (located at $\rm
z=0.0085$), when properly scaled by their differing redshifts (Dar 1999b;
Reichart 1999; Galama et al.~2000a). Possible evidence of similar
associations was found for GRB 990712 (Bjornsson et al.~2001; Dado et
al.~2002a), GRB 980703 (Holland et al.~2001; Dado et al.~2002a), GRB
000418 (Dar \& De R\'ujula 2000a;  Dado, Dar \& De R\'ujula 2000a), GRB
991208 (Castro-Tirado et al.~2001;  Dado, Dar \& De R\'ujula 2002a), GRB
970508 (Sokolov~2001; Dado et al.~2002a), GRB 000911 (Lazzati et
al.~2001; Dado et al.~2002 unpublished), GRB 010921 (Dado, Dar \& De
R\'ujula 2002d), GRB 011121 (Dado et al.~2002b; Garnavich et al.~2003;
Bloom et al.~2002)  GRB 020405 (Price et al.~2002; Dado et al.~2003b),
and GRB 021211 (Dado et al.~2003e).

In the absence of precise spectroscopic information on late-time AGs, the
identification of a SN contribution requires a reliable model for
extrapolating the early-time AG to later times, as well as reliable
information on the extinction in the host galaxy. Using the CB model, Dado
et al.~(2002a)  have shown that the optical AGs of {\it all} GRBs with
known redshift $ z<1.1$ contain either evidence for a SN1998bw-like
contribution (in the GRBs 980425, 970228, 990712, 991208, 000911, 012111,
010405, 021211 and 030329) or clear hints (in the cases of GRBs 970508,
980613, 980703 000418 and 010921, for which the scarcity of data and/or
the lack of spectral information and multi-colour photometry and/or
uncertain extinction in the host galaxy prevented a firmer conclusion). In
the more distant GRBs ($z>1.1$) the ansatz standard candle could not be
seen, and it was not seen.

Naturally, truly ``standard candles'' do not exist, but SN1998bw
made such a good job at it that it gave us enough confidence to {\it 
predict}
the SN contribution to the late-time afterglow of all recent cases of
early detection of the AGs of near-by GRBs (000911, 010921, 010405,
012111, 021211 and 030329). In {\it all} these cases, from a fit to the
early-time afterglow, the CB model {\it correctly predicted} the
late-time appearance of a SN in the late-time colour light curves 
(Dado et al.~2002b,c; 2003e,f). Besides the  GRB 980425--SN1998bw pair,  
the most convincing  GRB/SN associations were provided by the 
recent spectroscopic
discoveries of a SN in the optical afterglows of GRBs 030329
(Stanek et al.~2003; Hjorth et al.~2003) and GRB 021211 (Della Valle
et al.~2003). 

The observability of a SN signal depends on the size of the
AG ``background''.
The predictions of ``the future'' of AGs reviewed in the previuos
paragraph are in stark
contrast with the corresponding ones made in the realm of
FM models, not only at optical frequencies, but also in the radio 
(as we discussed in the previous Appendix in connection with AG ``break'' times,
and with GRB 991208; respectively) and in the X-ray domain (e.g., compare the
observations of Pian et al.~2003 with the predictions of Dado et al.~2002a; 2003a).

Additional indirect evidence relating long-duration GRBs to 
core-collapse SNe follows from the association of well localized
GRBs with massive star-formation regions in their host galaxies 
(e.g.~Paczynski 1998; Fruchter et al.~1999; Holland \& Hjorth 1999;
Bloom, Kulkarni \& Djorgovsky 2002), 
where most of these SNe are expected to take place, as well as
from the statistics of the types of galaxies that host GRBs (e.g.~Hogg
and Fruchter 1999).

A model that explicitely links GRBs to SNe is the
``supranova'' model of Vietri \& Stella (1999), in which it is assumed that a
GRB occurs weeks or months after the SN. This is
inconsistent with the measured delays between GRB times and the
estimated time at which their observed associated SNe exploded.

In an abridged version of the long history of the GRB/SN association we
have presented, Stanek et al.~(2003b) and Hjorth et al.~(2003)  attribute
its theoretical prediction to Woosley (1993a,b). In these papers, it is
proposed that GRBs are produced by the collapse of very massive stars into
a black holes in {\it failed} SNe: implosions that do {\it not} result in
a SN signal. The ``collapsar" model ---wherein massive stars collapse into
black holes in Type Ib or Ic SN events--- was proposed by MacFadyen \&
Woosley (1998, 1999) and by Woosley \& MacFadyen (1999) after the
discovery of the association between GRB980425 and SN1998bw. This latter
idea ---in the realm of the CB model and for quite ordinary SNe, as
opposed to hypothetical supranovae,
collapsars or hypernovae--- is, in our opinion,
the only one still alive and doing well.

%\end{document}

\begin{table}
      \caption[]{\bf Typical parameters of a CB$^{\mathrm{a}}$, the
      early SN luminosity, the ``wind'' and the ``ambient light''.}
\vskip -0.2cm
            $$ 
            \begin{array}{ p{0.13\linewidth}l p{0.28\linewidth}l}
            \hline
            \noalign{\smallskip}
            Parameter     &  \rm Value & Definition \\
            \noalign{\smallskip}
            \hline
            \noalign{\smallskip}
 $\;\;\;\;\;\:\;\gamma$      & \approx 10^3 & Lorentz factor \\
 $ \;\;\;\;\;\:\;\theta $ & \sim 10^{-3} & Observer's viewing angle\\
 $\;\;\;\;\;\:\;\delta$ & \sim 10^{3} & Doppler factor     \\
 $\;\;\;\;\;\:\;\beta_s$ & \sim 1 & Initial expansion 
velocity\\
 $\;\;\;\;\;\:\; N_{_{B}}$ & {\cal{O}}(10^{50}) & Baryon  number \\
  $\;\;\;\;\;\:\; E_i$ & \sim 1 \;{\rm eV} & Ambient-light energy\\ 
 $\;\;\;\;\;\:\; L_{_{SN}}$ & \sim 5\times 10^{42}\;{\rm erg\,s^{-1}} & 
  Early SN luminosity \\
 $\;\;\;\;\;\:\; \rho\,r^2$ & \sim10^{16}\;{\rm g\,cm^{-1}} & Wind's 
   surface density \\
   $\;\;\;\;\;\:\; {\cal{E}}'_{_{CB}}$ & \approx 10^{44}\;{\rm erg} & Rest 
  radiation energy\\
\hline
\hline
         \end{array}
     $$ 
\begin{list}{}{}
\item[$^{\mathrm{a}}$] {\bf Comments:} The various symbols reflect
the spread of the fitted values for CBs, ranging from ``$\approx$'' for a narrow
distribution to ``${\cal{O}}$'' for a rough estimate. The parameters
$\delta$ and ${\cal{E}}'_{_{CB}}$ are deduced, not used as inputs.

\end{list}
   \end{table}

\clearpage

\begin{deluxetable}{llccccc}
\tablewidth{0pt}
\tablecaption{\bf GRBs of known redshift for which we have performed
a CB-model fit to the AG. Listed are redshifts, viewing angles
$\theta$ (in mrad), initial Lorentz and
Doppler factors $\gamma$ and $\delta$,
and the quantity $\sigma$ defined in Eq.~(\ref{boosting}).
Two AGs were fitted with two CBs.}
\tablehead{
\colhead{GRB}  &\colhead{$z$} &\colhead{$\theta$} 
&\colhead{$\gamma$} 
&\colhead{$\delta$} &  \colhead{$\gamma\, \theta$} & 
\colhead{$\sigma$}}  
%\startdata
\startdata
970228 & 0.695 & 1.69   &  540  &  590  & 0.91 & 0.376    \\
970508 & 0.835 & 2.51   & 1123  &  325  & 2.82 & 0.398    \\
970828 & 0.958 & 0.86   & 1153  & 1163  & 1.00 & 1.370    \\
971214 & 3.418 & 0.71   &  999  & 1331  & 0.71 & 0.600    \\
980425 & 0.0085& 7.83   &  495  &   62  & 3.88 & 0.060    \\
980703 & 0.966 & 0.95   &  779  & 1004  & 0.74 & 0.795    \\
990123 & 1.600 & 0.46   & 1204  & 1630  & 0.55 & 1.509    \\
990510 & 1.619 & 0.26   & 1009  & 1889  & 0.26 & 1.455    \\ 
990712 & 0.434 & 0.75   &  948  & 1259  & 0.71 & 1.664    \\         
991208 & 0.700 & 0.11   & 1034  & 2041  & 0.11 & 2.482    \\
991216 & 1.020 & 0.40   &  906  & 1598  & 0.36 & 1.433    \\
000131 & 4.500 & 0.10   & 1200  & 2365  & 0.12 & 1.032    \\
000301c& 2.040 & 2.32   & 1061  &  300  & 2.46 & 0.209    \\
000418 & 1.119 & 2.06   & 1241  &  329  & 2.55 & 0.385    \\
000911 & 1.060 & 0.29   &  800  & 1516  & 0.23 & 1.177    \\
000926 & 2.066 & 0.74   &  787  & 1521  & 0.58 & 0.781    \\
010222 & 1.474 & 0.47   & 1178  & 1813  & 0.55 & 1.727    \\
010921 & 0.451 & 0.15   & 1013  & 1980  & 0.15 & 2.765    \\
011121 & 0.360 & 0.10   & 1222  & 2405  & 0.12 & 4.322    \\
011211 & 2.141 & 1.16   &  824  &  862  & 0.95 & 0.452    \\
020405 & 0.69  & 0.42   &  645  & 1201  & 0.27 & 0.917    \\
020813 & 1.225 & 0.58   & 1128  & 1587  & 0.65 & 1.558    \\
021004 & 2.330 & 1.47   & 1403  &  542  & 2.06 & 0.457    \\
       & 2.330 & 1.47   & 1259  &  576  & 1.85 & 0.436    \\
021211 & 1.006 & 1.76   &  262  &  431  & 0.48 & 0.113    \\
030226 & 1.989 & 1.14   &  824  &  876  & 0.94 & 0.482    \\
030329 & 0.168 & 2.25   & 1652  &  222  & 3.73 & 0.629    \\
       & 0.168 & 2.10   & 1037  &  362  & 2.18 & 0.644    \\
       &       &        &       &       &      &          \\

\enddata \end{deluxetable}

\clearpage

%
%\ref{CBGlory}, CBGlory

%\ref{figCB}, figCB

%\ref{f1}, f1

%\ref{f2}, f2

%\ref{figslab}, figslab

%\ref{figsphere}, figsphere

%\ref{fig3}, fig3

%\ref{fig3pulses}, fig3pulses

%\ref{figband}, figband

%\ref{figspectrum}, figspectrum

%\ref{Noft1}, Noft1

%\ref{figSpectra1bis}, figSpectra1bis

%\ref{Noft2}, Noft2

%\ref{figSpectra2}, figSpectra2

%\ref{fig425spectra}, fig425spectra

%\ref{fig425}, fig425

%\ref{fig329red}, fig329red

%\ref{fig329NC}, fig329NC

%\ref{FMshape}, FMshape

%\ref{Pictor}, Pictor

%\ref{figLogWLogE},    figLogWLogE

%\ref{figEpFp},  figEpFp

%\ref{figEpFot}, figEpFot

%\ref{figLogALogW}, figLogALogW

%\ref{figalphabeta},  figalphabeta

%\ref{figEpInt},  figEpInt

%\ref{figEpVar},   figEpVar

%\ref{figCostas}, figCostas

%\ref{figXray}, figXray

\begin{figure}[]
\begin{center}
\vspace{.3cm}
\vbox{\epsfig{file=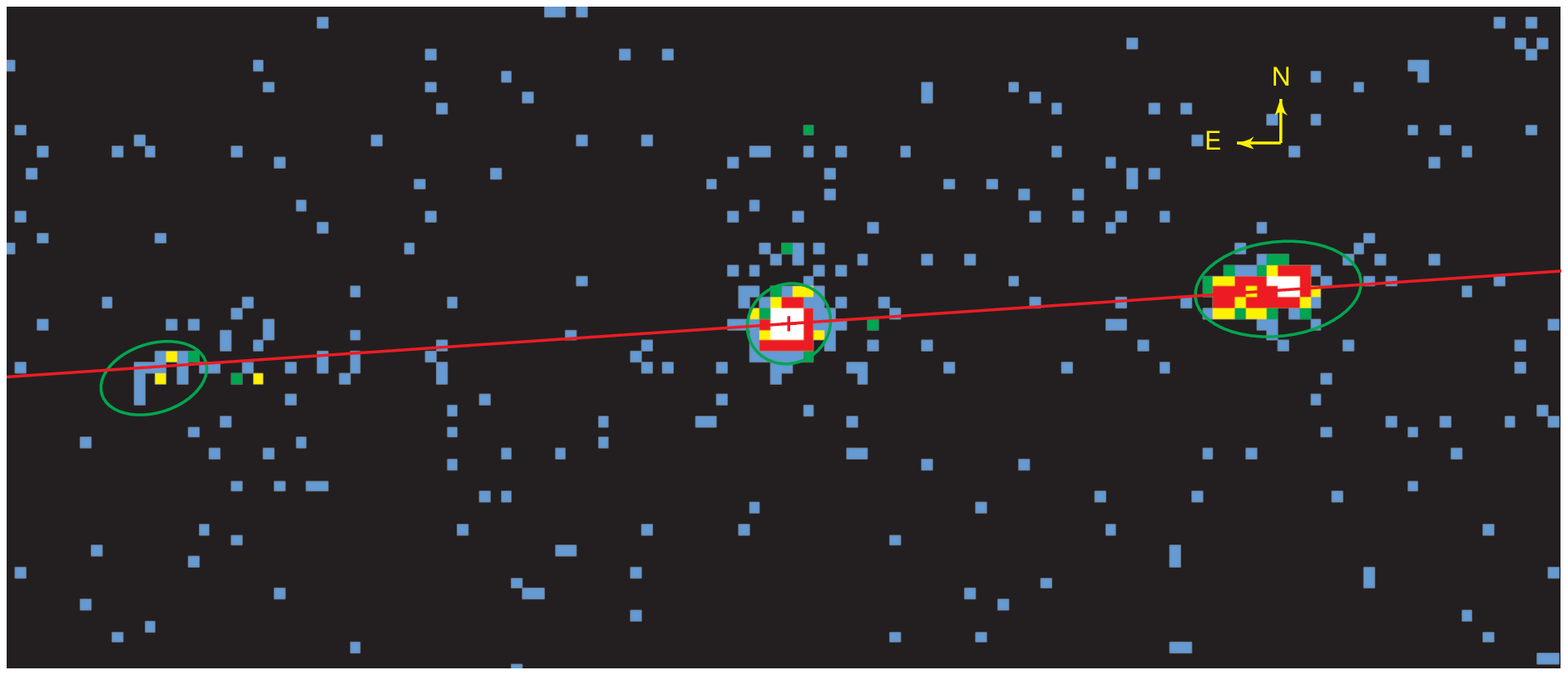,width=13cm}}
\vspace{+1cm}
\vbox{\epsfig{file=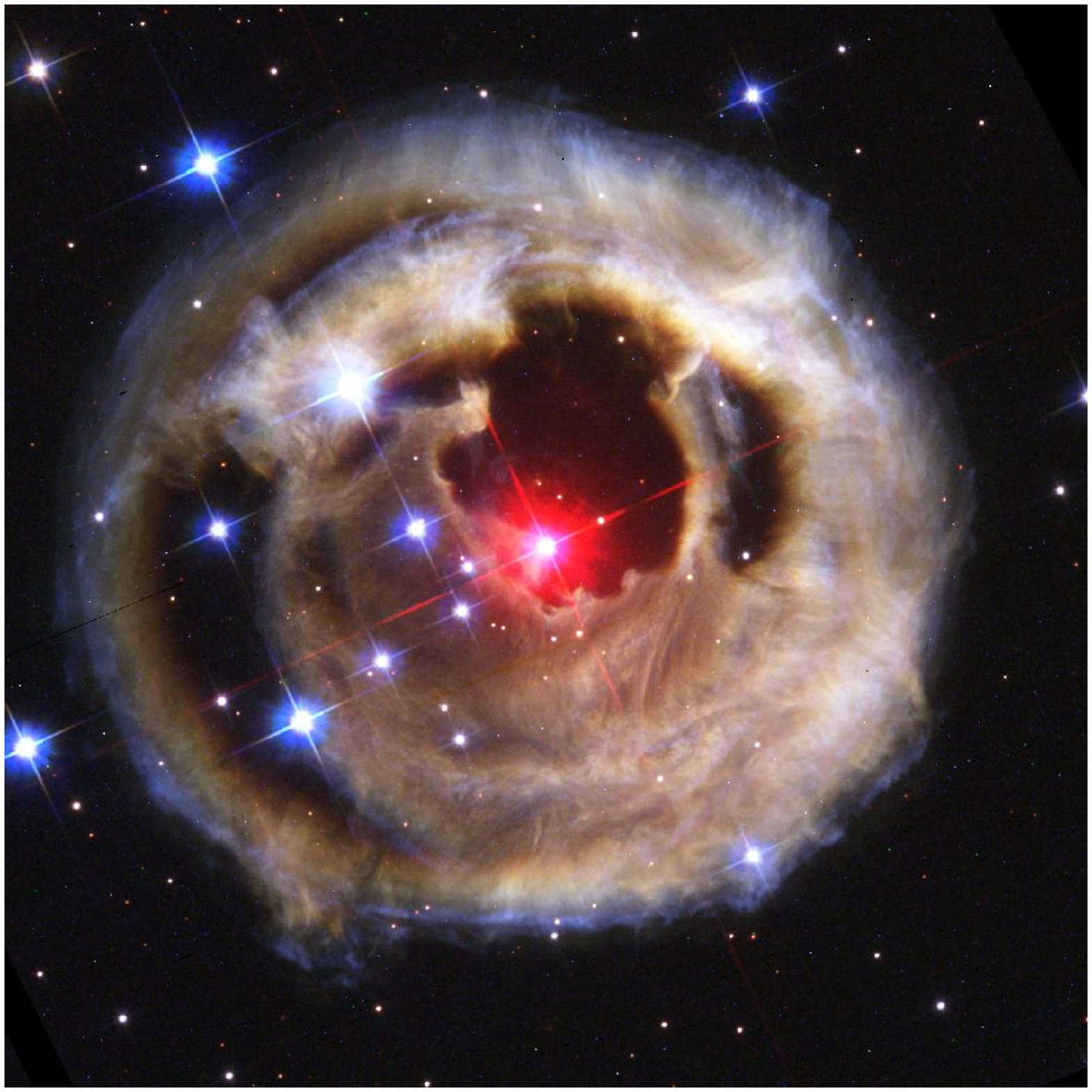, width=13cm}}
%\vspace{-1cm}
\end{center}
 \caption{Upper panel: Two relativistic CBs emitted in opposite directions
 by the microquasar XTE J1550-564, seen in X-rays by Corbel et
 al.~2002. Lower panel: 
 HST picture from 28 October 2002 of the {\it glory}, or light echo, 
of the  stellar outburst of the red supergiant V3838 Monocerosis
in early January 2002. The light echo was formed by scattering off
dust shells from previous ejections (Bond et al.~2003).}
 \label{CBGlory}
\end{figure}

\clearpage

\begin{figure}
\vskip -1cm
\hskip 2truecm
\begin{center}
\epsfig{file=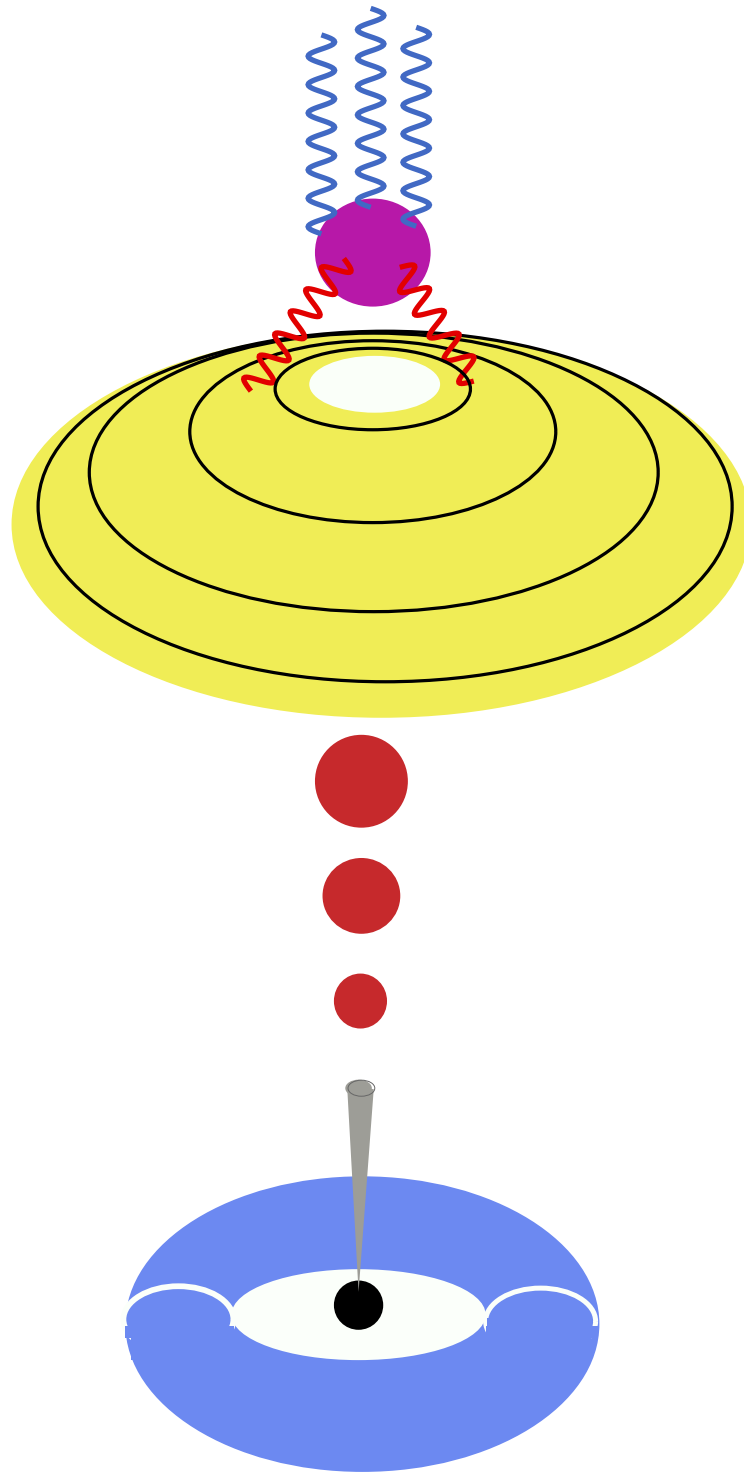, width=15cm}
\end{center}
\vspace{-.7cm}
\caption{A fewer-kbyte version of Fig.~(3) of Dar \& De R\'ujula 
(2000a) showing an
 ``artist's view'' (not to scale) of the CB model
of GRBs and their afterglows. A core-collapse SN results in
a compact object and a fast-rotating torus of non-ejected
fallen-back material. Matter (not shown) abruptly accreting
into the central object produces
a narrowly collimated beam of CBs, of which only some of
the ``northern'' ones are depicted. As these CBs move through
the ``ambient light'' surrounding the star, they Compton up-scatter
its photons to GRB energies.}
\label{figCB}
\end{figure}

%\newpage
\clearpage

\begin{figure}
\hskip 2truecm
\begin{center}
\epsfig{file=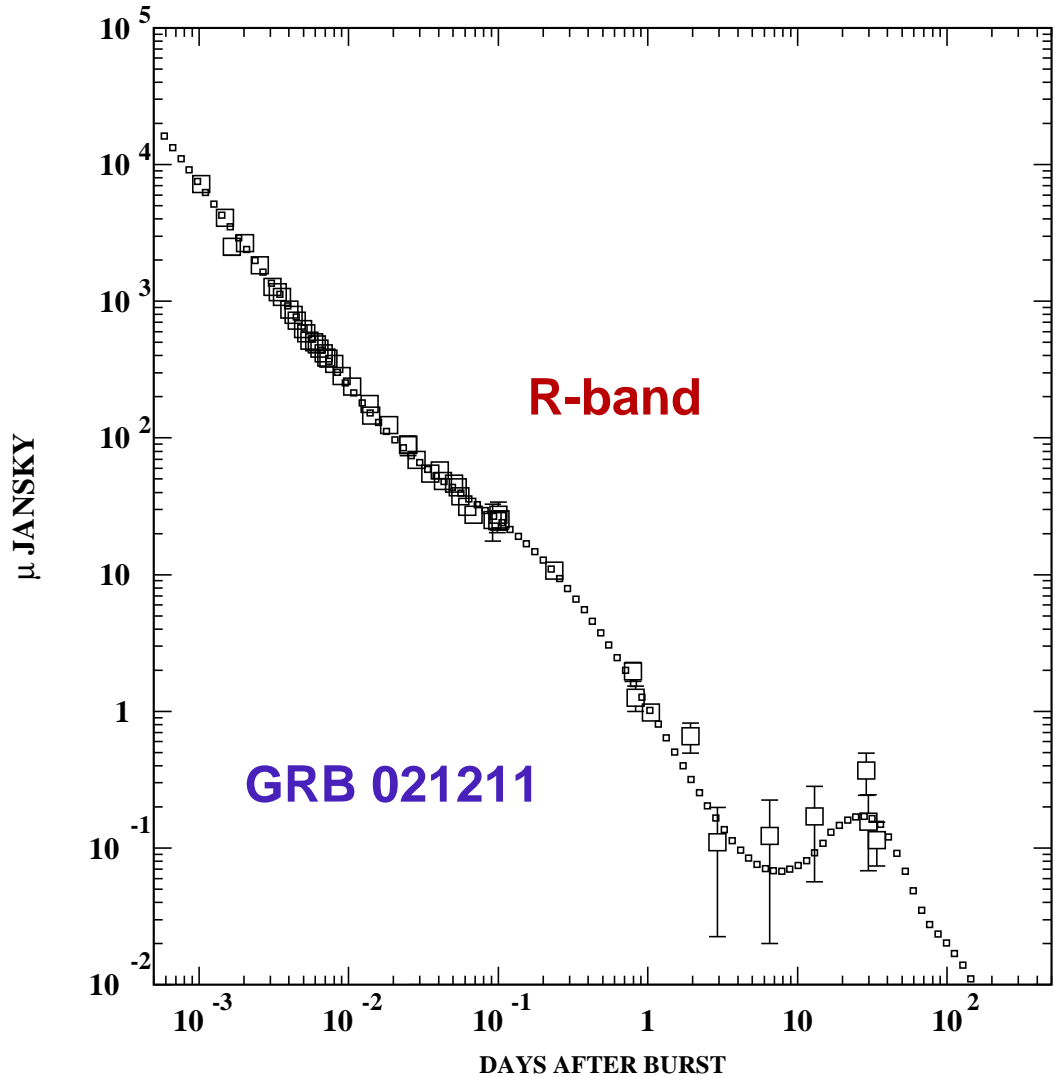, width=16cm}
\end{center}
\caption{Observations of the R-band
AG of GRB 021211, and their CB-model fit.  The ISM density
is a constant plus a ``wind'' contribution decreasing as the inverse square
of the distance.  The two density
contributions are equal at $\bar x\simeq 1.2$ pc, a
distance reached by the CBs in an observer's time $\bar t\simeq 0.025$ days
after burst (Dado et al.~2003e).    The
contribution of a SN1998bw-like SN at the GRB position
is discernible as the bump at late times.
The  host galaxy's contribution, which was also fitted, 
is subtracted in this plot.}
\label{f1}
\end{figure}

%\newpage
\clearpage

\begin{figure}[]
%\plotone{ag346nro021211.eps}
\vskip -4.5cm
\hskip 2truecm
\begin{center}
\epsfig{file=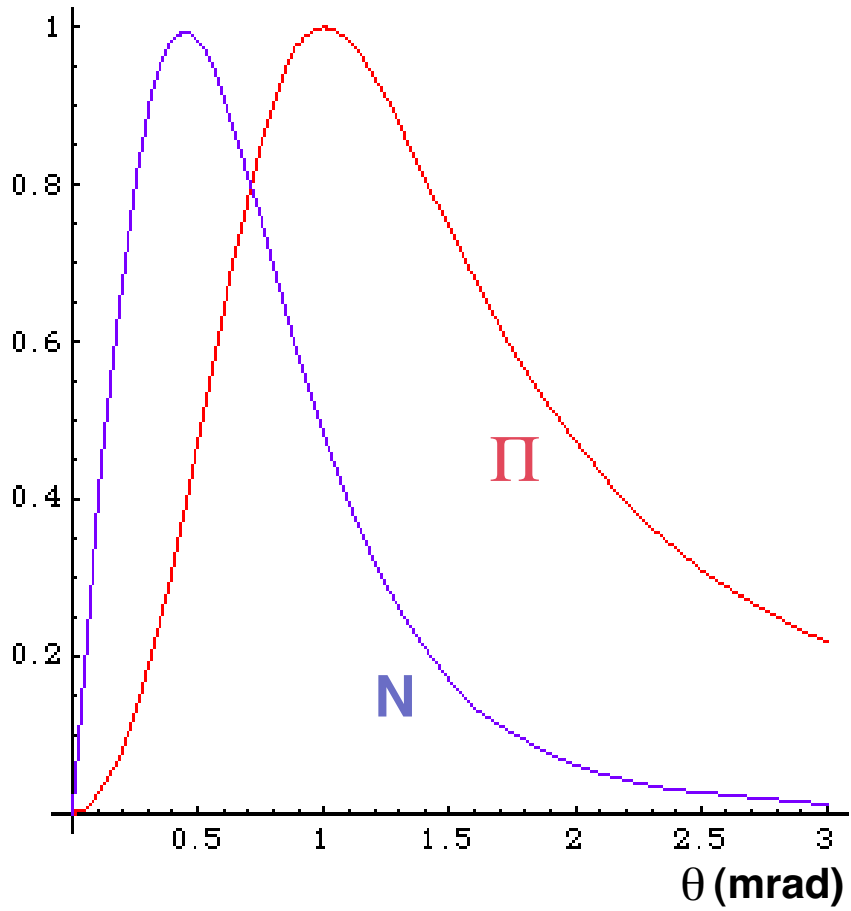, width=18cm}
\end{center}
\caption{The predicted polarization $\Pi(\theta,\gamma)$ of 
Eq.~(\ref{polSN}) and the (arbitrarily normalized) rough expectation 
$dN_{GRB}/{d\theta\,dt}\propto \theta\, \delta^3$ for the rate of photons
detected above a certain threshold, 
both as functions of $\theta$ (in mrad), at a typical $\gamma=10^3$.
The observation of a GRB with a measured $\Pi=80\pm20\%$ is
seen to be very probable.}
\label{f2}
\end{figure}

\clearpage

\begin{figure}[]
\vspace{-10cm}
\hspace{-1cm}
\epsfig{file=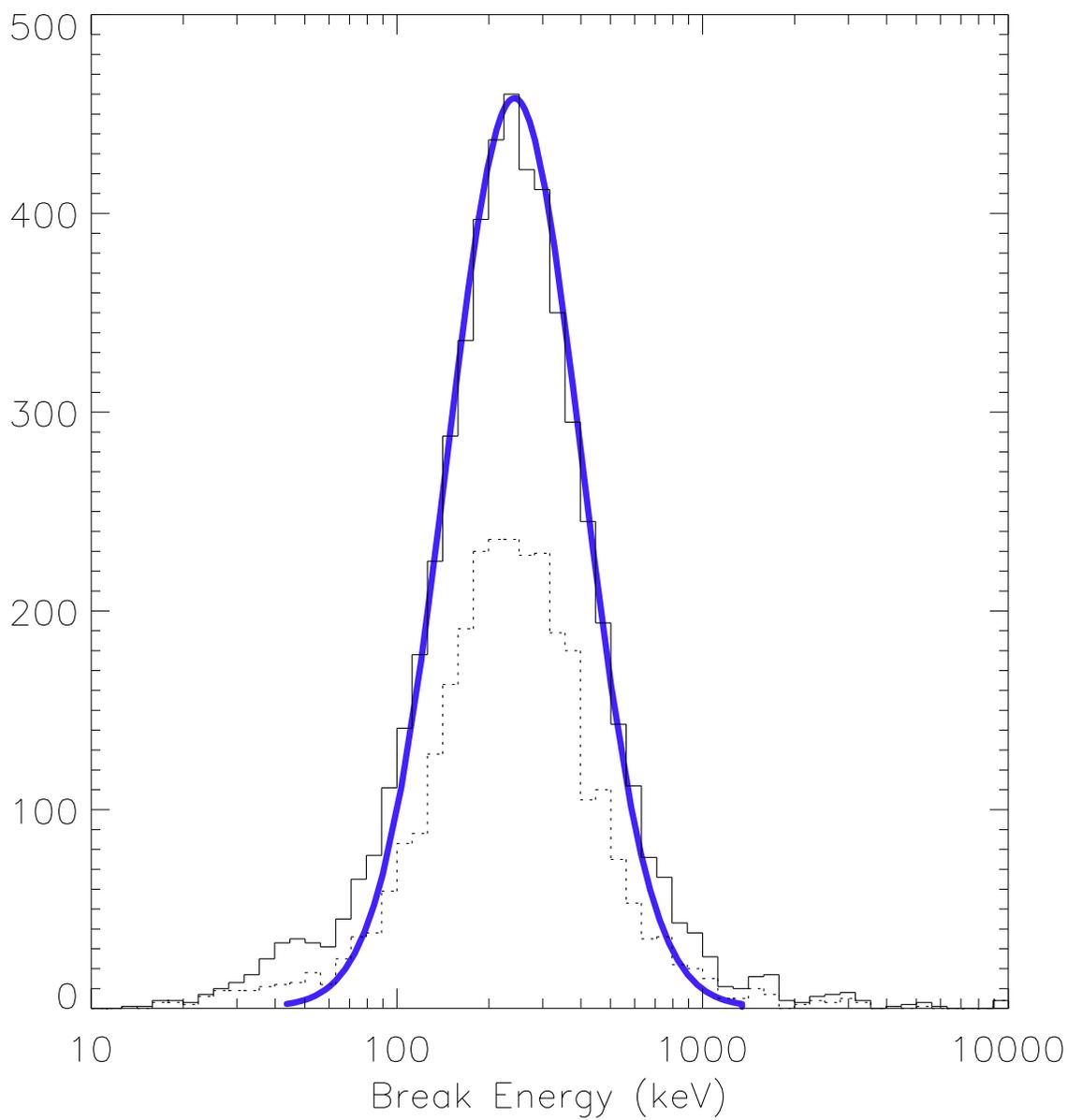,width=20cm}
%\hspace{-1cm}
 \caption{The break energy, or $E_p$ distribution
 of an ensemble of BATSE GRBs (Preece et al.~2000).
 The continuous line is borrowed from Fig.~(\ref{figEpth}).}
 \label{figEpobs}
\end{figure}

\clearpage

\begin{figure}[]
\begin{center}
\vspace{.5cm}
\epsfig{file=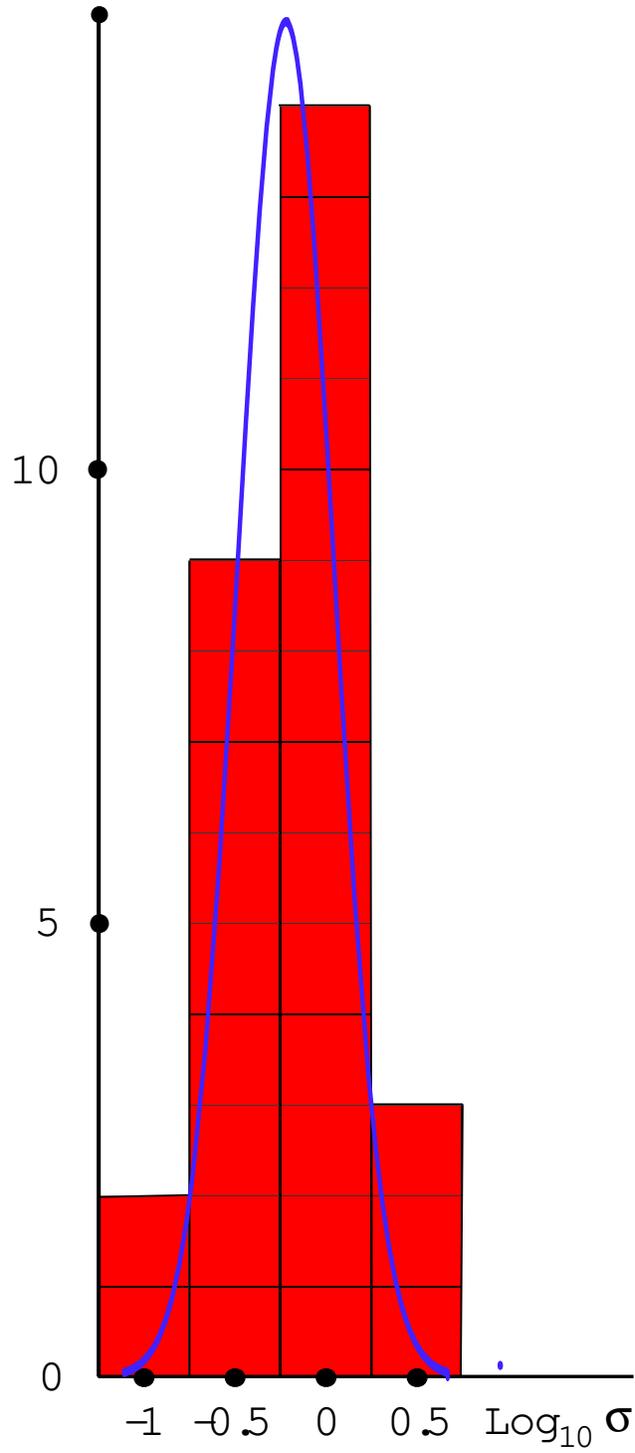, width=15cm}
%}
\vspace{-1cm}
\end{center}
 \caption{The distribution of the values of the quantity $\sigma$
 defined in Eq.~(\ref{boosting}) and listed in Table 2.
 The line is a log-normal Kolmogorov-Smirnov fit to the
 unbinned $\sigma$ distribution.}
 \label{figEpth}
\end{figure}

\clearpage

\begin{figure}[]
\vskip -13cm
%\hskip 2truecm
%\begin{center}
\hskip -1cm
\epsfig{file=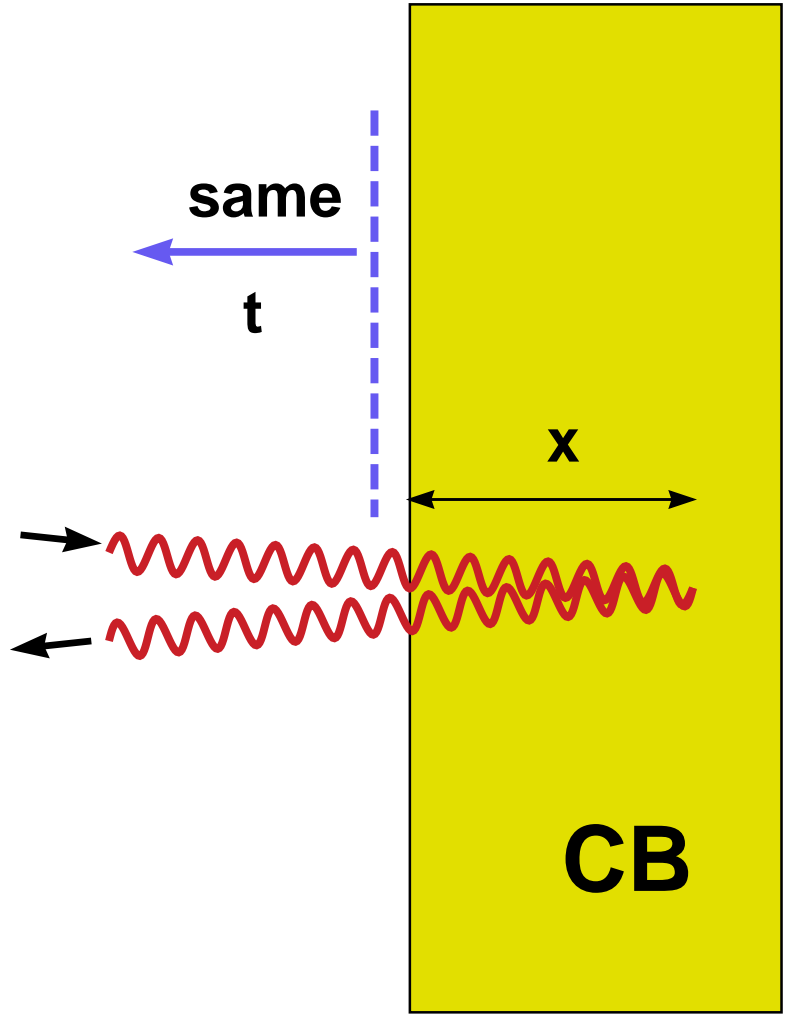, width=20cm}
%\end{center}
\vspace*{- .4cm}
%\hspace*{-0.2cm}
\caption{The ``slab'' CB geometry described in the text for the sake
of illustration. Photons exiting
the CB at the isochronous ``same $t$'' plane (the dashed line) may have 
suffered
backward Compton scattering at various depths, $x$, corresponding to
different times of entry of the incoming photon into the CB.}
\label{figslab}
\end{figure}

%\newpage
\clearpage

\begin{figure}[]
\vskip -10cm
\hskip 2truecm
\begin{center}
\epsfig{file=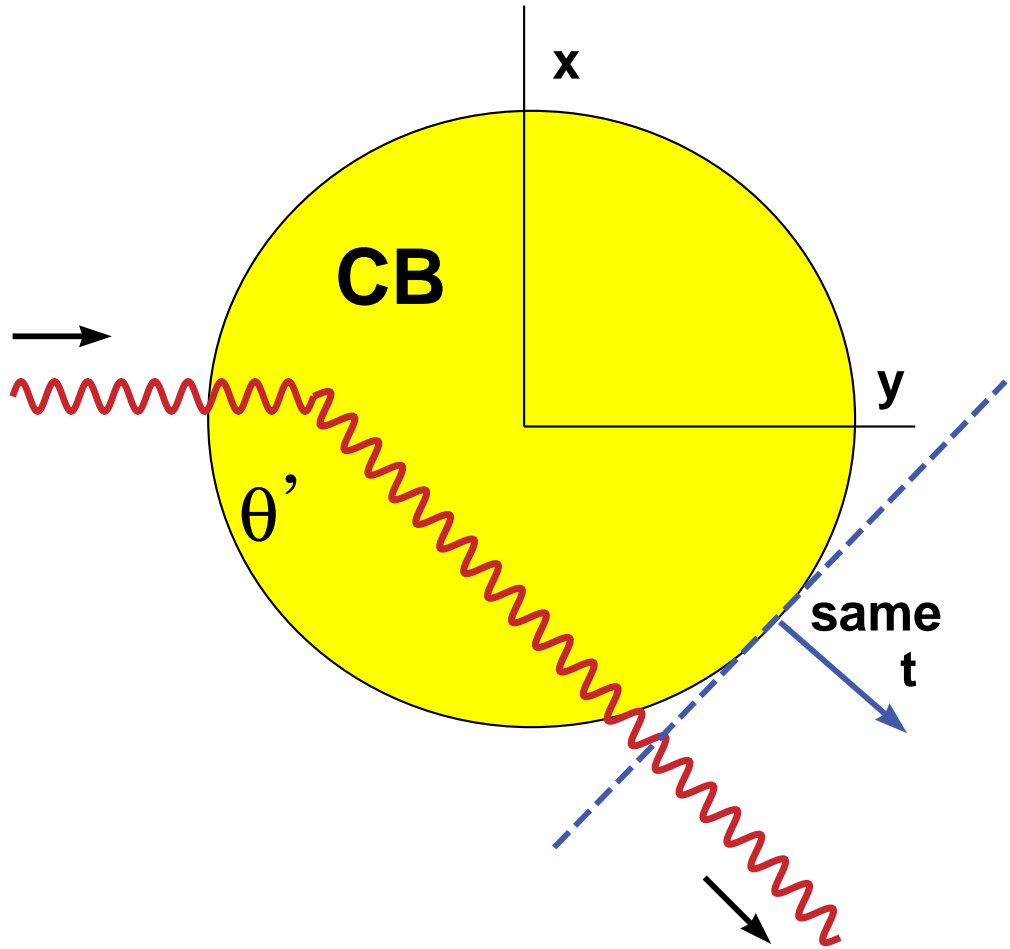, width=17cm}
\end{center}
\vspace*{- .4cm}
%\hspace*{-0.2cm}
\caption{The spherical CB geometry described in the text. Photons exiting
the CB at the isochronous ``same $t$'' plane (the dashed line) may have 
suffered
Compton scattering at a given angle ($\theta'$ in the CB's rest
frame) at various positions within the sphere. We have not attempted to
draw this figure in its full 3-D detail.}
\label{figsphere}
\end{figure}
%\newpage
\clearpage

\begin{figure}[]
%\vskip -1.5cm
\hskip 2truecm
%\vspace*{0.3cm}
\vspace*{- .4cm}
%\hspace*{-0.2cm}
\begin{center}
\epsfig{file=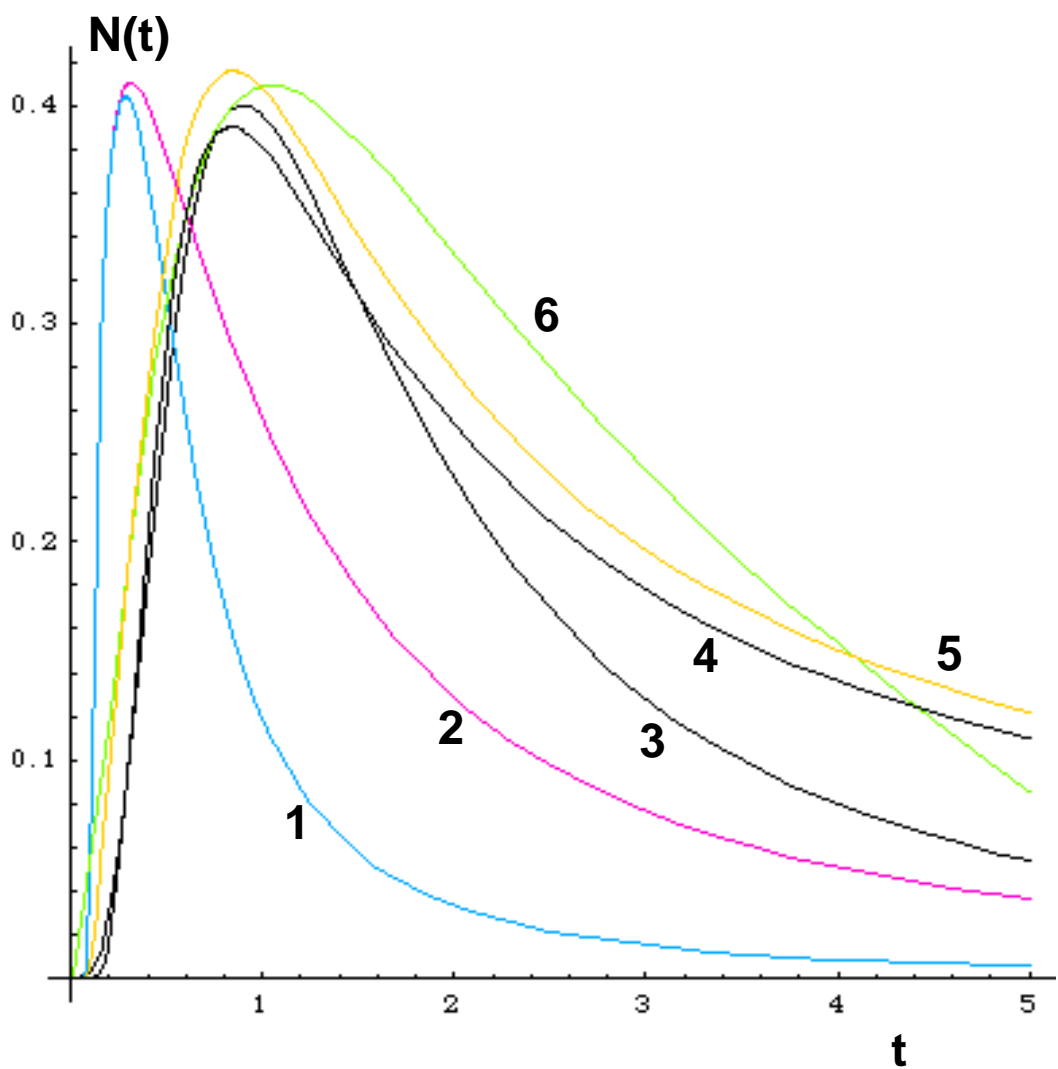, width=14cm}
\end{center}
\caption{The shapes of GRB pulses for various CB geometries,
described in the text.}
\label{fig3}
\end{figure}

\clearpage

\begin{figure}[]
\vskip -16.5cm
\hskip -2truecm
%\vspace*{0.3cm}
%\hspace*{-0.2cm}
\epsfig{file=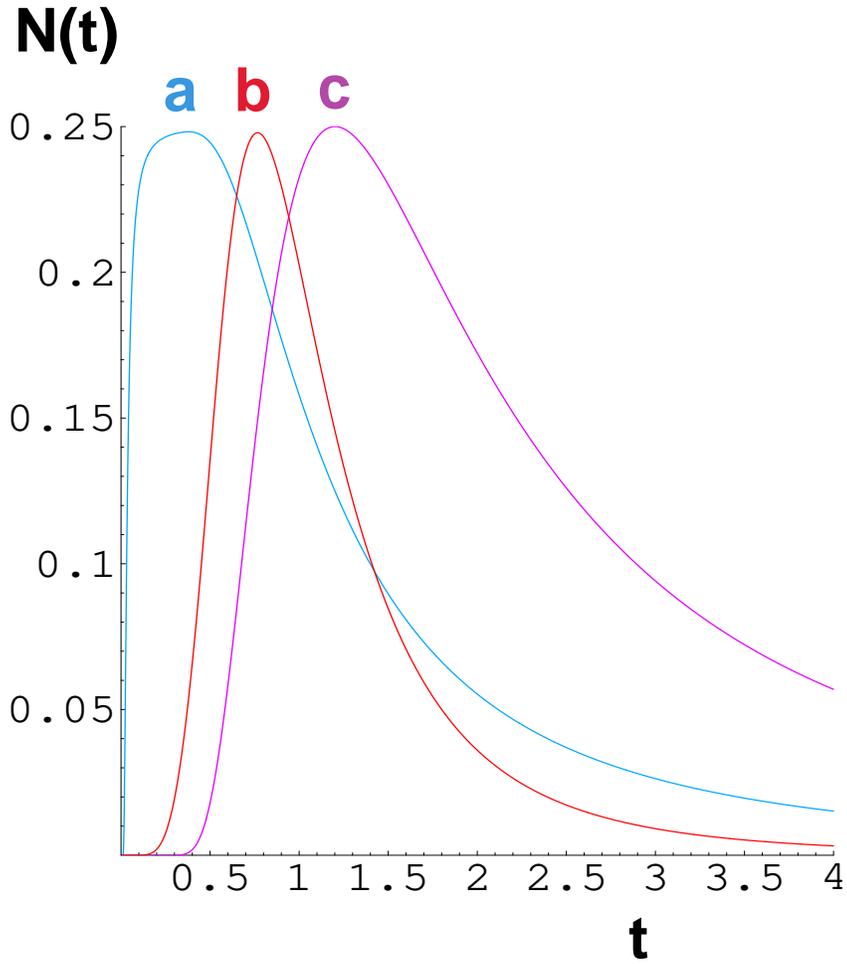, width=20cm}
\vspace*{- 1cm}
\caption{Three shapes of GRB pulses, as described by
Eq.~(\ref{pheno}). a) is for $t_{1}=0.03\,t_2$ and $m=n=2$.
 b) has $t_{1}=t_2$, $m=2$ and $n=4$.
 c) is the ``typical'' CB-model pulse shape with
$t_{1}=t_2$ and $m=n=2$.
These shapes describe GRB pulses integrated over energy.}
\label{fig3pulses}
\end{figure}

\clearpage
\begin{figure}[]
\vskip -20cm
\hskip -3truecm
%\vspace*{-5cm}
%\hspace*{-0.2cm}
%\begin{center}
\epsfig{file=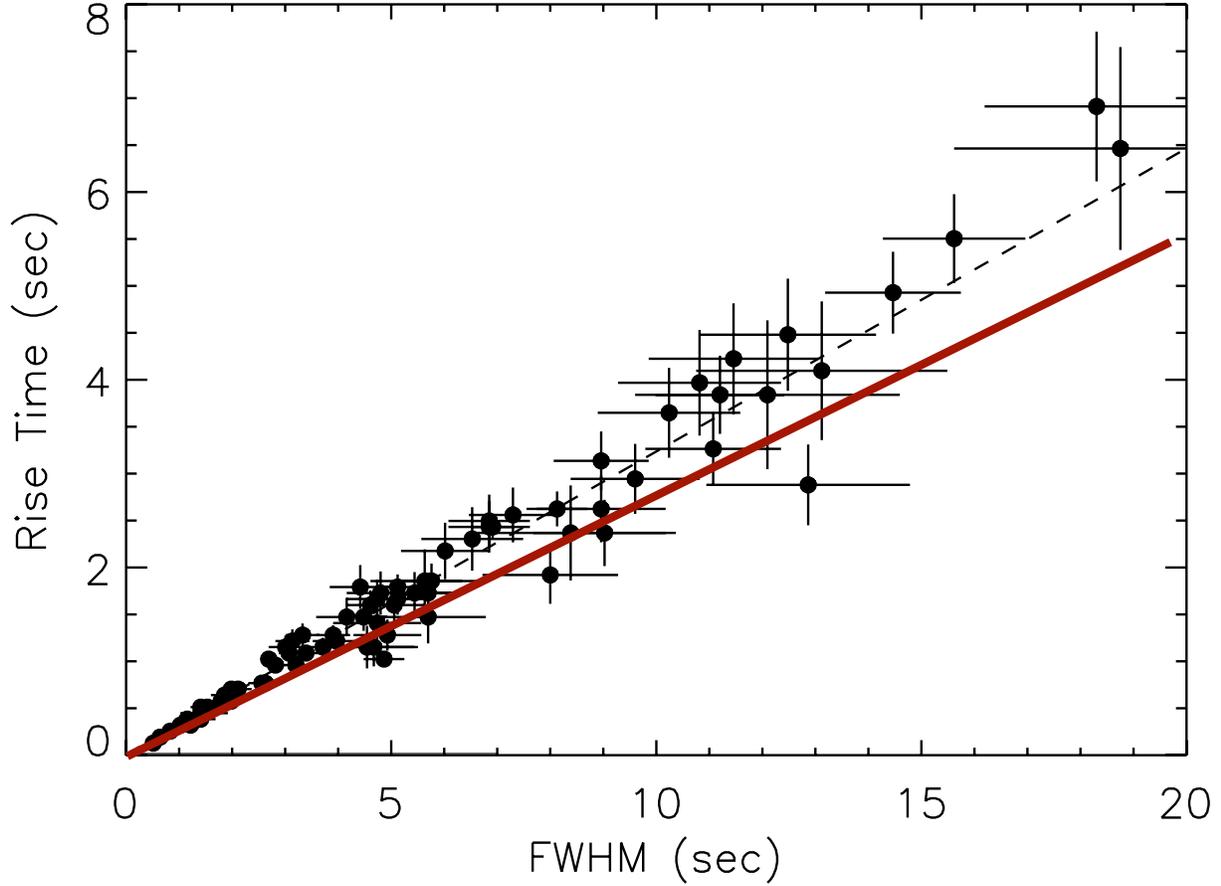, width=22cm}
\vskip -2cm
%\end{center}
%\vspace*{- 9cm}
\caption{Rise-time from half-maximum to maximum versus full 
width at half-maximum of an ensemble of GRB single pulses
(Kocevski et al.~2003). The
data are from pulses of bright BATSE GRBs, the 
theoretical prediction (the continuous line)
is from the naive pulse shape of Eq.~(\ref{naivepulse}).
The dotted line is the best linear fit.}
\label{figFWRise}
\end{figure}

%\newpage
\clearpage

\begin{figure}[]
\vskip -9cm
%\hskip 1truecm
%\vspace*{0.3cm}
%\vspace*{- .4cm}
\hspace*{-1.3cm}
%\begin{center}
\epsfig{file=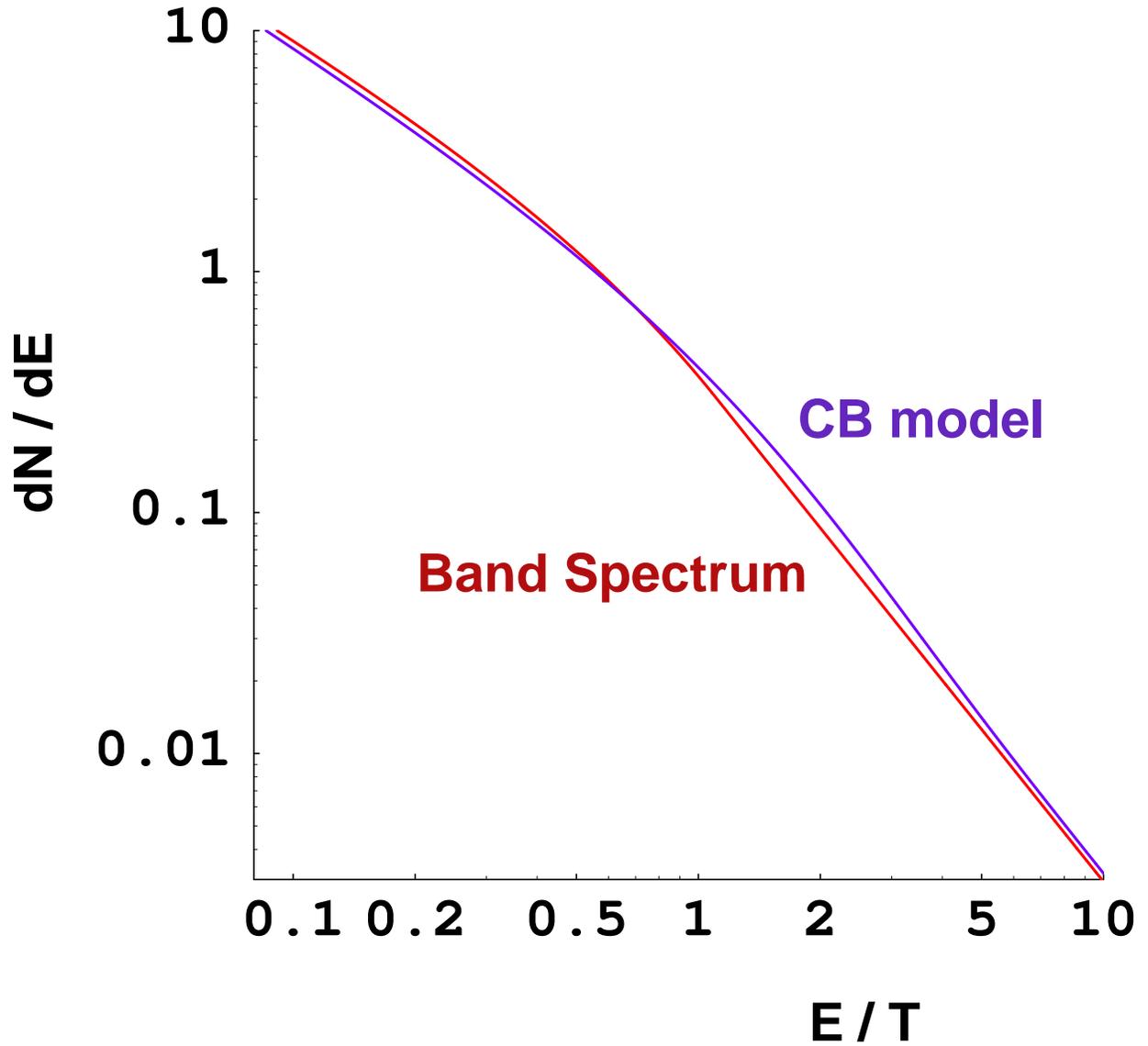, width=19cm}
%\end{center}
%\vspace*{- 2.4cm}
\caption{Two shapes of GRB spectra, 
the number of photons per energy interval
$dN/dE$. One is the prediction of the CB model,
Eq.~(\ref{totdist}). The other is the successful phenomenological
Band spectrum of Eq.~(\ref{band}) (Band et al.~1993);
$T$ stands for the bend energy in the Band's case. Considering that the
prediction is based exclusively on first principles, the agreement is rather
satisfying. }
\label{figband}
\end{figure}

\clearpage

\begin{figure}[]
\begin{center}
\vspace{-6.5cm}
\vbox{\epsfig{file=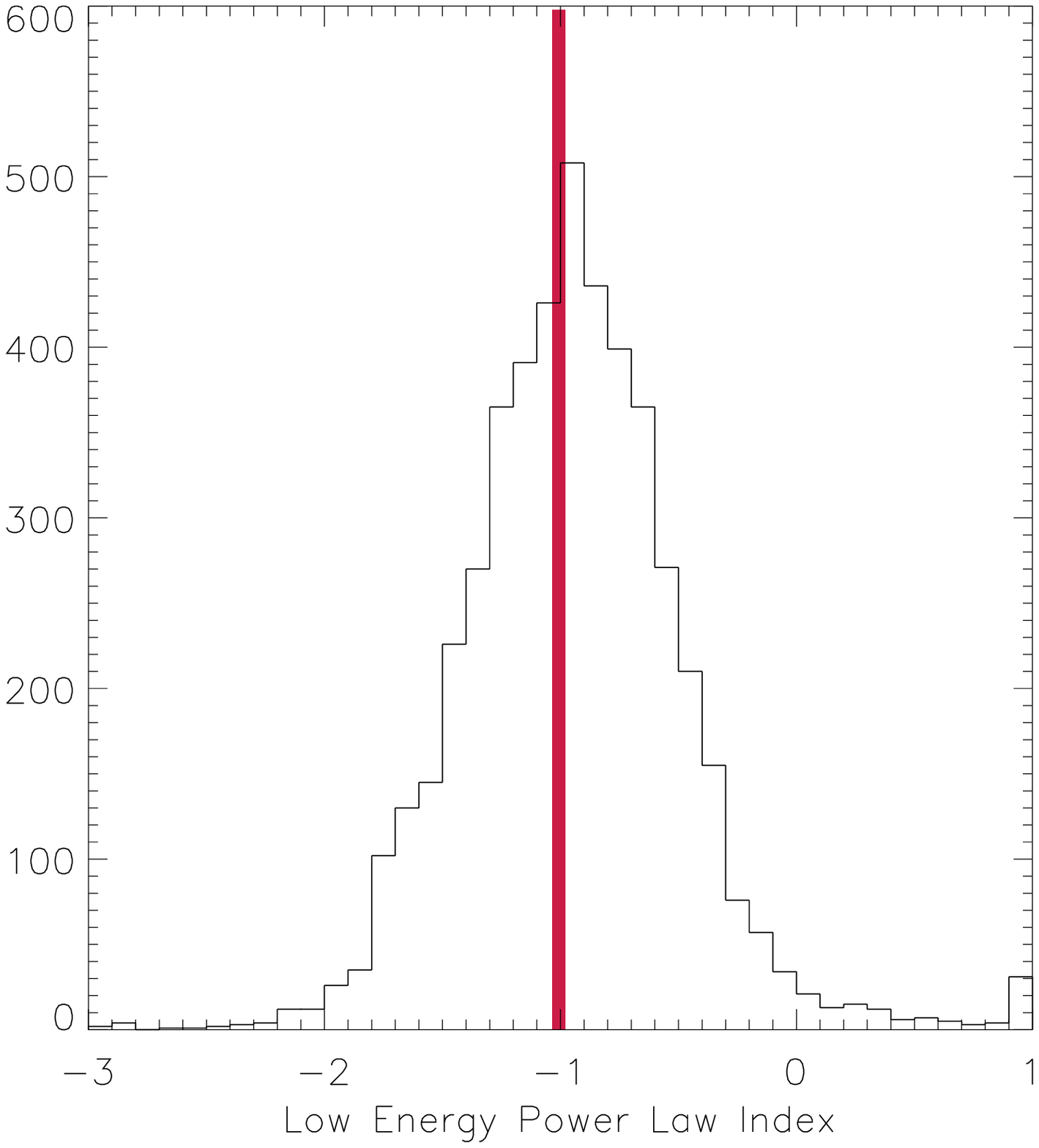,height=17cm,width=15cm}}
\vspace{-7cm}
\vbox{\epsfig{file=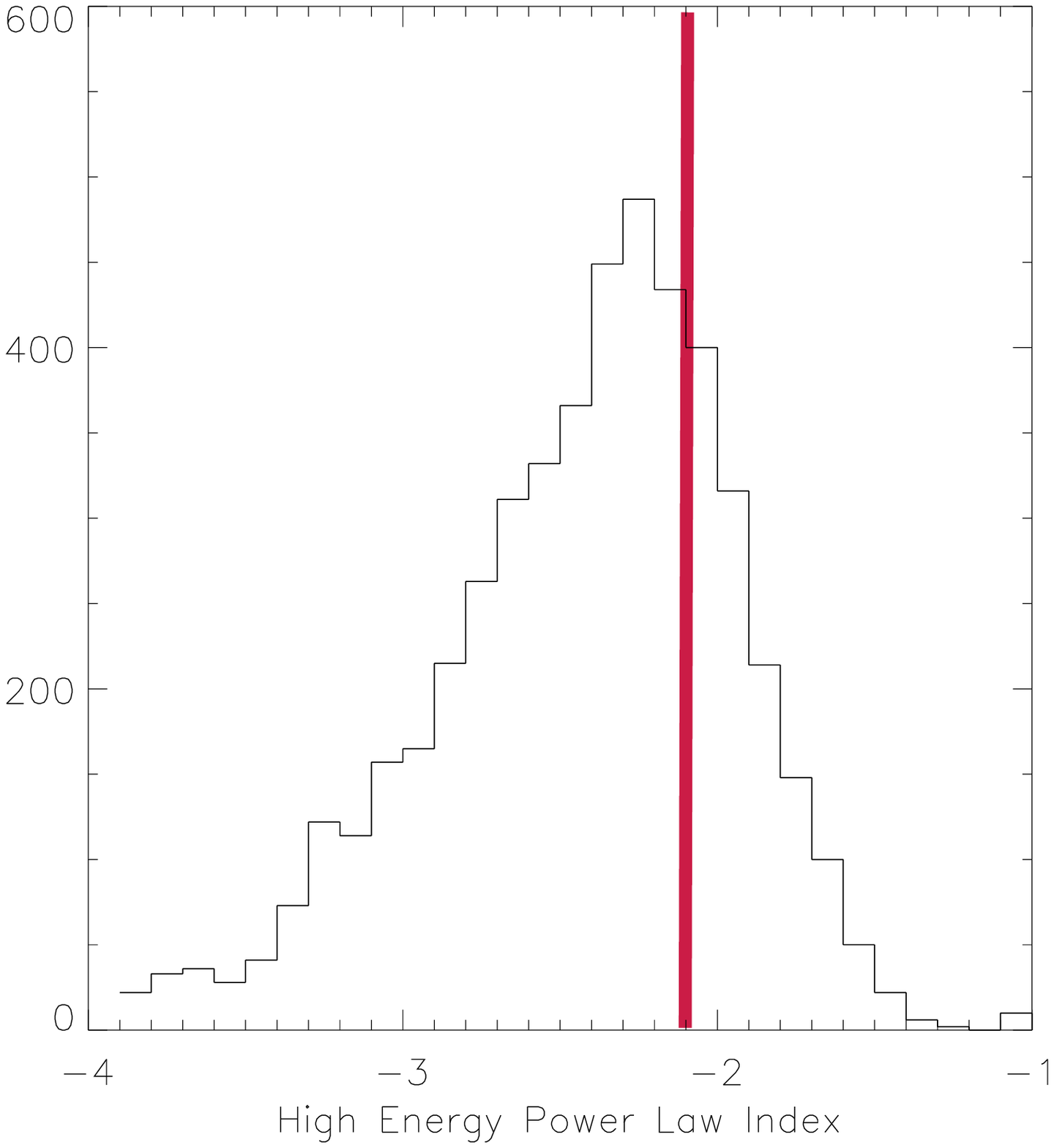,height=17cm,width=15cm}}
\vspace{-1.3cm}
\end{center}
 \caption{
 Upper panel: The observed distribution of the
 index $-\alpha$ of Eq.~(\ref{totdist}), compared with
 its central predicted value (the line). Lower panel:
 The same for the index $-\beta$. The data analysis is that
 of Preece et al.~(2000) and, in the $\beta$ distribution.
 We have eliminated their bin at $\beta=4$, which contains
 events without a determined $\beta$. The prediction
 for the central $\alpha$-value does not depend on the
 adopted index of the accelerated electron distribution;
 the one for $\beta$ does. We interpret the events with
 $\beta>2.1$ as due in part to the effect of the energy
 cutoff of the knocked-on electrons, Eq.~(\ref{dndgamma}).}
 \label{figalphabeta}
\end{figure}

%\newpage
\clearpage

\begin{figure}[]
\begin{center}
\vspace{-0.5cm}
\vbox{\epsfig{file=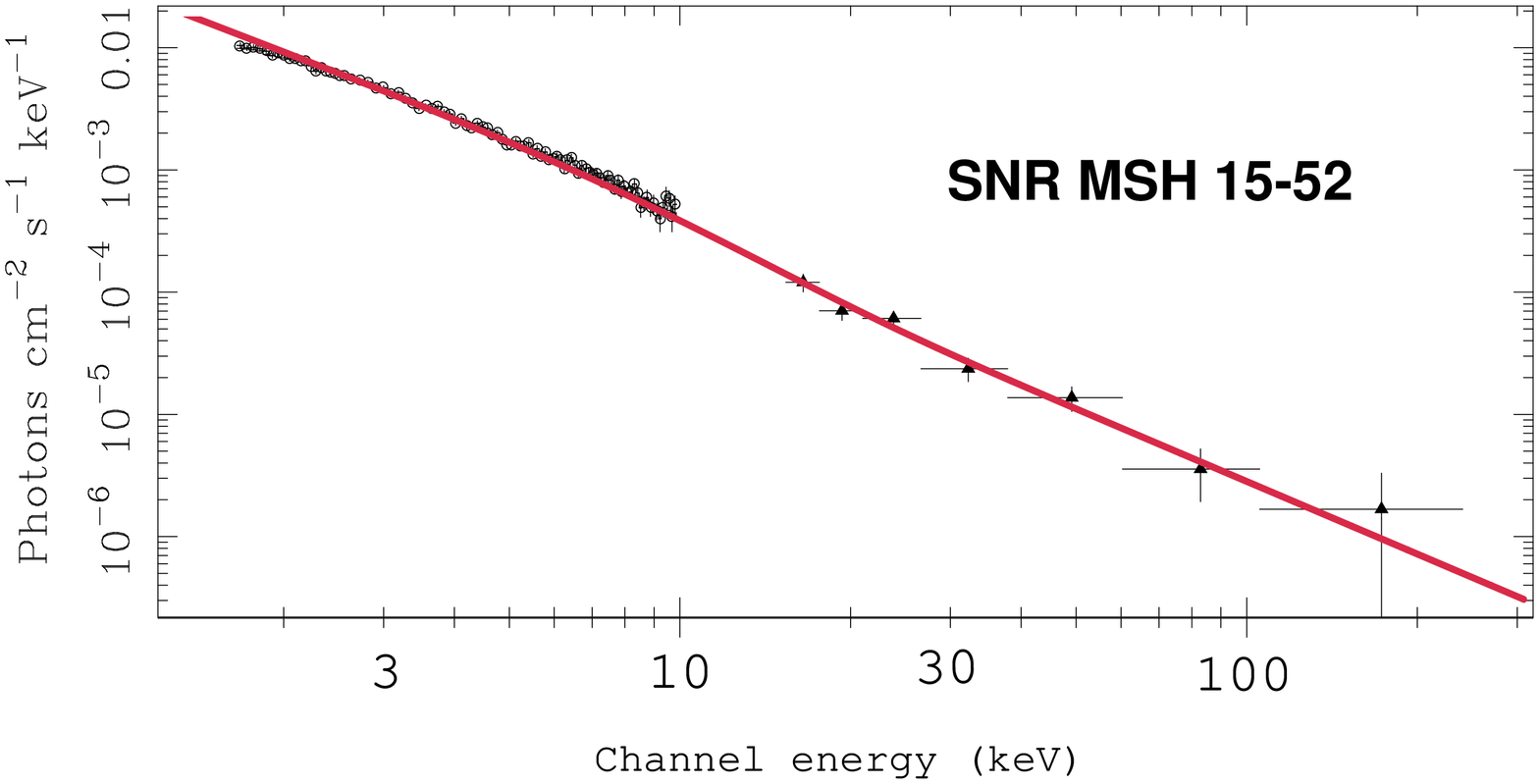,height=16cm,width=14.cm,angle=-90}}
\vspace{-4.5cm}
\vbox{\epsfig{file=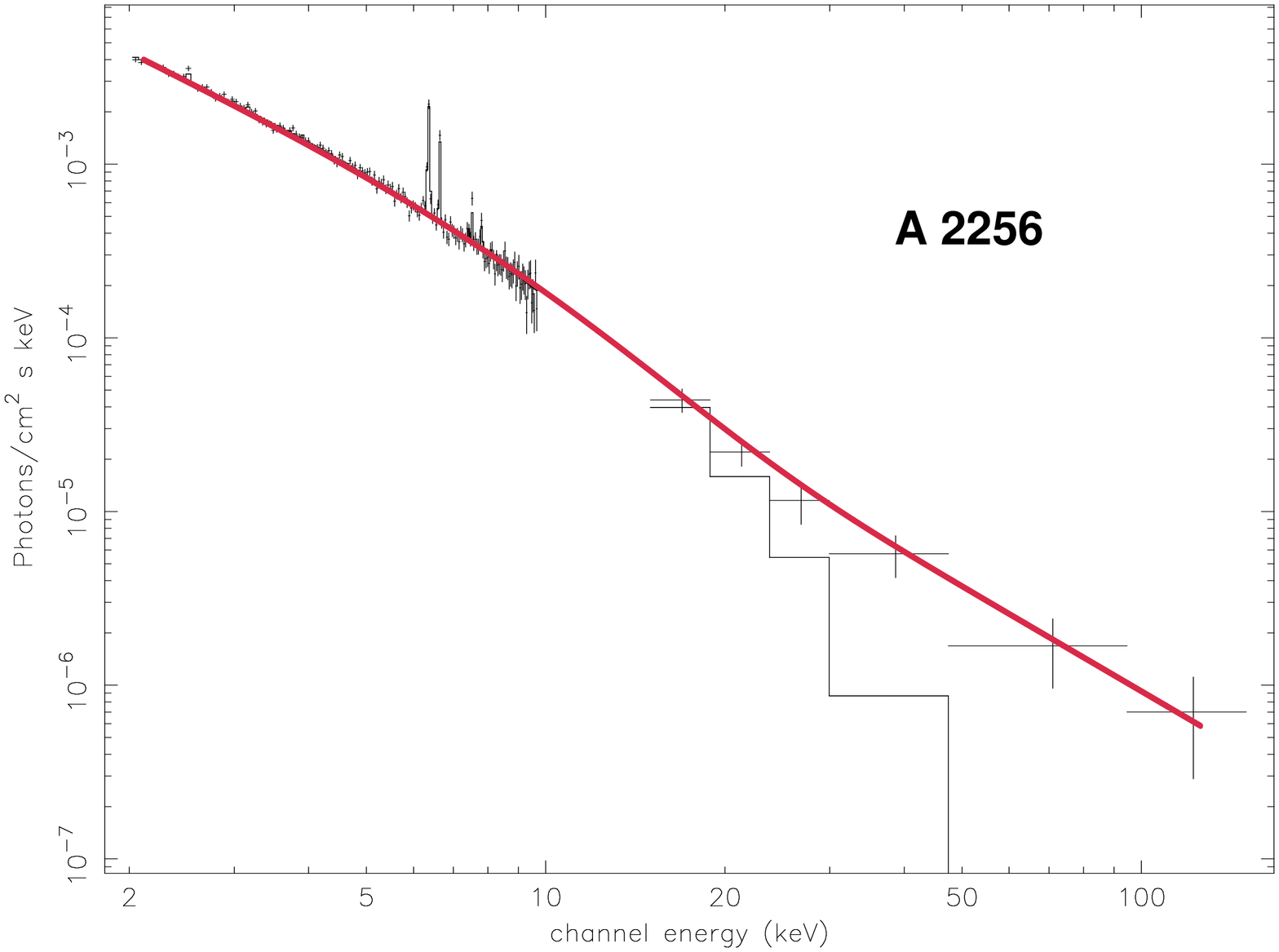,height=16cm,width=12.cm,angle=-90}}
\vspace{-1cm}
\end{center}
 \caption{{Two figures from Colafancesco, Dar \& De R\'ujula (2003).
 Upper panel: The thick (red) line is a
``$\chi$-by-eye'' fit of Eq.~(\ref{totdist}) to the thin thermal bremsstrahlung
emission from the SN remnant SNR MSH 15-52. The data and figure are from Mineo
et al.~(2001). Lower panel: The same for the cluster A2256. The data and figure
are from Fusco-Femiano et al.~(2000). The thin (black) line is their
binned, purely thermal extrapolation.}}
 \label{figspectrum}
\end{figure}

%\newpage
\clearpage

\begin{figure}[]
\vskip -8.5cm
\hskip 2truecm
%\vspace*{0.3cm}
%\vspace*{- .4cm}
%\hspace*{-0.2cm}
\begin{center}
\epsfig{file=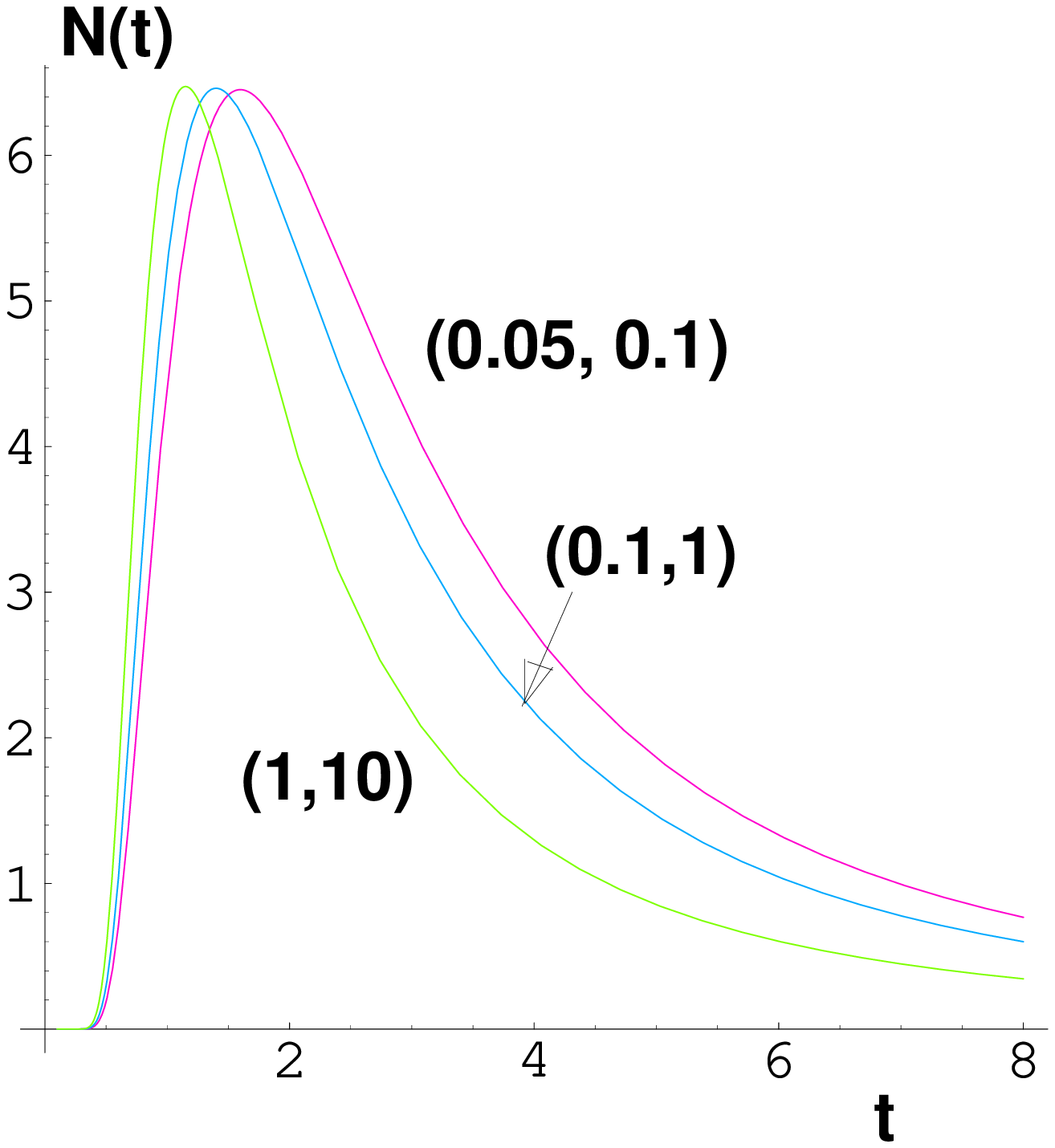, width=20cm}
\end{center}
\caption{Pulse shapes (arbitrarily normalized for presentation)
at three energy intervals, given in parenthesis in units of $T$
and corresponding to Eqs.~(\ref{pevol}, \ref{dEdt1}).
The time is in units of $t_{tr}$. }
\label{Noft1}
\end{figure}

%\newpage
\clearpage

\begin{figure}[]
%\vskip -4.5cm
\hskip 2truecm
%\vspace*{0.3cm}
\vspace*{- .4cm}
%\hspace*{-0.2cm}
\begin{center}
\epsfig{file=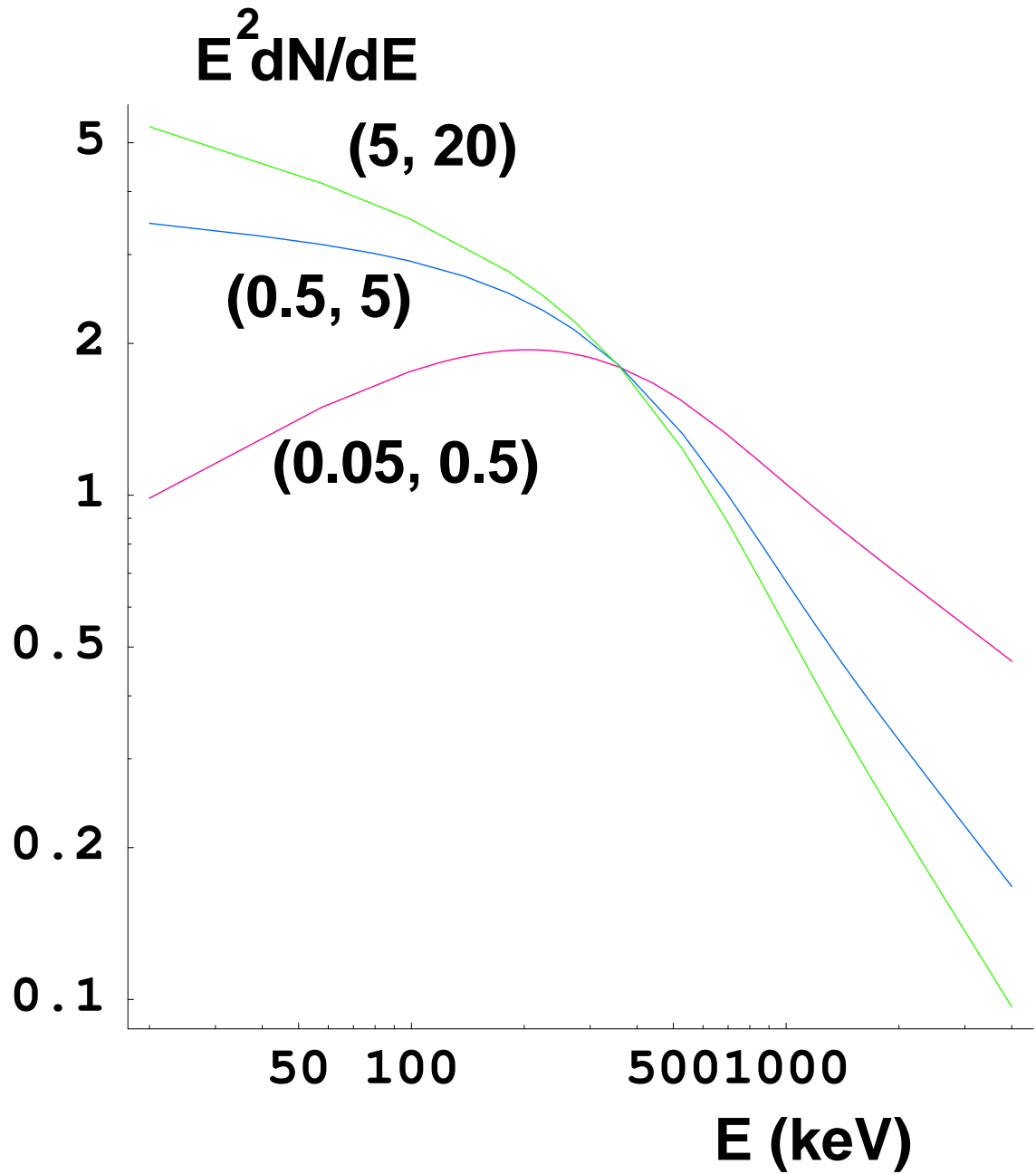, width=15cm}
\end{center}
\caption{Spectral shapes $E^2\,dN/dE$, predicted by 
Eqs.~(\ref{pevol}, \ref{dEdt1})
at three time intervals within a pulse, shown in parenthesis 
in units of $t_{tr}$.}
\label{figSpectra1bis}
\end{figure}

%\newpage
\clearpage

\begin{figure}[]
\vskip -8.5cm
\hskip 2truecm
%\vspace*{0.3cm}
%\vspace*{- .4cm}
%\hspace*{-0.2cm}
\begin{center}
\epsfig{file=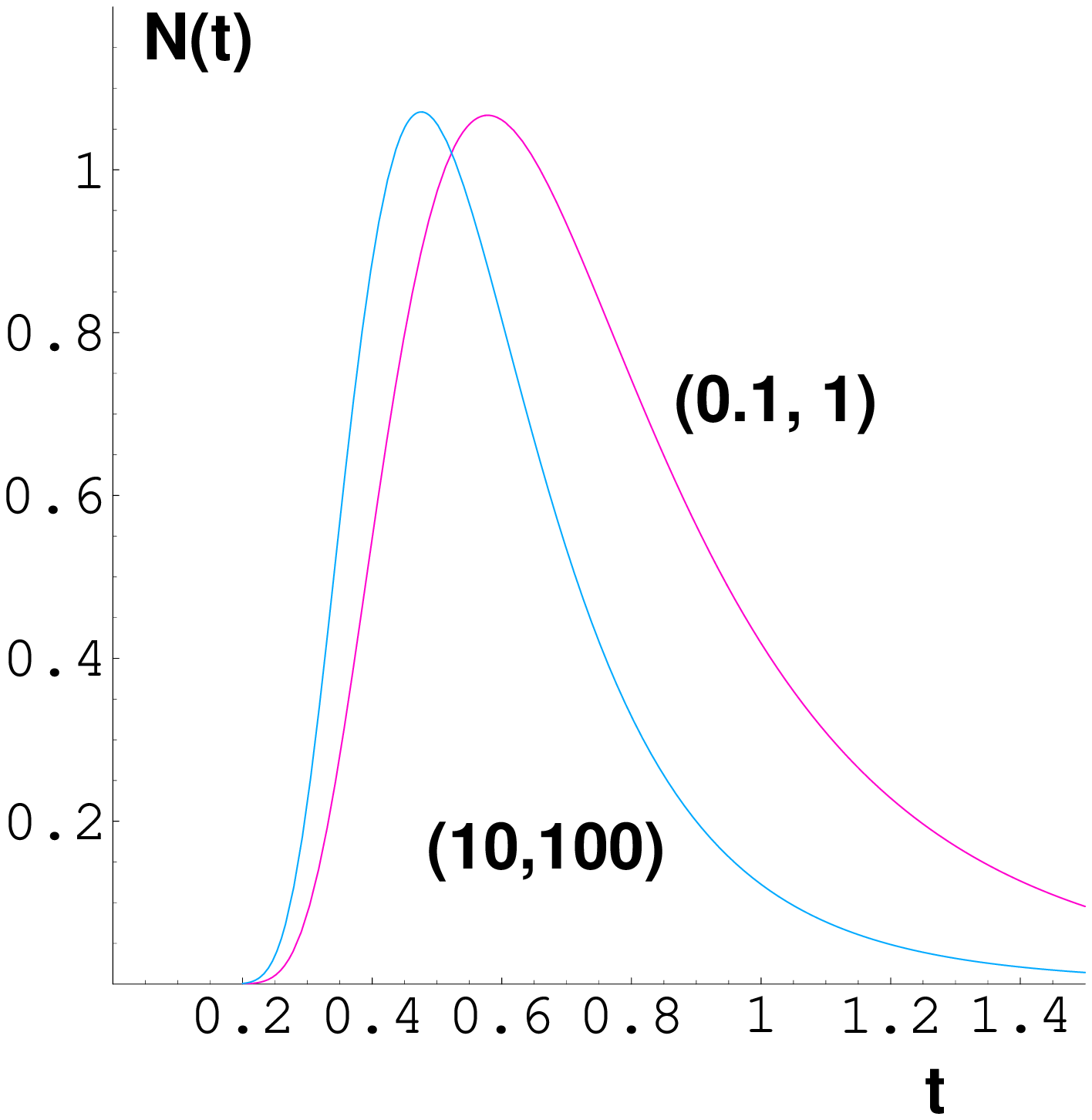, width=20cm}
\end{center}
\caption{Pulse shapes $dN/dt$ (arbitrarily normalized for presentation)
at two energy intervals, given in parenthesis in units of $T$
and corresponding to Eqs.~(\ref{Tevol}, \ref{dEdt}).
The time is in units of $t_{tr}$. }
\label{Noft2}
\end{figure}

%\newpage
\clearpage

\begin{figure}[]
\vskip -8.5cm
\hskip 2truecm
%\vspace*{0.3cm}
%\vspace*{- .4cm}
%\hspace*{-0.2cm}
\begin{center}
\epsfig{file=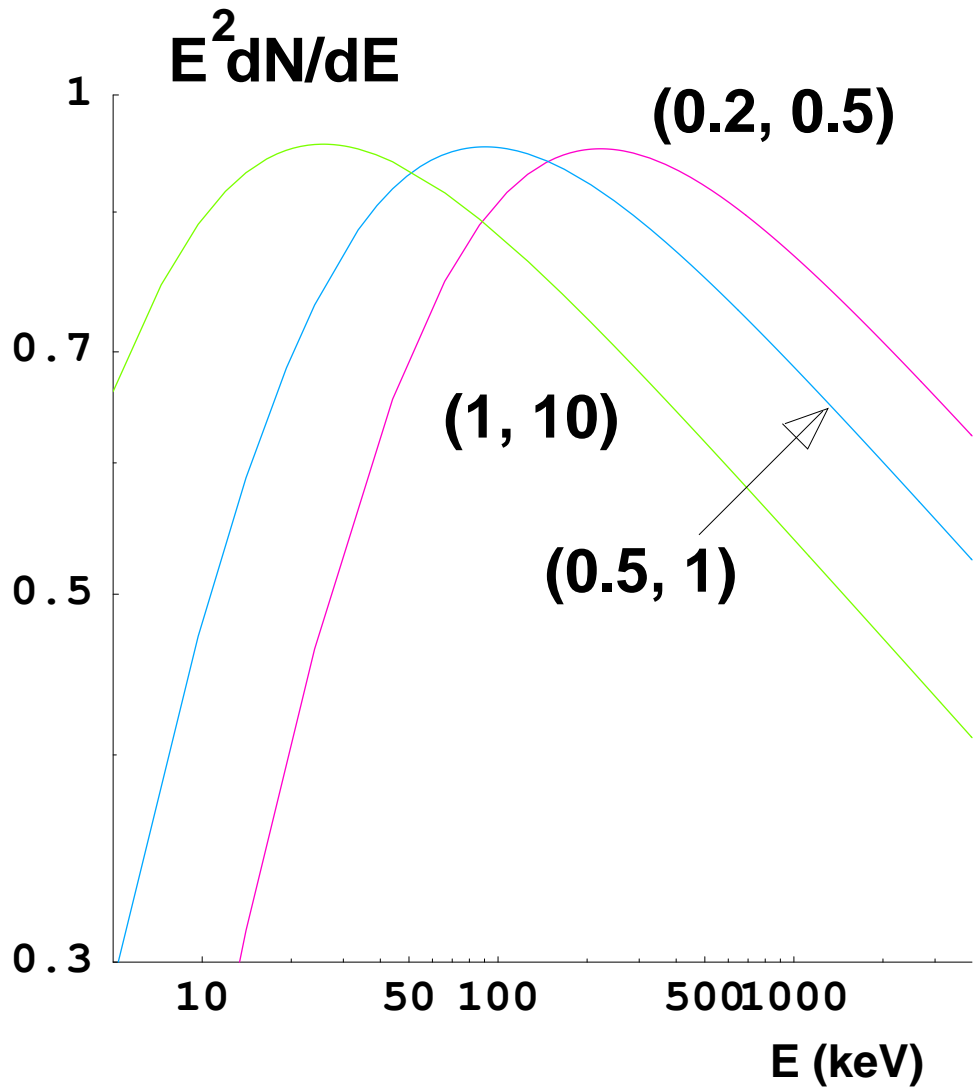, width=20cm}
\end{center}
\vskip -1cm
\caption{Energy distribution $E^2\,dN/dE$, at various times in a pulse,
 given in parenthesis
in units of $t_{tr}$, and corresponding to Eqs.~(\ref{Tevol}, \ref{dEdt}).}
\label{figSpectra2}
\end{figure}

\clearpage

\begin{figure}[]
\vskip -6.5cm
%\vspace*{0.3cm}
\vspace*{- 3.4cm}
\hspace*{-1.2cm}
%\begin{center}
\epsfig{file=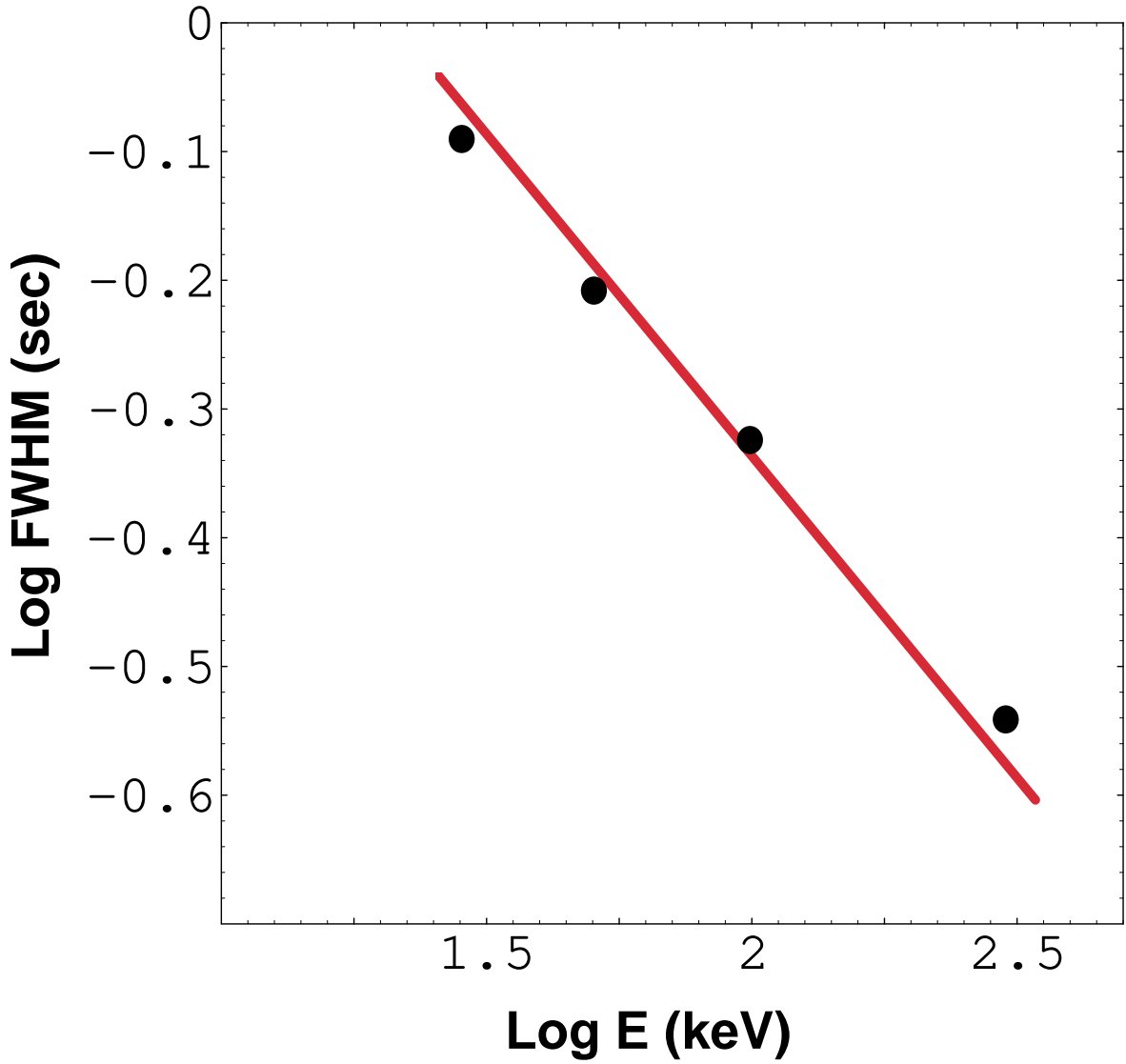, width=18cm}
%\end{center}
\vspace{-1cm}
\caption{Average FWHM for the four BATSE energy channels,
as reported by Fenimore et al.~(1995) and Norris et al.~(1996)  
compared with the prediction
of Eq.~(\ref{fenimore}). A more elaborate theoretical analysis would result
in $\kappa \,\lsim\, 0.5$, a slightly less steep prediction.}
\label{figLogWLogE}
\end{figure}

\clearpage

\begin{figure}[]
\vskip -10.5cm
\hskip 2truecm
%\vspace*{0.3cm}
\vspace*{- 3.4cm}
%\hspace*{-0.2cm}
\begin{center}
\epsfig{file=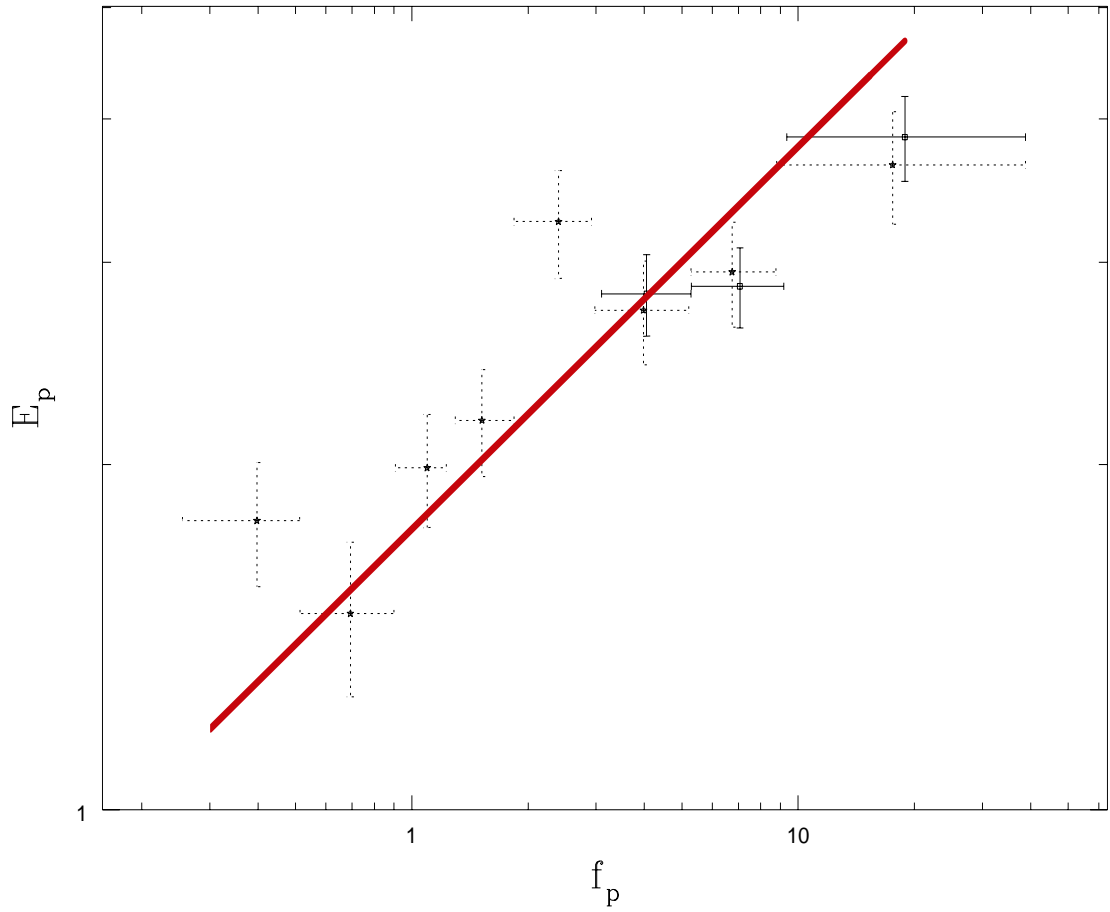, width=18cm}
\end{center}
\vspace{-1cm}
\caption{Average peak energy $E_p$, versus 
peak photon intensity $f_p$ in bins
containing  20 GRBs of similar peak intensity (Lloyd et al.~2000).
The line is the prediction of Eq.~(\ref{epint}).}
\label{figEpFp}
\end{figure}

\clearpage

\begin{figure}[]
\vskip -13.5cm
\hskip 2truecm
%\vspace*{0.3cm}
\vspace*{- 3.4cm}
%\hspace*{-0.2cm}
\begin{center}
\epsfig{file=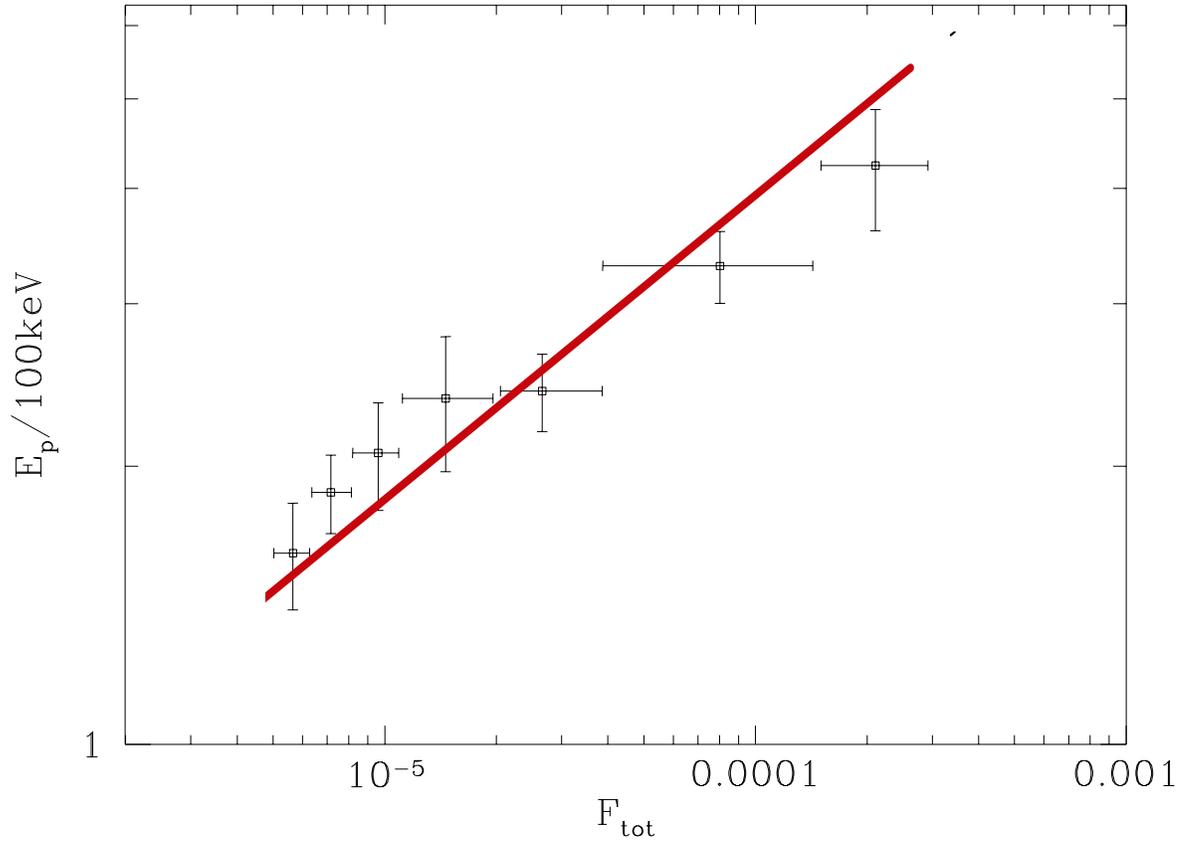, width=18cm}
\end{center}
\vspace{-1cm}
\caption{Averaged peak energy, $E_p$, versus total fluence, $F_{tot}$,
in bins containing  20 GRBs of similar fluence (Lloyd et al.~2000). 
The line is the prediction of Eq.~(\ref{epflu}).}
\label{figEpFot}
\end{figure}

\clearpage

\begin{figure}[]
\vskip -8.5cm
\hskip -2truecm
%\vspace*{0.3cm}
\vspace*{- 3.4cm}
%\hspace*{-0.2cm}
\epsfig{file=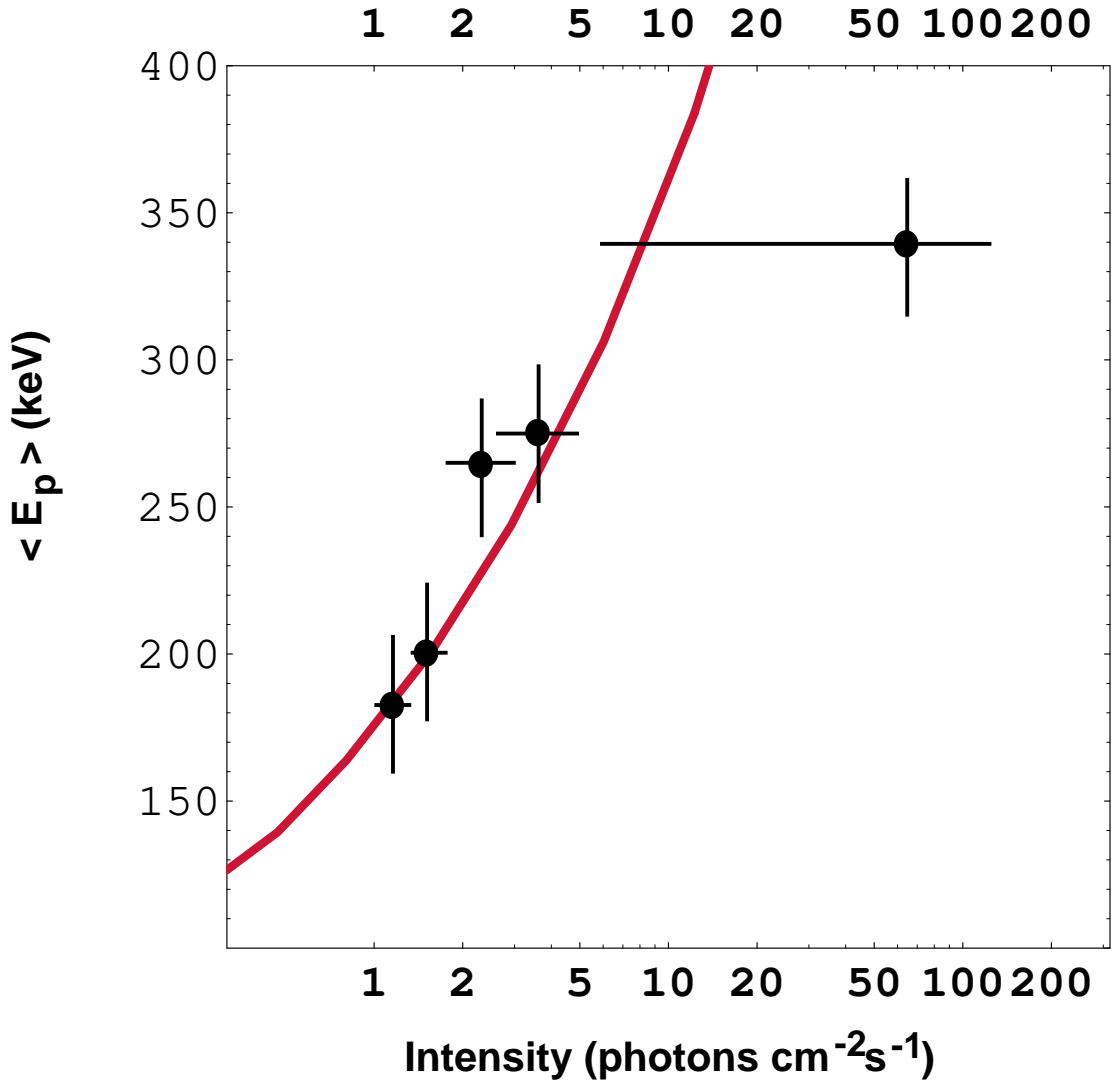, width=18cm}
\vspace{2cm}
\caption{
Averaged peak energy, $E_p$, versus peak photon
intensity, $f_p$, from an analysis by Mallozzi et al.~(1995). 
The line is the prediction of Eq.~(\ref{epint}).}
\label{figEpInt}
\end{figure}

\clearpage

\begin{figure}[]
\vskip -21.5cm
\hskip -2truecm
%\vspace*{0.3cm}
\vspace*{- 3.4cm}
%\hspace*{-0.2cm}
\epsfig{file=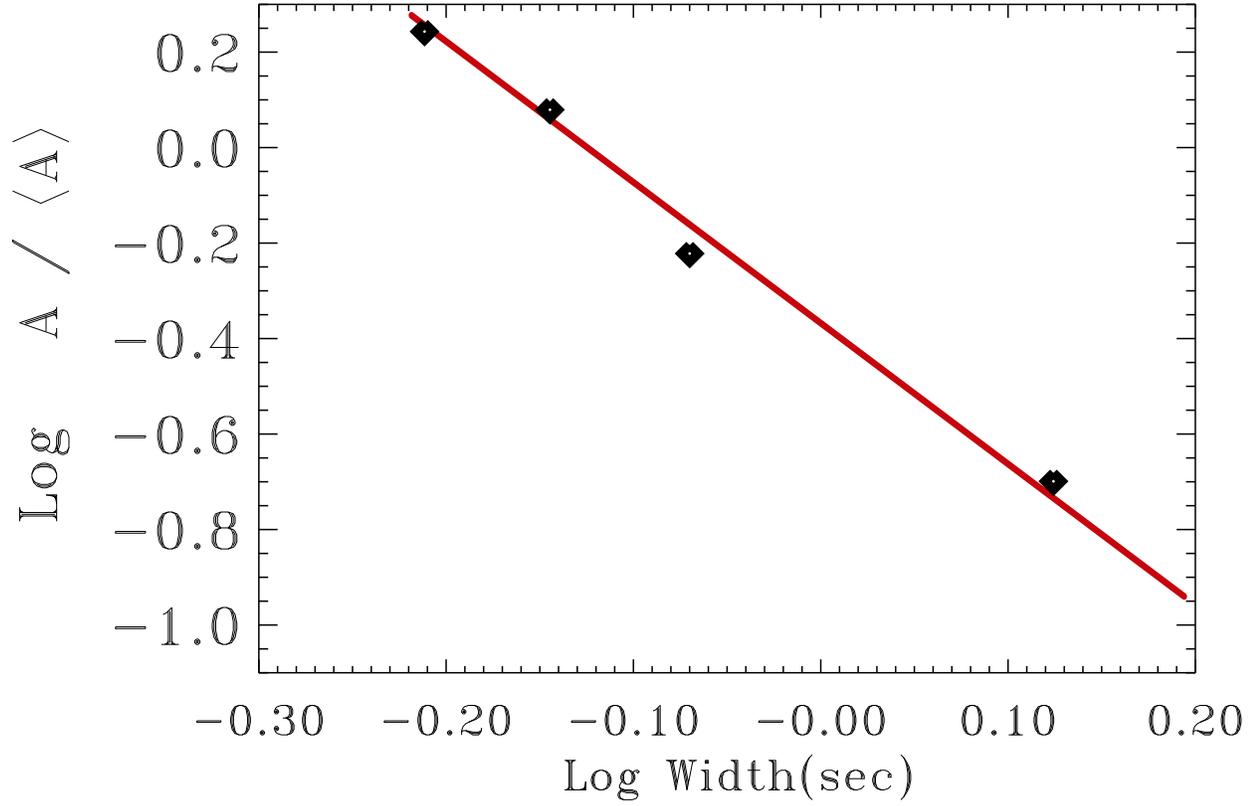, width=21cm}
\vspace{2cm}
\caption{The ratio $A / {\langle A\rangle}
\equiv f_p/{\langle f_p\rangle}$ 
versus the FWHM of GRB pulses, from
Ramirez-Ruiz \& Fenimore~(2000). The line is the prediction of
Eq.~(\ref{epint}).}
\label{figLogALogW}
\end{figure}

\clearpage

\begin{figure}[]
\vskip -4cm
%\hskip -1truecm
%\vspace*{-44cm}
%\hspace*{-2.2cm}
%\begin{center}
%\hspace*{-.8cm}
\epsfig{file=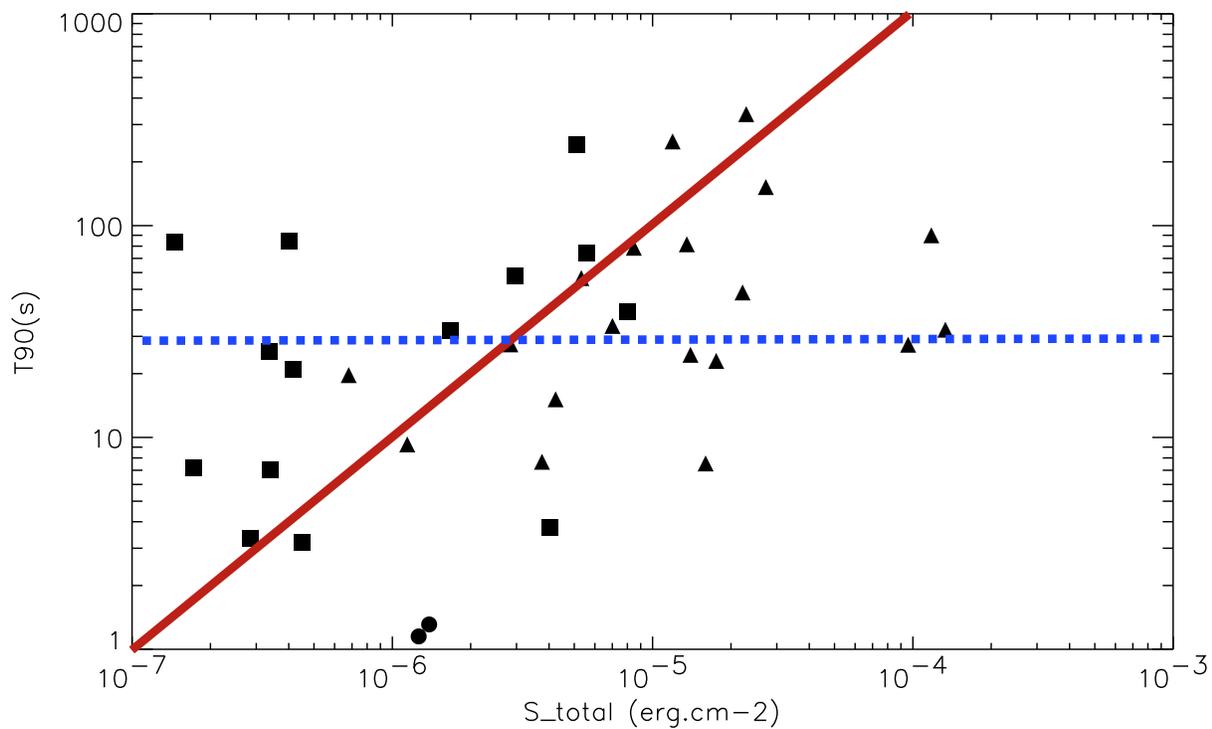, width=14cm}
%\end{center}
\vspace*{1cm}
\caption{Observed GRB duration ($t_{90}$) and total energy fluence
in the 7-400 keV band of 35 GRBs that were measured by HETE II
(Barraud et al. 2003). The theoretical continuous line is the average trend 
expected in CB model for the width of the individual
pulses. The theoretical dashed line is the expectation for the time
intervals between pulses.}
\label{figT90}
\end{figure}

\clearpage

\begin{figure}
\vspace*{-9cm}
\hskip -.5truecm
%\begin{center}
\epsfig{file=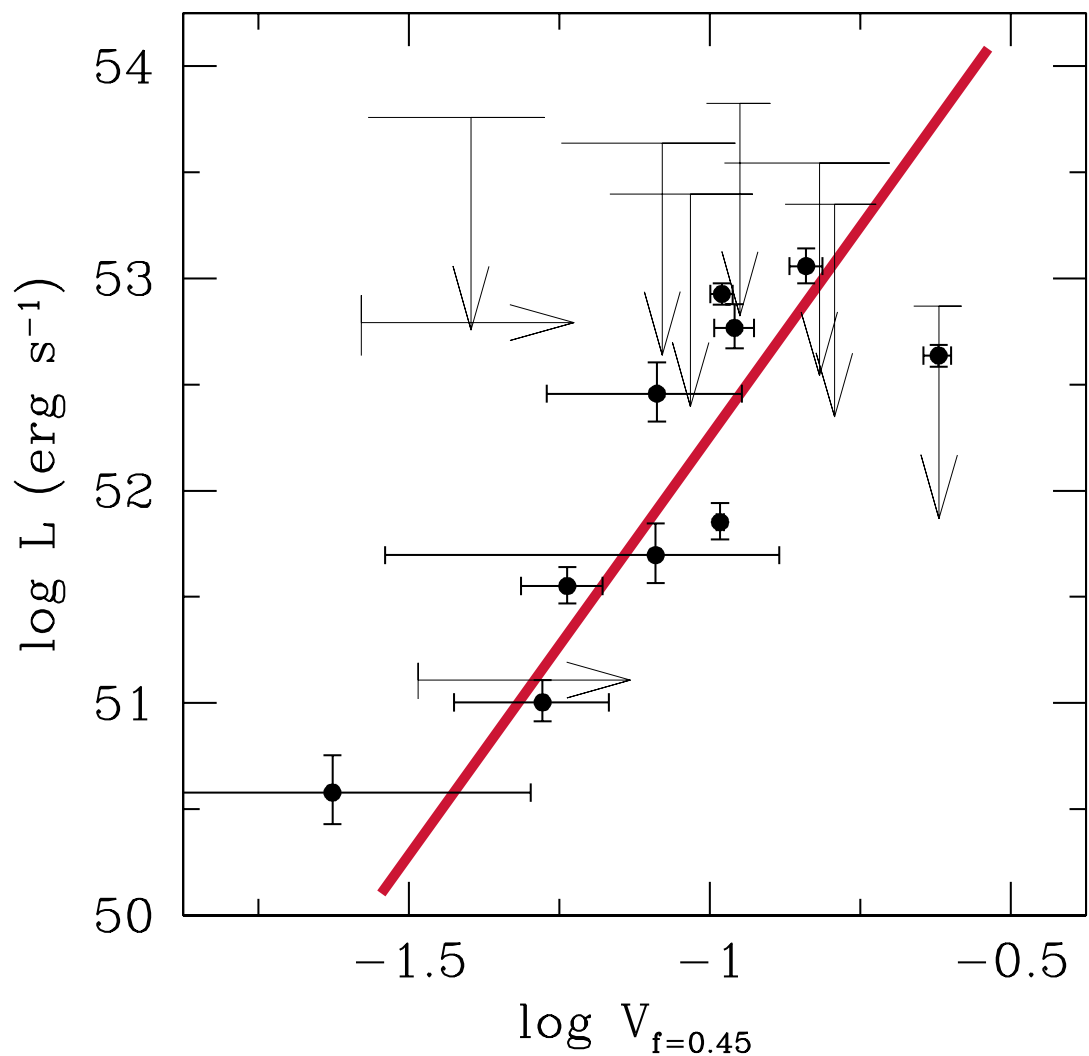, width=16cm}
%\end{center}
\caption{The luminosity--variability correlation. The data analysis is 
from
Reichart et al.~(2001). The line is the prediction of Eq.~(\ref{epilum}),
applied to complete GRBs, rather than single peaks.}
\label{figLV}
\end{figure}

\clearpage

\begin{figure}
\vspace*{-16cm}
\hskip -.5truecm
%\begin{center}
\epsfig{file=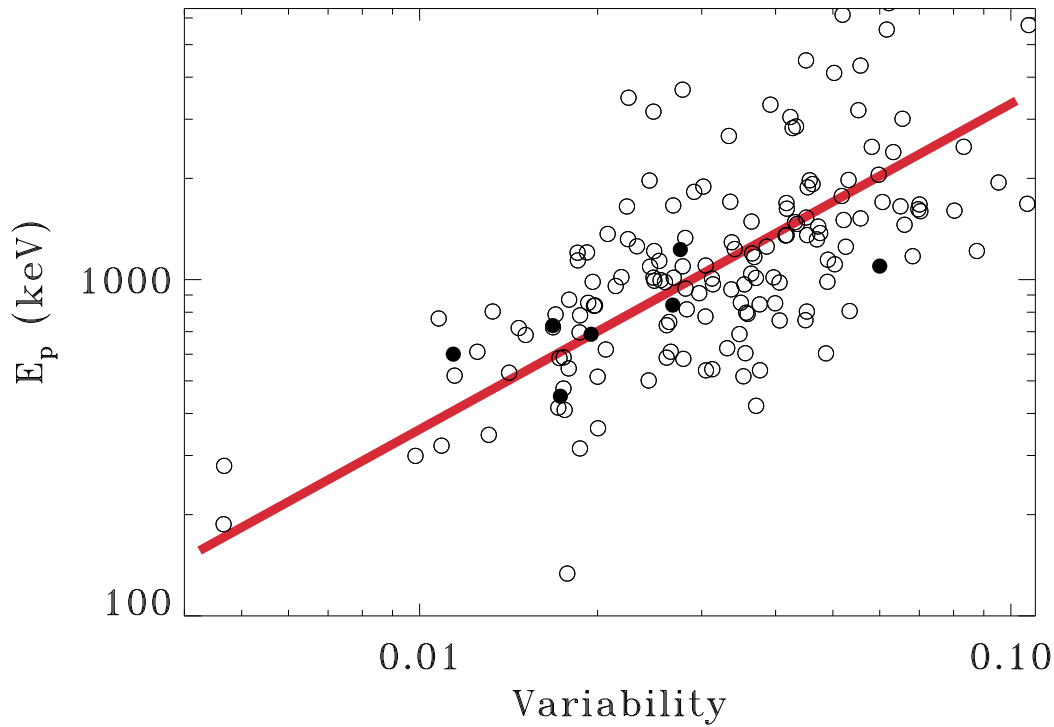, width=16cm}
%\end{center}
\caption{The peak energy--variability correlation.  The data analysis
is from
Ramirez-Ruiz \& Lloyd-Roning (2002), the filled circles are GRBs of
known redshift. The line is the prediction of Eq.~(\ref{epwidth}),
for complete GRBs, as opposed to individual pulses.}
\label{figEpVar}
\end{figure}

\clearpage

\begin{figure}[]
\vskip 1cm
\begin{center}
\epsfig{file=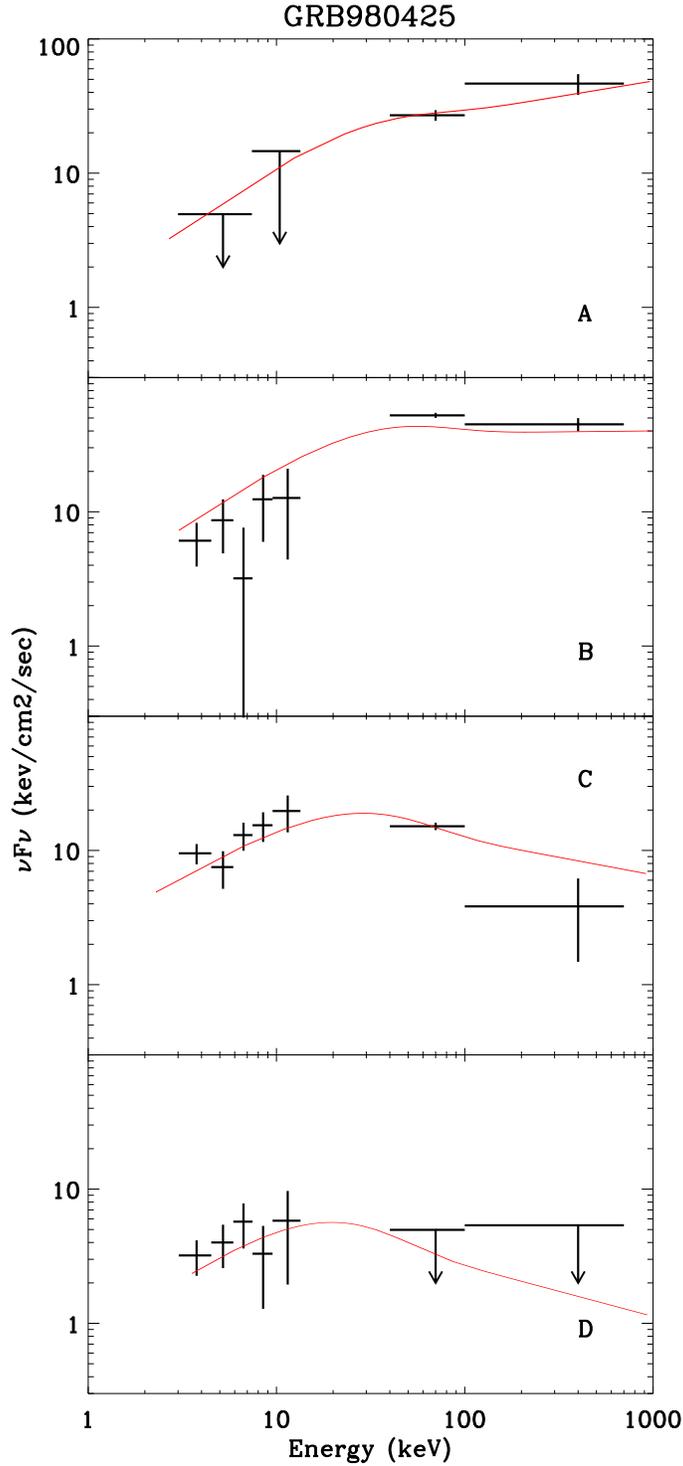, width=9cm}
%\vspace{-3cm}
\end{center}
 \caption{The distributions $\nu\,F_\nu\equiv E^2\,dN/dE$
 of GRB 980425 in the time intervals (in seconds after the onset
 of the burst) A = (0,5), B = (5,10),
 C = (10,30) and D = (30,50), measured with Beppo-Sax by Frontera et 
 al.~(2000).
 The lines are not fits, but ``descriptions'' (with all parameters but $t_c$
 at their reference values, and $\beta_2$ at the $\beta$ peak value, 2.3, as in 
Fig.~(\ref{figalphabeta}) made using Eqs.~(\ref{pevol}, \ref {dEdt1},
 \ref{Tevol},  \ref{dEdt}).} 
 \label{fig425spectra}
\end{figure}

\clearpage

\begin{figure}[]
\vskip -16cm
\hskip -3cm
%\begin{center}
\epsfig{file=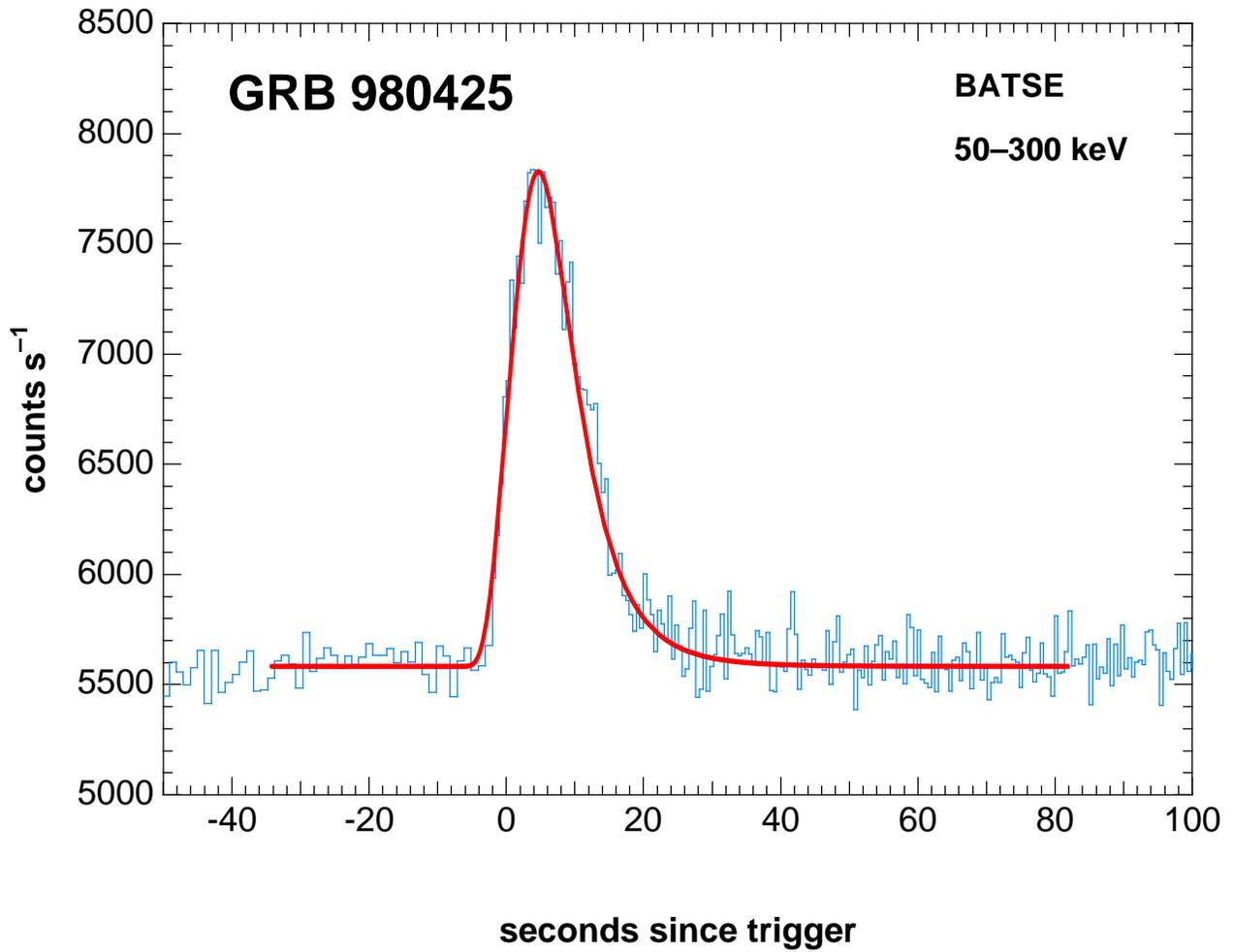, width=21cm}
\vspace{-1cm}
%\end{center}
 \caption{The single pulse in the light curve $dN/dt$ of GRB 980425,
 as seen by BATSE at energies in the 50--300 keV bracket (Kippen 1998) 
 and its description in the CB model.}
 \label{fig425}
\end{figure}

\clearpage

\begin{figure}
\vspace*{-3.5cm}
\hspace*{-1.9cm}
%\begin{center}
\epsfig{file=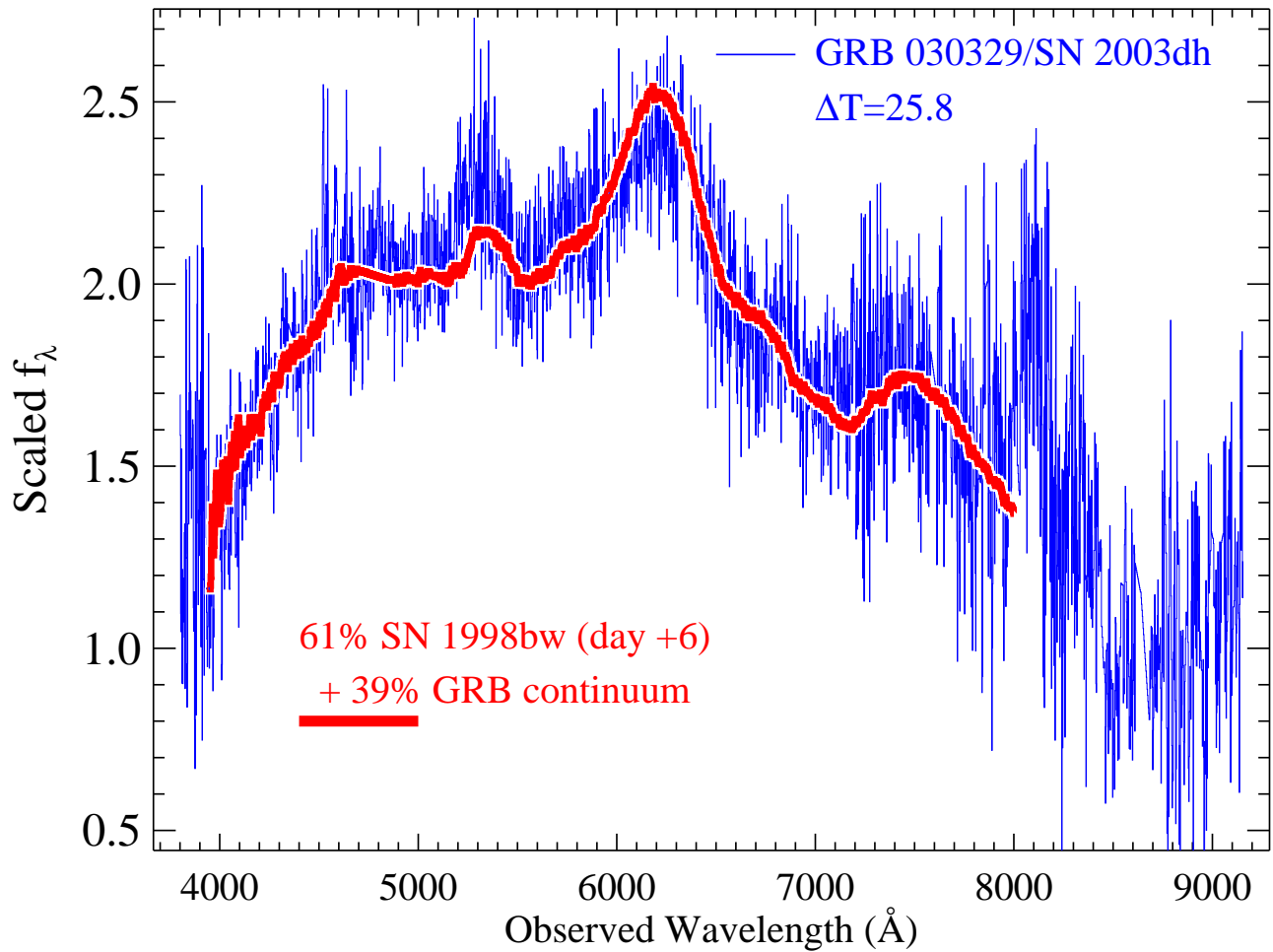, width=14cm,angle=90}
%\end{center}
\hskip .5truecm
\caption{Comparison of the optical spectra of SN2003dh and SN1998bw
(Matheson et al. 2003).}
\label{figSN2003dh}
\end{figure}

\clearpage

\begin{figure}[]
\vskip -18cm
%\hskip -2cm
%\begin{center}
\epsfig{file=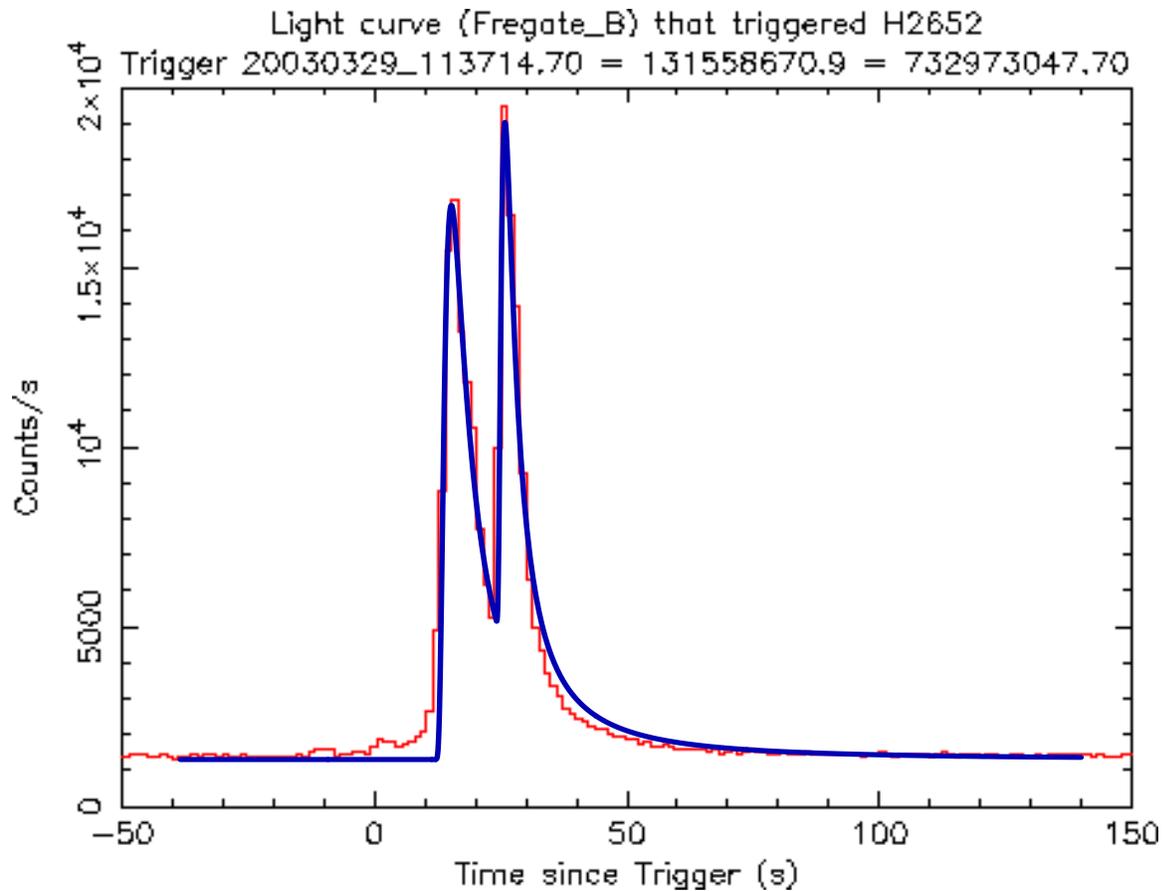, width=20cm}
%\end{center}
%\vspace{-8cm}
\caption{The $\gamma$-ray number count $dN/dt$ of GRB 030329,
as measured by HETE II (Vanderspek et al.~2003, 
http://space.mit.edu/HETE/Bursts/GRB030329) showing two dominant 
pulses, or CB contributions. The theoretical curve is not a 
full-fledged fit, but a ``description'' in terms of the naive pulse shape of 
Eq.~(\ref{naivepulse}), wherein only the pulses' heights, widths and 
relative delay have been adjusted.}
 \label{fig329NC}
\end{figure}

\clearpage

\begin{figure}[]
\vskip -.5cm
%\hskip -2cm
\begin{center}
\vbox{\epsfig{file=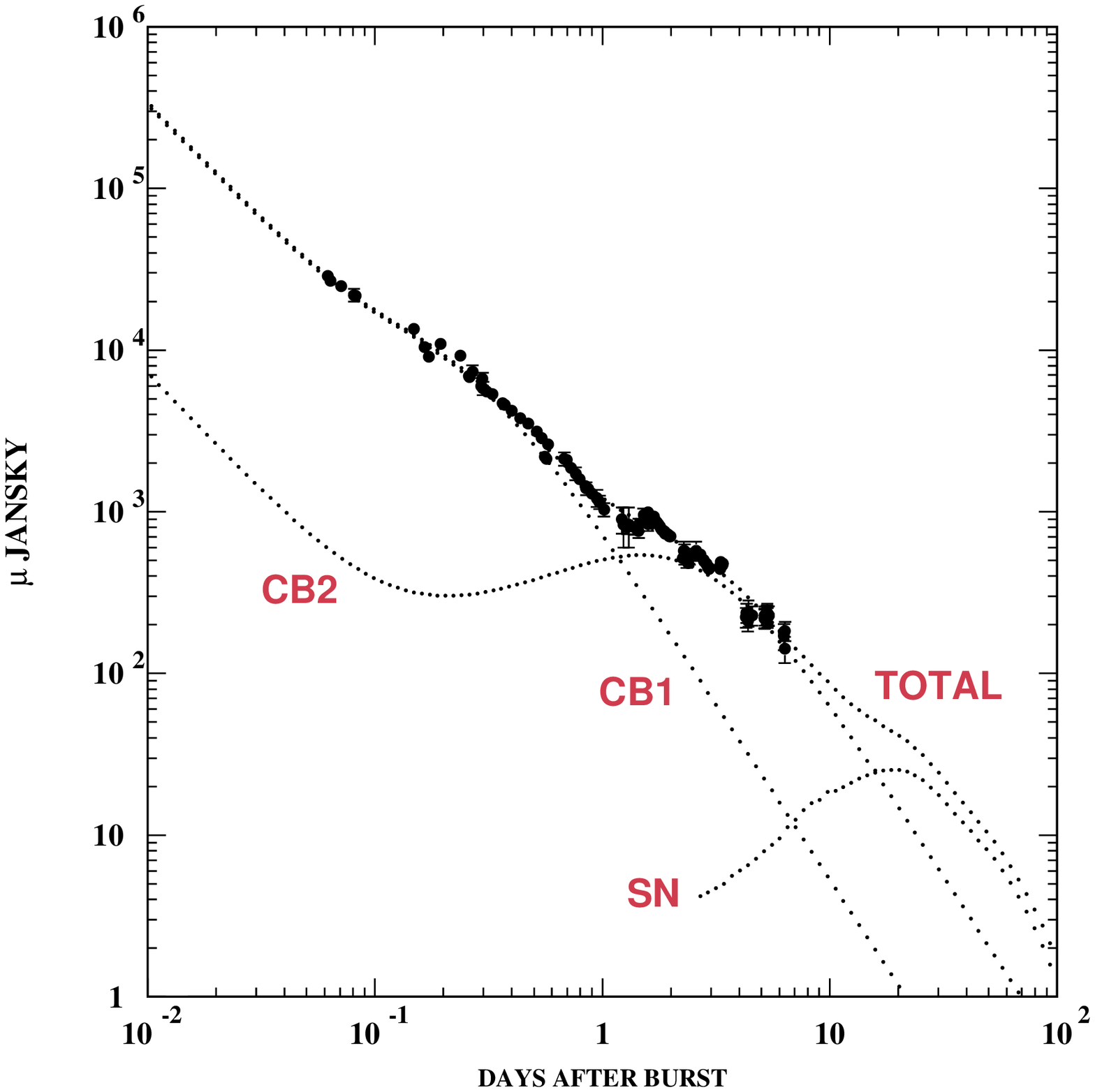,width=10.5cm}}
\vspace{-5.5cm}
\vbox{\epsfig{file=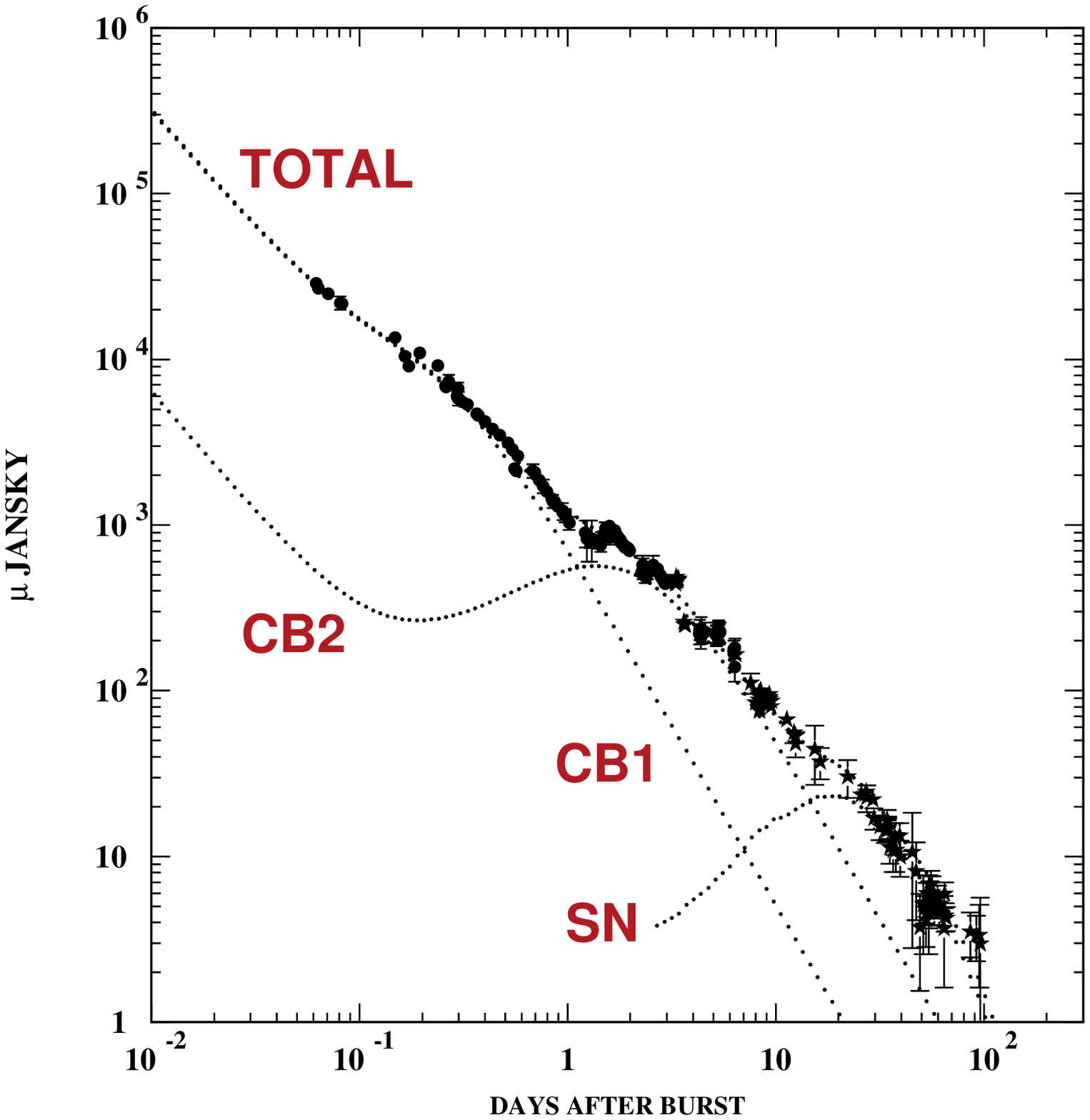,width=11cm}}
\vspace{-1.4cm}
\end{center}
%\vspace{-8cm}
\caption{Upper panel: The R-band AG of GRB 030329, 
used along with other optical data to
predict, in the CB model, the presence of a SN akin to SN1998bw
(Dado et al.~2003f). Lower panel:
The subsequent data (the $\star$ symbols) are added.}
 \label{fig329red}
\end{figure}

\clearpage

\begin{figure}
\vspace*{-7cm}
\hspace*{-10cm}
%\begin{center}
\epsfig{file=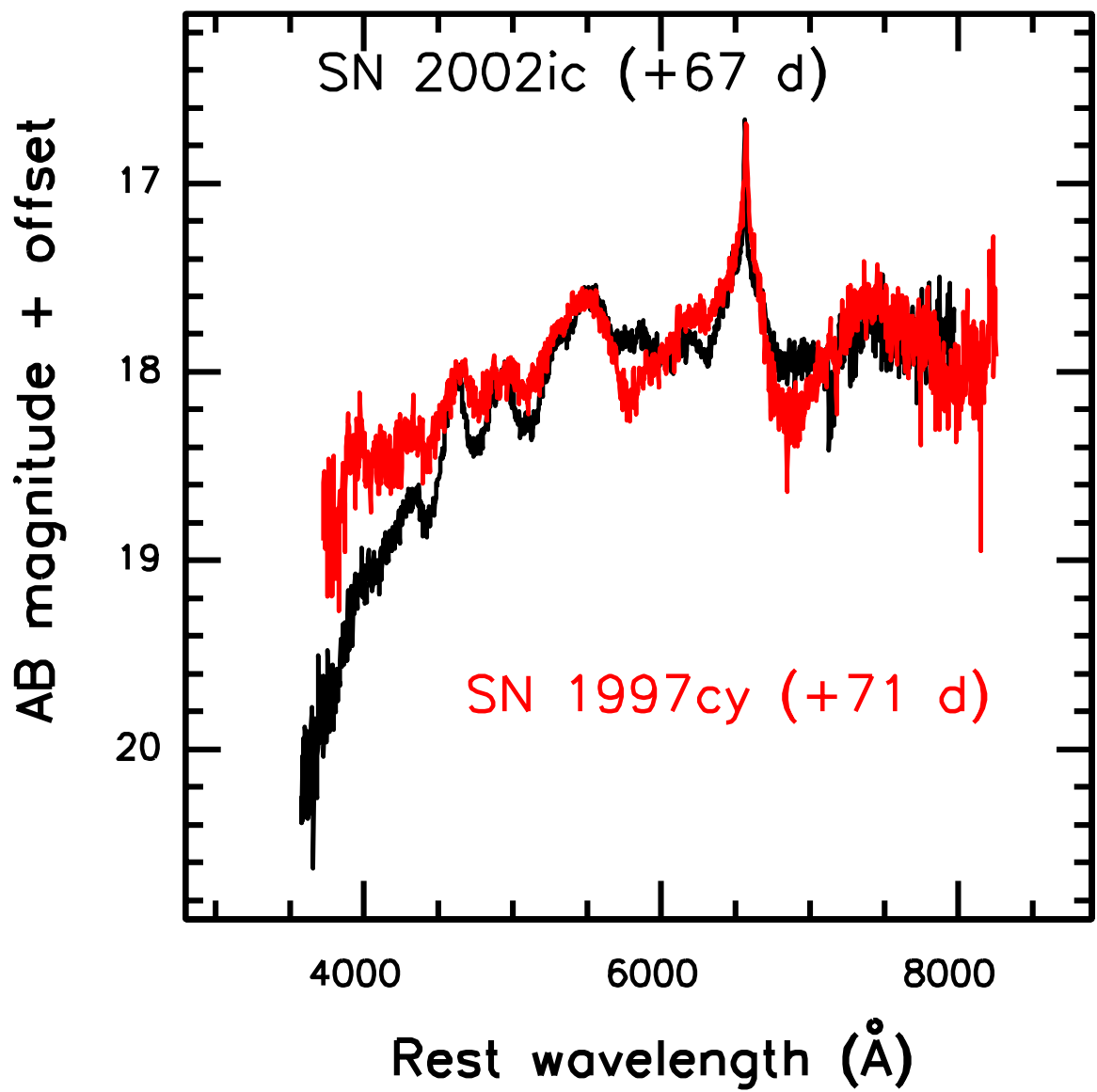, width=35cm}
%\end{center}
\vspace*{-10cm}
\caption{Comparison of the optical spectra of SN2002ic and SN1997cy 
(Hamuy et al.~2003).}
\label{figSN2002ic}
\end{figure}

\clearpage

\begin{figure}[]
%\vskip -12cm
%\hskip -1truecm
\vspace*{-9cm}
\hspace*{-2.2cm}
\begin{center}
\hspace*{-1.8cm}
\epsfig{file=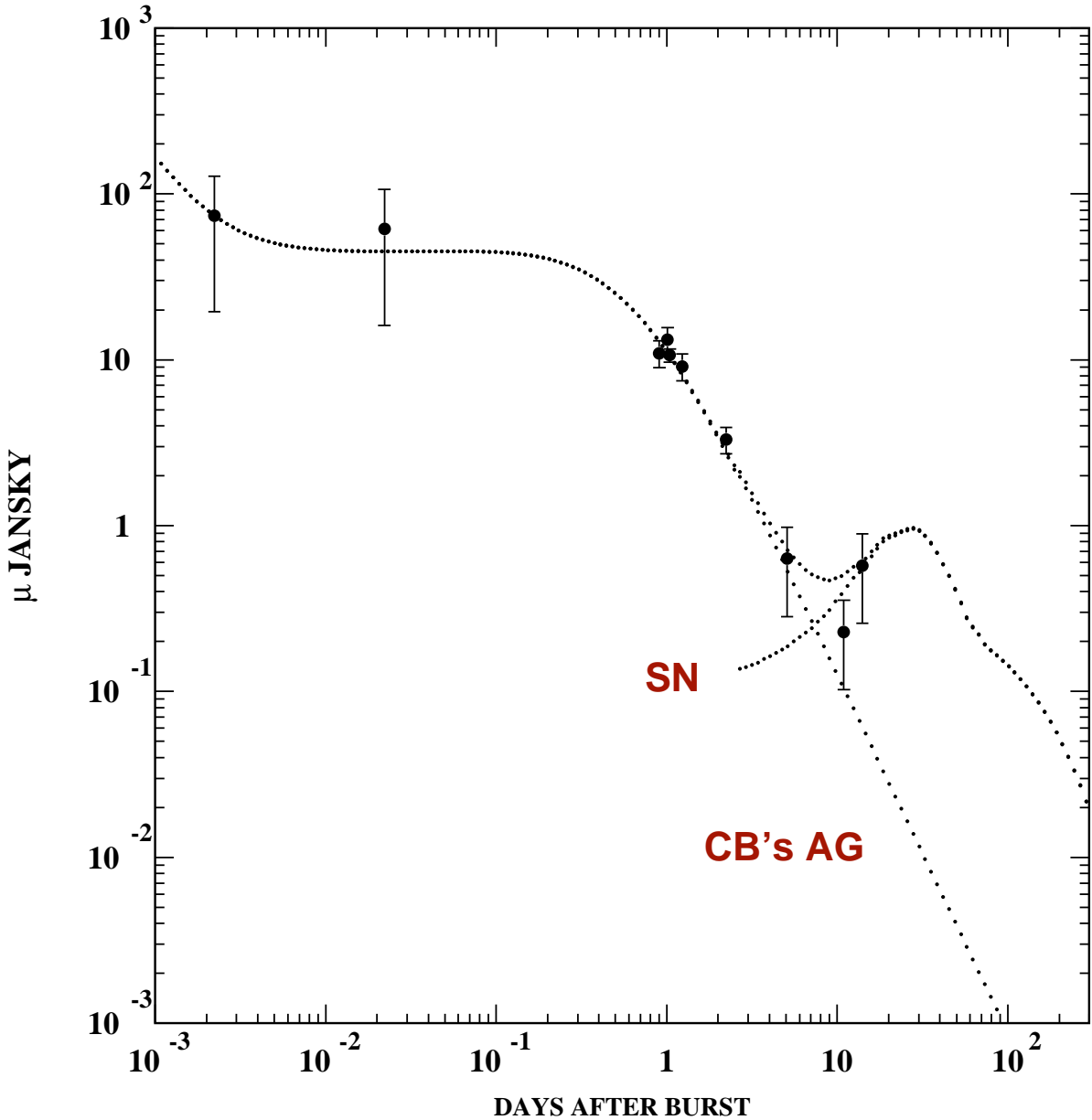, width=19cm}
\end{center}
\vspace*{- 1.5cm}
\caption{The optical AG of XRF 030723 (Prigozhin et al.~2003;
Fox et al.~2003b; Dullighan et al.~2003a,b; Smith et al.~2003; Bond et 
al.~2003) showing
a ``rebrightening'' 14 days after the XRF 
that may be due to the contribution of a SN
(Fynbo et al.~2003). The first two points have been deduced from the 
unfiltered measurements of  Smith 
et al.~(2003), assuming an early  $F_\nu\sim \nu^{-0.5}$ (Dado et al.~2003a). 
All errors were  multiplied 
by a factor 2 to account for cross-calibration uncertainties. 
The fit is a preview of
Dado et al.,~in preparation.
The redshift of the 1998bw-like SN contribution has been
adjusted to the normalization of the late-time points;
 without extinction corrections in the host galaxy or ours,
it corresponds to a redshift of $z\sim 0.75$.}
\label{figXRF}
\end{figure}

\clearpage

\begin{figure}[]
%\vskip -1.5cm
\hskip 2truecm
%\vspace*{0.3cm}
\vspace*{- .4cm}
%\hspace*{-0.2cm}
\begin{center}
\epsfig{file=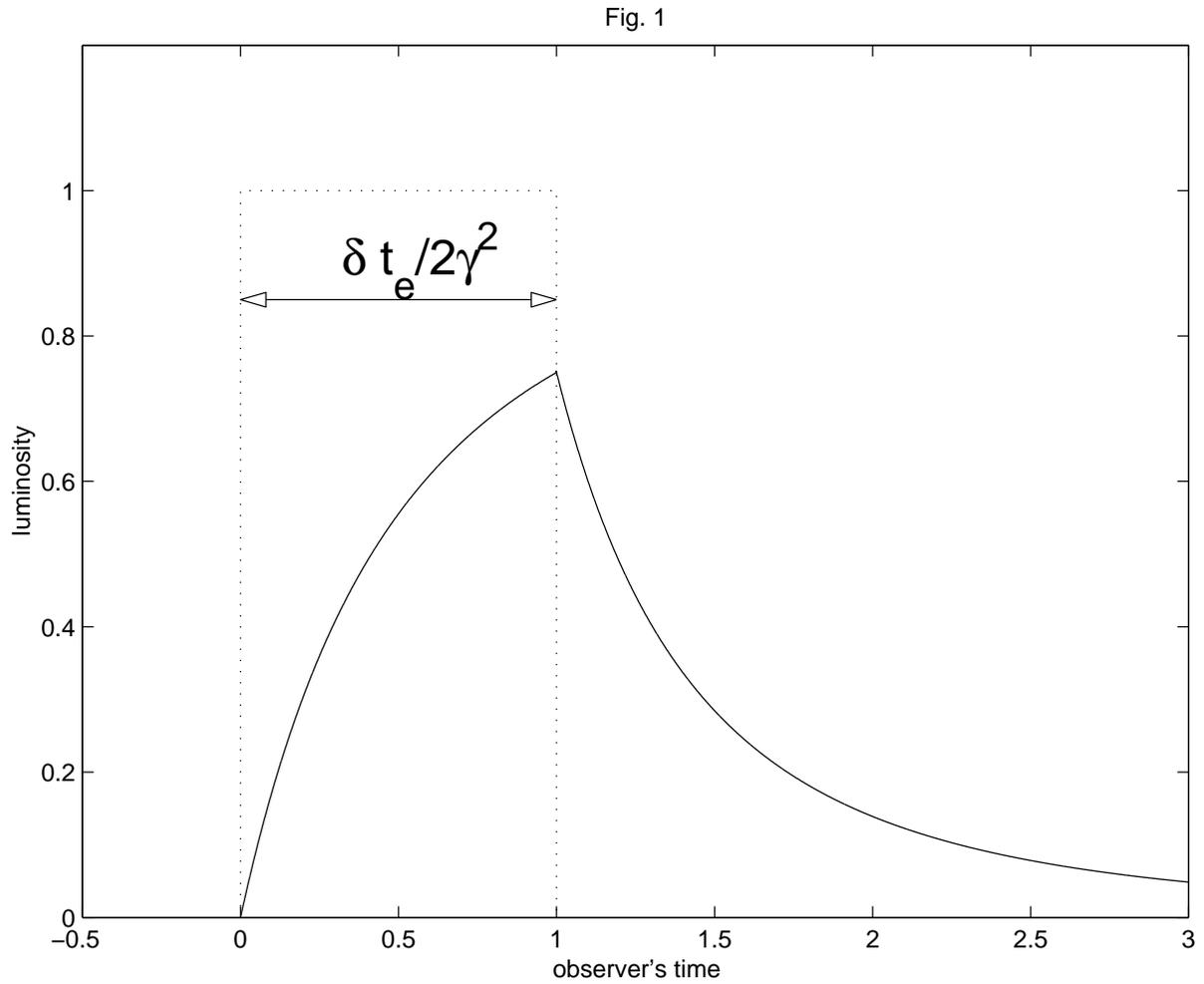, width=16cm}
\end{center}
\vspace{2cm}
\caption{The shape of the {\it luminosity}, $E\,dN/dt$, of a GRB in an
FB model (Kobayashi et al.~1997).}
\label{FMshape}
\end{figure}

\clearpage

\begin{figure}[]
%\vskip -12cm
%\hskip -1truecm
%\vspace*{-2cm}
\hspace*{-2.2cm}
\begin{center}
\hspace*{-.8cm}
\epsfig{file=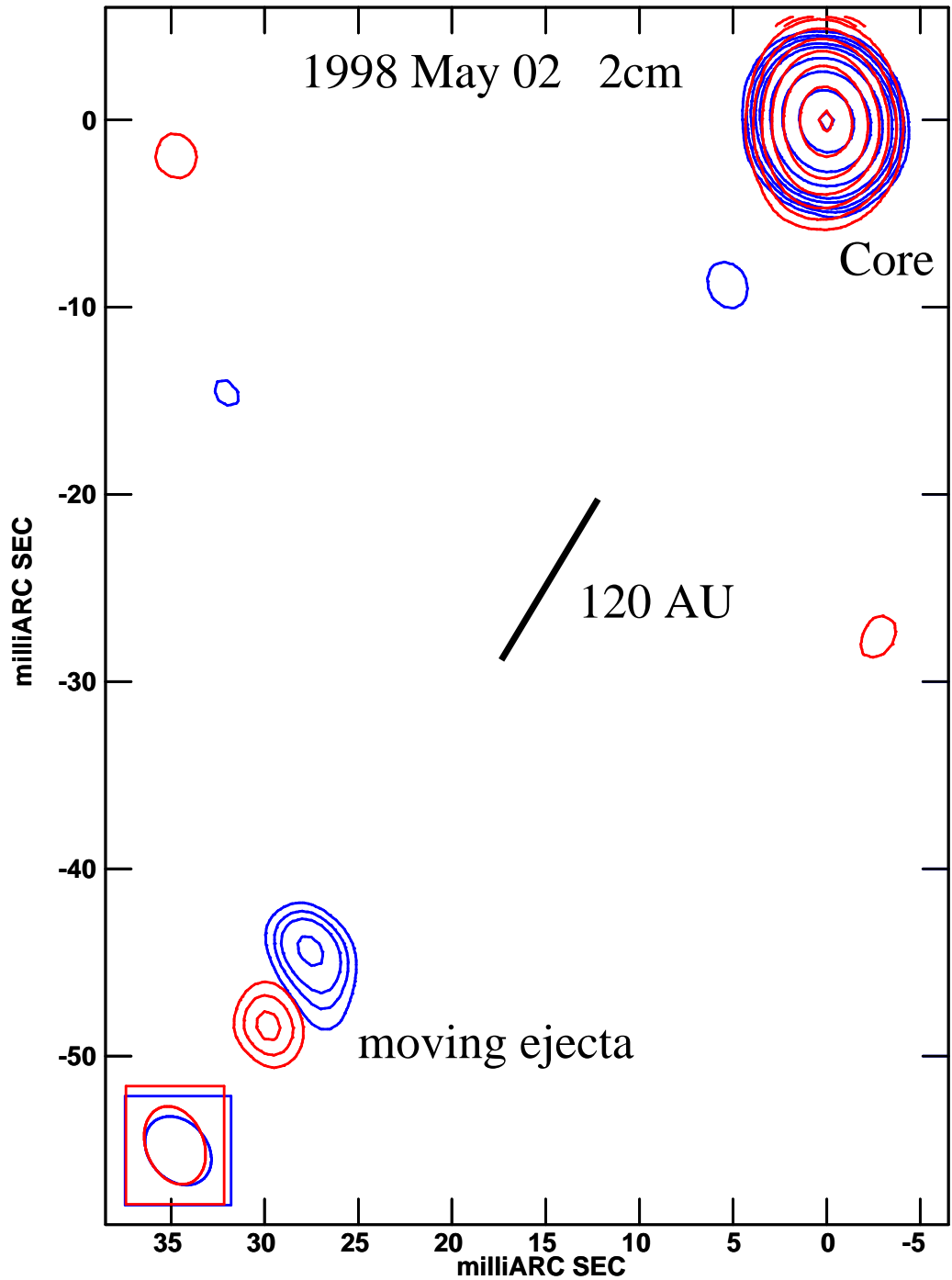, width=14cm}
\end{center}
%\vspace*{- 3cm}
\caption{The May 1998
VLBA 2-cm radio images of the microquasar GRS 1915+105
and the superluminal CB it ejected $67 \pm 7$ h before,
at 75 AU resolution. Contours are at $2$, 2.3, 4, 6, 
8, 16, 32, 64, and 96\% of the peak intensity. The blue and red contours show 
time-resolved images 4.5 h apart.}
\label{figFelix}
\end{figure}

\clearpage

\begin{figure}
\vspace*{-8cm}
\begin{center}
\epsfig{file=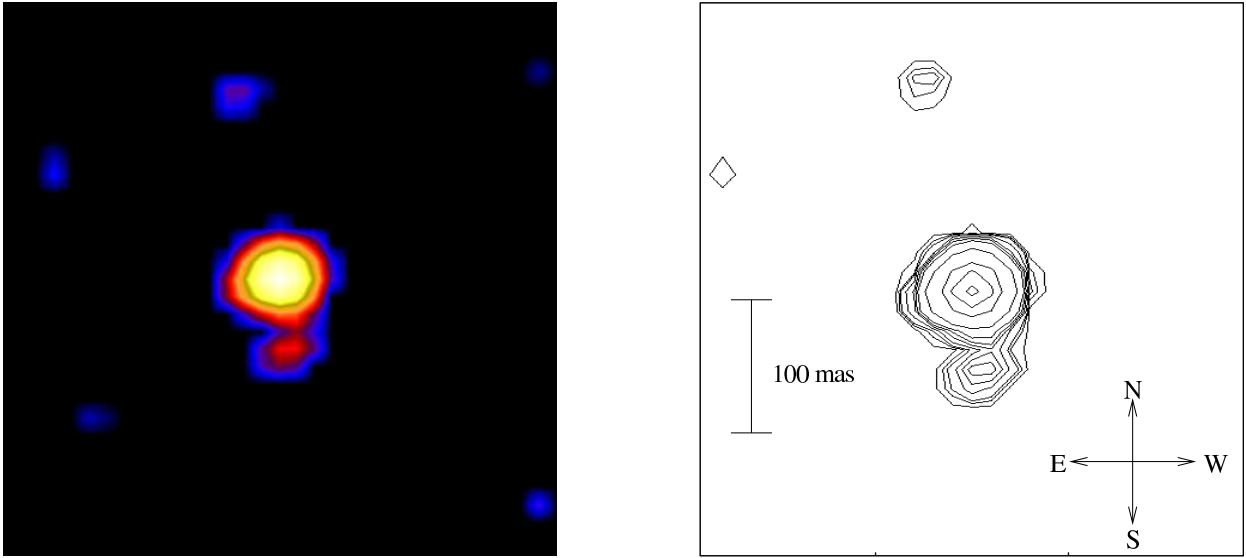, width=16.5cm}
\end{center}
\hskip .5truecm
\caption{The two CBs emitted by SN1987A in opposite axial
directions (Nisenson \& Papaliolios 2000). The northern and southern
bright spots are compatible with being jets of CBs emitted at
the time of the SN explosion and travelling at a velocity equal,
within errors, to $c$. One of the {\it apparent} velocities is superluminal.
The corresponding GRBs were not pointing in our direction, which
may have been a blessing (Dar \& De R\'ujula 2002).}
\label{figCostas}
\end{figure}

\clearpage

\begin{figure}[]
\begin{center}
\vspace{-9.0cm}
\vbox{\epsfig{file=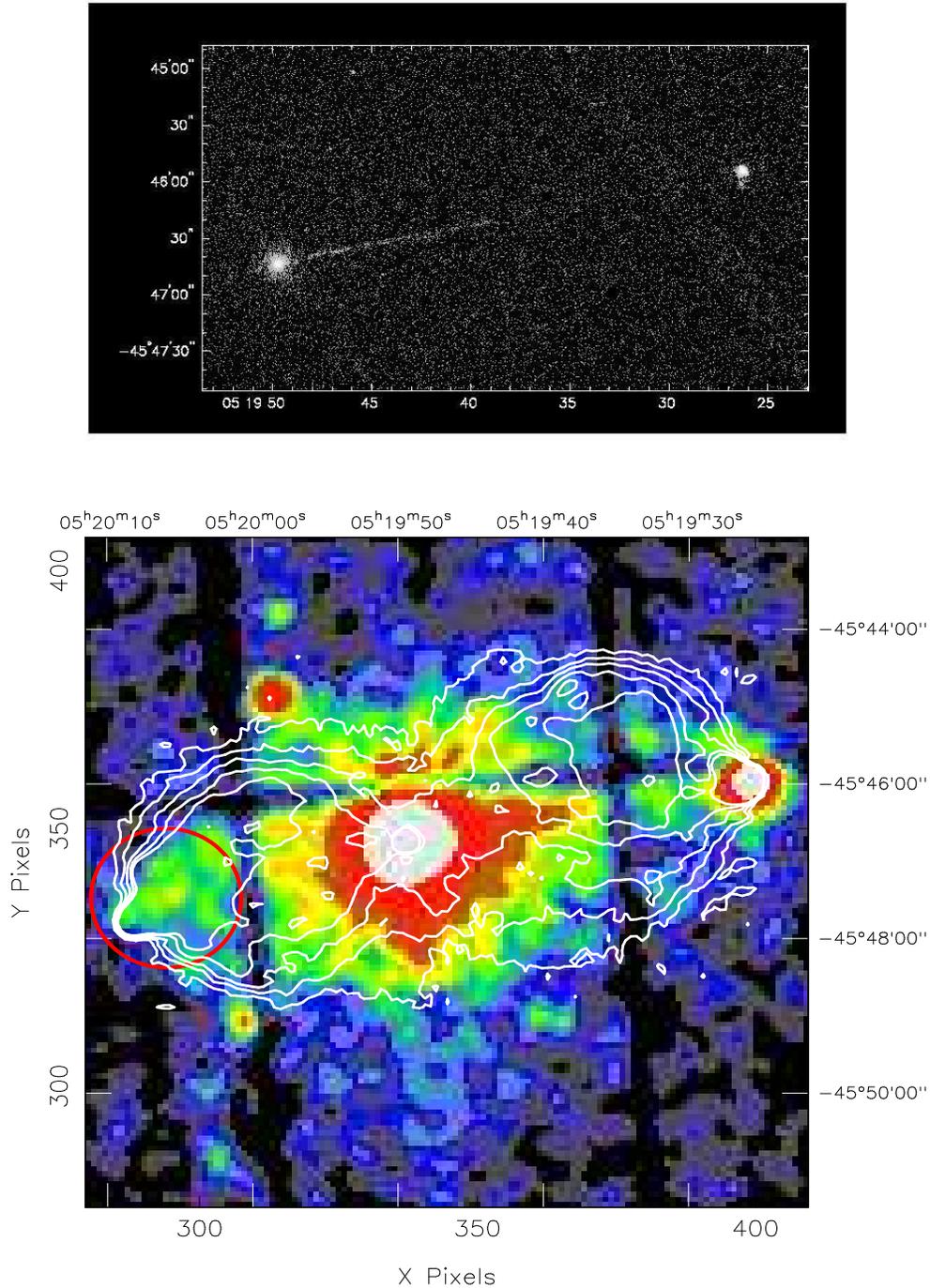,height=17cm,width=12cm}}
\vspace{-.5cm}
\vbox{\epsfig{file=PictorRadio1.ps,height=13cm,width=11.cm,
angle=-90}}
%\vspace{-1cm}
\end{center}
 \caption{Upper panel: Chandra X-ray image of the radio galaxy 
Pictor A (Wilson, Young  \& Shopbell 2001), showing a 
non-expanding jet that emanates from the centre
of the galaxy and extends
across 360 thousand light years towards a brilliant hot spot at least 800
thousand light years away from where the jet originates. Lower panel: 
XMM/p-n image of Pictor A in the 0.2--12 keV energy interval,
centred at the position of the leftmost spot in the upper panel, and
superimposed on the radio
contours from a 1.4 GHz radio VLA map (Grandi et al.~2003).}
 \label{Pictor}
\end{figure}

\clearpage

\begin{figure}[]
\begin{center}
\epsfig{file=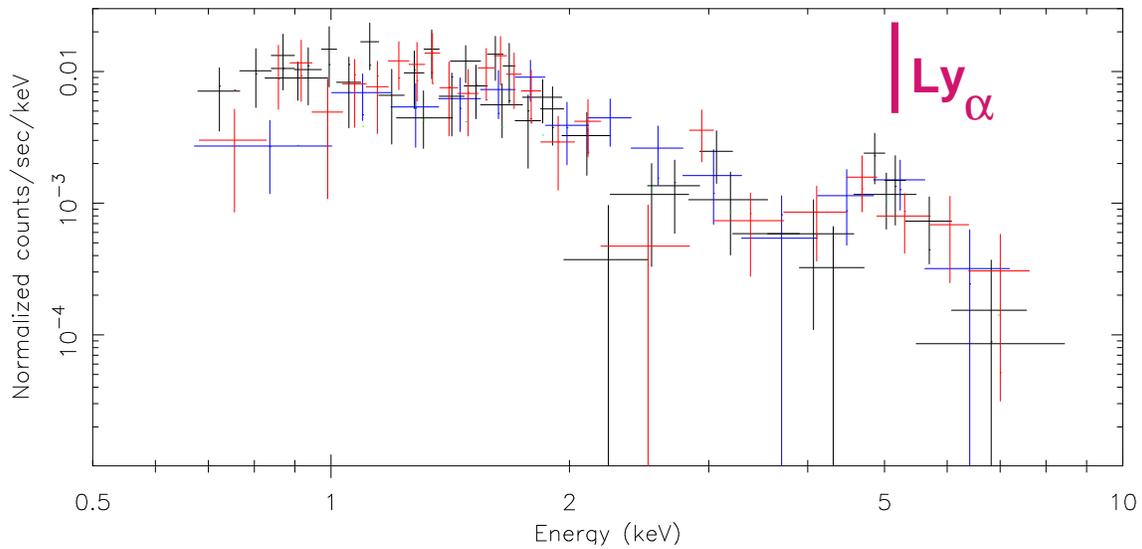, width=15cm}
\vspace{2cm}
\end{center}
 \caption{{The X-ray spectrum of GRB 970828 in the 
intermediate time-period
in which a putative line feature was observed (Yoshida et al.~2001).
The vertical line is at the position (at that time)
{\it predicted} in the CB model for the highly Doppler-boosted Hydrogen
$\rm Ly\alpha$ transition (Dado et al.~2003b).}}
 \label{figXray}
\end{figure}

%\newpage

%\newpage

\end{document}